\documentclass[onecolumn,authoryear]{els-mrw}

\usepackage{amsmath,amssymb,amsfonts,amsthm,makeidx,graphicx}
\usepackage{txfonts}
\usepackage{helvet}
\usepackage{braket}
\usepackage{environ}
\usepackage{bbm}
\usepackage{isotope}

\newcommand{\sic}{\textit{sic}}

\newcommand{\ii}{\mathrm{i}}
\newcommand{\ee}{\mathrm{e}}
\newcommand{\dd}{\mathrm{d}}

\newcommand{\abs}[1]{|#1|}

\renewcommand{\vec}[1]{\boldsymbol{#1}}

\newcommand{\Id}{\mathbbm{1}}
\newcommand{\OO}{\mathcal{O}}

\NewEnviron{subalign}[1][]{%
\begin{subequations}\begin{align}
  \BODY
\end{align}\label{#1}\end{subequations}
}

\NewEnviron{spliteq}{%
\begin{equation}\begin{split}
  \BODY
\end{split}\end{equation}
}

\newcommand{\cphase}[1]{\ee^{{#1}\ii\theta}}

\begin{document}

\chapter{What is a resonance? And why does it matter?}

\author[1]{Sebastian König}

\author[2,3]{Kévin Fossez}

\author[4]{Rimantas Lazauskas}

\address[1]{
 \orgdiv{Department of Physics and Astronomy},
 \orgname{North Carolina State University},
 \orgaddress{Raleigh, NC 27695, USA}
}
\address[2]{
 \orgname{Florida State University},
 \orgdiv{Department of Physics},
 \orgaddress{Tallahassee, FL 32306, USA}
}

\address[3]{
 \orgname{Argonne National Laboratory},
 \orgdiv{Physics Division},
 \orgaddress{Lemont, IL 60439, USA}
}

\address[4]{
 \orgname{IPHC, CNRS/IN2P3, Université de Strasbourg},
 \orgaddress{23 rue de Loess, 67037 Strasbourg, France}
}

\maketitle

\begin{abstract}
The resonance phenomenon is of central importance in many areas of physics,
with particular significance in the study of nuclear structure and reactions.
Starting from the classical framework of damped driven oscillations, this
text introduces and analyzes quantum-mechanical resonances in a pedagogical and
systematic fashion, with emphasis on applications in nuclear physics.
Building on the formal theory of resonances, the text elucidates the
relationship between experimental observations, phenomenological insights, and
computational methods used to characterize and describe resonant states.
The discussion encompasses the diverse manifestations of nuclear resonances,
ranging from few- to many-body systems, all the way to collective phenomena and
to exotic systems that appear near the limits of nuclear stability.
References to the relevant literature are provided to assist readers
who wish to explore specific topics in more depth.
\end{abstract}

\section{Introduction}
\label{sec:Intro}

Resonances are  ubiquitous in nuclear physics and, more generally, throughout
physics.
When resonances are discussed in a nuclear-physics context, the term implicitly
refers to the \emph{quantum-mechanical} resonance phenomenon.
Before analyzing this quantum-mechanical case in detail in the present chapter,
it is instructive to recapitulate the corresponding concept in a purely
classical context.

A classic (\sic) example of a classical resonance is provided by a driven
(or forced) harmonic oscillator.
Systems of this type occur in a wide variety of physical situations, and most
readers will be familiar with at least one everyday realization, such as a
playground swing.\footnote{
Describing the person-swing system as a harmonic oscillator is, of course, an
idealization.}
In this particular case, the damping is typically sufficiently large that --
fortunately, from a safety standpoint -- it is difficult to reach a regime of
strongly pronounced resonant behavior.
For this reason, we set aside this intuitive but quantitatively less
controllable example and instead consider a more idealized and controlled setup
in order to recall the equations that govern classical resonance phenomena.

\subsection{The damped driven oscillator}
\label{sec:DampedDrivenOsc}

\begin{figure}
 \centering
\includegraphics[width=0.6\textwidth]{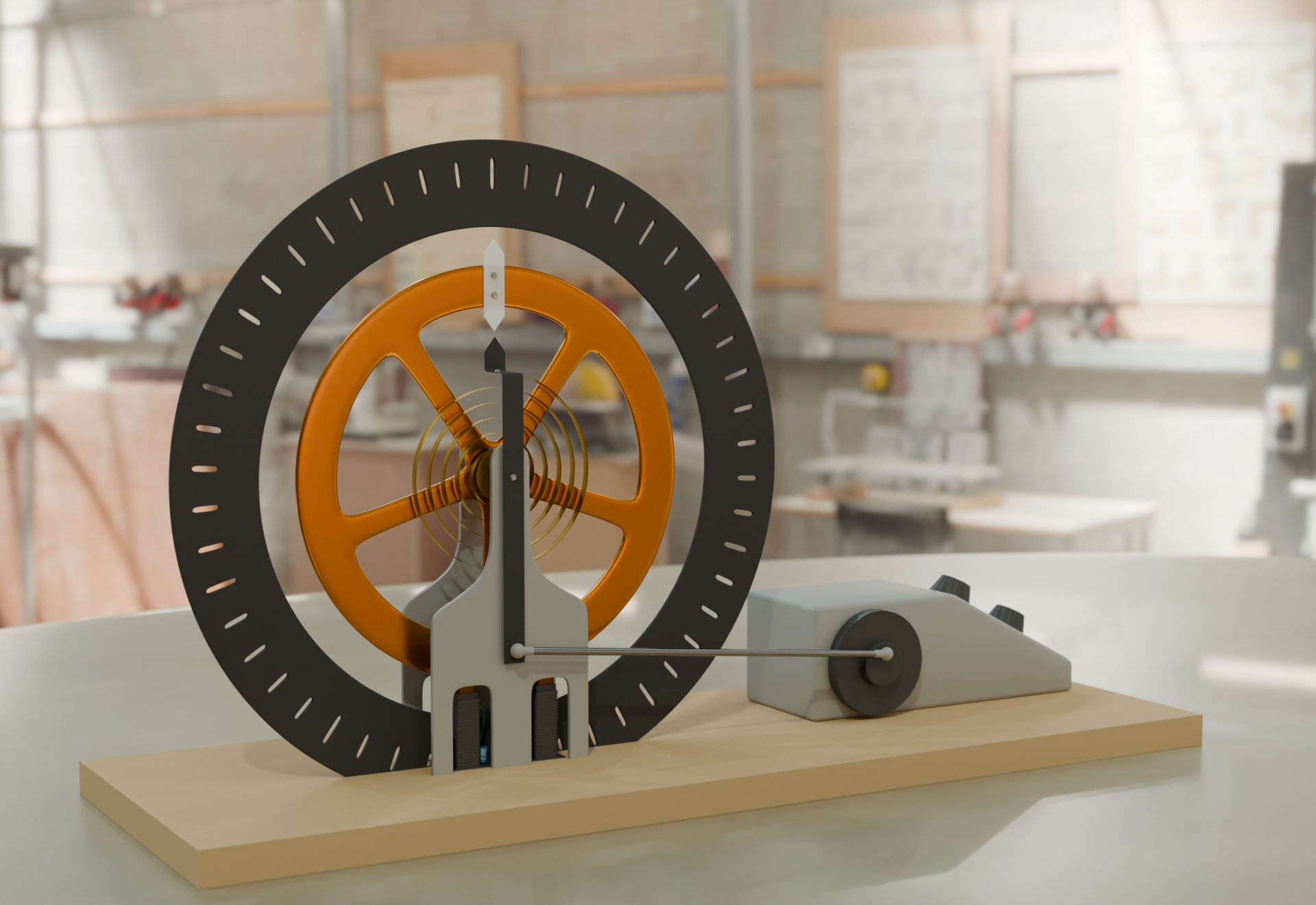}
\caption{%
 A torsion pendulum, with a motor to drive the system
 and eddy current brakes to control the damping~\citep{Pohl:2017}.
\label{fig:TorsionPendulum}
}
\end{figure}

The damped driven oscillator is one of the most important topics in classical
mechanics, as it can be realized in many different ways.
One somewhat construed, yet pedagogically very valuable, realization is the
driven torsion pendulum shown schematically in
Fig.~\ref{fig:TorsionPendulum}.
A flat copper wheel is mounted to an axle, to which a spiral spring is attached.
The wheel can rotate around the axle, with the spring providing a restoring
force.
To provide controlled damping, the wheel rotates through a variable
eddy-current brake installed at the bottom.
In addition, a motor with a lever is installed to provide a sinusoidal driving
force, with adjustable frequency.

If $\theta(t)$ denotes the angular displacement from the equilibrium position
($\theta = 0$), the equation of motion, in normalized form, can be written as
\begin{equation}
 \ddot{\theta}(t) + 2\gamma\dot{\theta}(t) + \omega_0^2 \theta(t)
 = \phi(t) \,,
\label{eq:dd-osc}
\end{equation}
where $\phi(t)$ is the driving force.
In this case, $\omega_0^2$ is determined by the ratio of the torsional constant
of the spring and the moment of inertia of the wheel, while $\gamma$
parametrizes the damping force relative to the wheel's moment of inertia.
Importantly, Eq.~\eqref{eq:dd-osc} governs \emph{any} damped driven oscillator,
so we could replace $\theta$ with some generic coordinate $q(t)$, but for
concreteness we keep the torsion pendulum in mind and stick with $\theta(t)$.
Otherwise, we will follow \citet{Nussenzveig:1972} to look at the
formal properties of the equation and to discuss how they relate to the motion
of the system.

Let us start with some simple observations.
In the absence of a driving term ($\phi\equiv0$), Eq.~\eqref{eq:dd-osc}, as a
second-order ordinary differential equation, has two linearly independent
solutions, and any general solution is given by a superposition of these.
Among the infinitely many ways to pick a basis pair of solutions, the
canonical choice
\begin{equation}
 \theta(t) = \theta_A \exp({-}\ii\omega_1 t) + \theta_B \exp({-}\ii\omega_2 t)
\label{eq:dd-osc-sol}
\end{equation}
stands out because it exposes that the motion is governed by two
frequencies, given as the solutions of
\begin{equation}
 \omega^2 + 2\ii\gamma\omega - \omega_0^2 = 0 \,.
\end{equation}
For $\gamma > 0$, these are given by the complex values
\begin{equation}
 \omega_{1,2} = \pm\sqrt{\omega_0^2 - \gamma^2} - \ii\gamma \,,
\label{eq:dd-osc-freqs}
\end{equation}
and inserting this into Eq.~\eqref{eq:dd-osc-sol} exhibits the damping: the
amplitude of the general solution will decay like $\exp({-}\gamma t)$ over time.
For $\gamma<0$, the system would be unstable, while for $\gamma = 0$, $\omega_1
= \omega_2 = \omega_0$, where $\omega_0$ is the \emph{natural frequency} of
the oscillator, and we get simple harmonic motion with constant
amplitude.

\begin{figure}
 \centering
 \begin{minipage}{0.5\textwidth}
  \centering
  \includegraphics[width=0.9\textwidth]{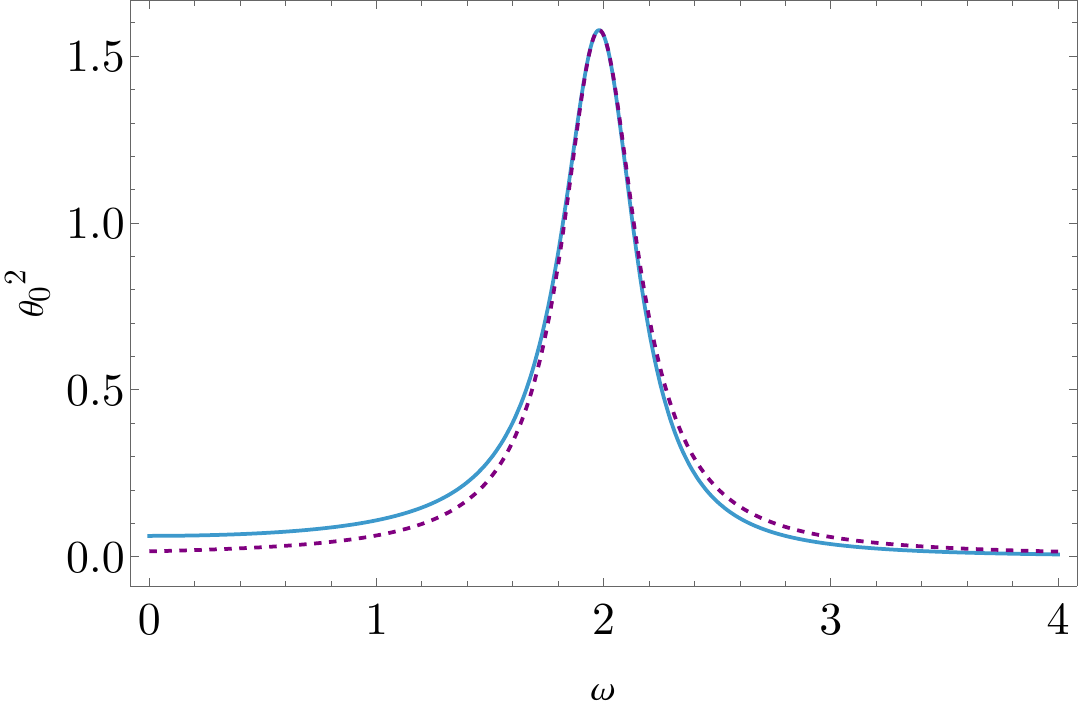}
 \end{minipage}\begin{minipage}{0.5\textwidth}
  \centering
  \includegraphics[width=0.9\textwidth]{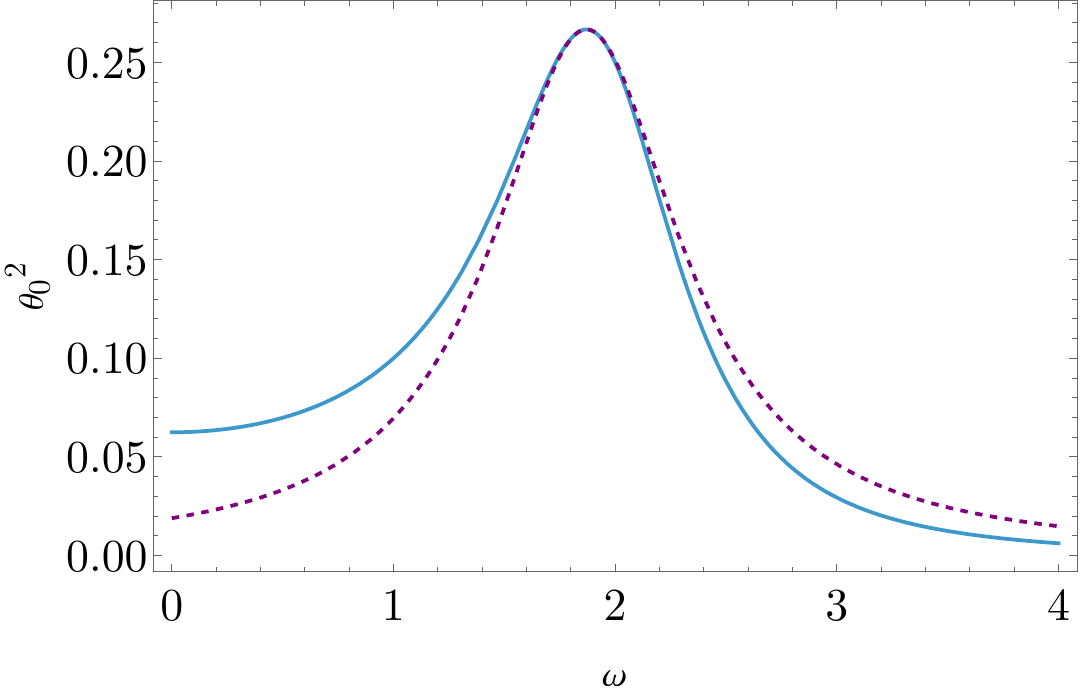}
 \end{minipage}
\caption{%
 Resonance curves for weak damping ($\gamma = 0.1\omega_0$, left panel) and
 stronger damping ($\gamma = 0.5\omega_0$, right panel).
 The solid lines represent the exact shape as given in Eq.~\eqref{eq:res-curve},
 while the dashed lines show the Lorentzian approximation,
 Eq.~\eqref{eq:res-curve-alt}.
\label{fig:ResonanceCurves}
}
\end{figure}

The resonance phenomenon can be observed most directly when the system is driven
harmonically, i.e., for
\begin{equation}
 \phi(t) = \phi_0 \exp({-}\ii\omega t) \,,
\label{eq:phi-harmonic}
\end{equation}
where the complex exponential form is chosen purely for mathematical convenience
and it is understood that in the actual experiment one would have a real
sinusoidal driving force.
In this case, for large times $t$ the system will simply follow the driving
term,
\begin{equation}
 \theta(t) = \theta_0(\omega) \exp({-}\ii\omega t) \,,
\end{equation}
but, importantly, the amplitude $\theta(\omega)$ with which it does this depends
on the driving frequency $\omega$.
It is an elementary textbook fact (see for example \citep{Taylor:2004}) that
this functional dependence is given by
\begin{equation}
 \theta_0(\omega)^2
 = \frac{\phi_0^2}{(\omega_0^2-\omega^2)^2+4\gamma^2\omega^2} \,.
\label{eq:res-curve}
\end{equation}
This famous \emph{resonance curve} assumes its maximum when the driving
frequency $\omega$ is such that the denominator in Eq.~\eqref{eq:res-curve} is
minimal, i.e., for $\omega_{\text{res}} = \sqrt{\omega_0^2-2\gamma^2}$.
For this driving frequency one says that the system is ``at resonance,'' and
$\omega_{\text{res}}$ is referred to as the \emph{resonance frequency}.
By Taylor-expanding the denominator around the minimum ($\omega =
\omega_{\text{res}}$), we can write
\begin{equation}
 \theta_0(\omega)^2
 \approx \frac{\phi_0^2/(4\omega_{\text{res}}^2)}{
  (\omega-\omega_{\text{res}})^2
  + \gamma^2\dfrac{\omega_{\text{res}}^2 + \gamma^2}{\omega_{\text{res}}^2}
 } \,.
\label{eq:res-curve-alt}
\end{equation}
This curve shape is referred to as a \emph{Lorentzian}.
The difference between the Lorentzian approximation and the exact form is
illustrated in Fig.~\ref{fig:ResonanceCurves}, for both weak damping (left)
and stronger damping (right).

In the case of negligible damping ($\gamma\ll\omega_0$), the resonance frequency
is simply the \emph{natural frequency} $\omega_0$ of the undamped oscillator,
and the resonance curve simplifies further since $\omega_{\text{res}} \approx
\omega_0$.

Let us now study this more formally and carefully.
Firstly, we can note that the denominator in Eq.~\eqref{eq:res-curve} has two
roots, which are simply the two complex ``frequencies'' $\omega_{1,2}$ given in
Eq.~\eqref{eq:dd-osc-freqs}.
These root gives rise to \emph{poles} in $\theta_0(\omega)$ when it is
regarded, by analytic continuation, as a function of complex $\omega$, and the
pole with positive real part, $\omega_1$, is responsible for the maximum
of the amplitude at $\omega_{\text{res}}$.
To distinguish these quantities, let us write
\begin{equation}
 \omega_1 \equiv \tilde{\omega}_{\text{res}} - \ii\gamma
 \quad,\quad \tilde{\omega}_{\text{res}} = \sqrt{\omega_0^2-\gamma^2} \,.
\end{equation}

\citet{Nussenzveig:1972} points out that the poles are in fact
features of the \emph{Green's function}
\begin{equation}
 G(\omega) = {-}\frac{1}{(\omega - \omega_1)(\omega - \omega_2)} \,,
\end{equation}
in terms of which $\theta_0(\omega) = G(\omega) \Phi(\omega)$ \,, where
$\Phi(\omega)$ is the Fourier transform of any generic driving force $\phi(t)$,
not necessarily of the simple harmonic form~\eqref{eq:phi-harmonic}.
In the time domain, the (inverse) Fourier transform $g(t)$ of $G(\omega)$ can be
used to calculate the solution $\theta(t)$ for given $\phi(t)$:
\begin{equation}
 \theta(t) = \int_{{-}\infty}^t \dd t' \ee^{{-}\gamma(t-t')}
 \frac{\sin\left[\tilde{\omega}_{\text{res}}(t-t')\right]}
 {\tilde{\omega}_{\text{res}}} \phi(t') \,.
\end{equation}
What we have therefore found is that ultimately the complex frequencies
$\omega_{1,2} = \pm\tilde{\omega}_{\text{res}}-\ii\gamma$ fully encode the
system's response to an arbitrary driving force, and that in particular an
instantaneous pulse will excite an oscillation at the oscillator's natural
frequency, with the amplitude ``decaying'' over a time scale $1/\gamma$.

\subsection{Quantum resonances}
\label{sec:QuantumResonances}

At this point we can establish the connection between the classical resonance
phenomenon and quantum resonances, both conceptually and formally.
On the conceptual level, a resonance of a quantum mechanical system can be
thought of as a ``state'' that can be excited by transferring energy to the system,
if this energy matches a natural resonance frequency of the system.
Just like in the case of the classical damped oscillator, this resonance state
will decay exponentially over time, characterized by a scale $\tau \sim
1/\Gamma$, where the resonance width $\Gamma$ is the direct analog of $\gamma$
in the classical system.
In fact, we can complete the analogy by describing the resonance overall with a
complex energy
\begin{equation}
 E = E_{\text{res}} - \ii\Gamma/2
 \equiv \hbar(\omega_{\text{res}} - \ii\gamma) \,,
\label{eq:E-res}
\end{equation}
where we have factored out $\hbar$ to expose the frequency interpretation even
more directly, and we set $\Gamma = 2\hbar\gamma$.
This factor of two (or $1/2$ in the imaginary part of $E$) is related to the
fact that on the classical side we are looking at how the \emph{amplitude}
decays over time, while in the quantum-mechanical regime it is more natural to
define $\Gamma$ based on the decay of the probability density.
Specifically, if we are looking at a stationary quantum state $\ket{\psi}$ with
energy as in Eq.~\eqref{eq:E-res}, then
\begin{equation}
 P(t) = \braket{\psi(t)|\psi(t)} \sim \ee^{{-}\Gamma t/\hbar}
\label{eq:P-res}
\end{equation}
since $\ket{\psi(t)} = \ee^{{-}\ii E t/\hbar} \ket{\psi}$.

If we consider the resonance being excited in a scattering process of two
quantum particles, where kinetic energy gets converted to form the metastable
state $\ket{\psi}$, then $\Gamma$ is the \emph{width}, in terms of energy, of a
corresponding peak in the cross section $\sigma$.
Specifically, in the simplest possible scenario (a single isolated resonance),
the cross section in the vicinity of the resonance energy assumes the famous
\emph{Breit-Wigner form}
\begin{equation}
 \sigma(E) \sim \frac{\Gamma^2}{(E-E_{\text{res}})^2 + \Gamma^2/4} \,.
\label{eq:sigma-BW}
\end{equation}
This looks similar to Eq.~\eqref{eq:res-curve-alt}, i.e., it is a Lorentzian
curve, and only the details regarding the width are a bit different.

Before we resolve this further, let us establish another important
connection.
Schematically, if our quantum system is described by a Hamiltonian $H = H_0 +
V$, with the free (kinetic) Hamiltonian $H_0$ and a potential $V$, it is
well-known from scattering theory that one can solve the Lippmann-Schwinger
equation,
\begin{equation}
 T(E) = V + V G_0(E) T(E) \quad,\quad G_0(E) = (E - H_0)^{{-}1} \,,
\end{equation}
and calculate the cross section $\sigma(E) \sim \abs{T(E)}^2$.
For a comprehensive discussion of scattering theory we recommend Taylor's
textbook~\citep{Taylor:1972}, while a compact summary (which we draw from in the
following) can be found in \citep{Gloeckle:1983}.

The $T$ matrix $T(E)$ is related to the $S$ matrix $S(E)$ via
\begin{equation}
 S(E) = 1 - 2\pi\ii T(E) \,,
\label{eq:S-T-E}
\end{equation}
where $E$ is understood to take any complex value as we are considering the
analytic continuation of these operators \emph{off-shell}, i.e., outside the
domain of real positive energies.
In terms of $S(E)$, $\sigma(E) \sim \abs{S(E) - 1}^2$, and the resonant
enhancement in the cross section (for real $E$) arises from
\begin{equation}
 S(E) \sim \frac{E-E_{\text{res}}-\ii\Gamma/2}{E-E_{\text{res}}+\ii\Gamma/2} \,,
\end{equation}
the parametrization of $S(E)$ in the vicinity of the resonance.
The key feature here is the pole at $E = E_{\text{res}} - \ii\Gamma/2$,
determined by the denominator, while the form in the numerator follows from
enforcing unitarity ($\abs{S(E)}=1$).
To complete the connection to the classical discussion, we can finally note that
in terms of the quantum-mechanical Green's function
\begin{equation}
 G(E) = (E - H)^{{-}1} \,,
\label{eq:G-E}
\end{equation}
$T(E) = V + V G(E) V$, so the resonance pole must indeed occur in $G(E)$,
exactly as in the classical case.
In other words, the behavior of the system -- near the resonance and in fact in
general -- is encoded in the analytic structure of $G(E)$.

From Eq.~\eqref{eq:G-E} we can furthermore see that indeed a resonance pole
in $G(E)$ corresponds directly to an eigenstate of $H$ with a complex energy as
given in Eq.~\eqref{eq:E-res}.
The obvious question now is: if $H$ is a Hermitian operator, as generally
assumed in quantum mechanics, how can this be?
The answer is that, because resonances as metastable states that decay are
inherently related to a dissipative process, there is a price to pay if one
tries to describe them within the framework of \emph{time-independent}
scattering theory, as we have done above.
This price is that one needs to let go of the assumption that $H$ is strictly
Hermitian and enter the regime of \emph{non-Hermitian quantum mechanics}.
This we will discuss further in the following.

\section{Complex-energy eigenstates}
\label{sec:States}

\subsection{Connection to scattering theory}
\label{sec:ScattTheory}

Resonances in nuclear physics show up experimentally in some kind of scattering
setup, so we will start our formal discussion by looking further at the relation
between scattering cross sections (the key quantity that can be directly
inferred from measurements), the scattering amplitude and phase shift, and the
$S$ matrix as the unifying theory construct.
For simplicity, we restrict the discussion here to a simple two-body system (of
spinless particles), and throughout this article we will remain in the regime of
nonrelativistic quantum mechanics.
For the low-energy nuclear systems we are interested in here, this is by and
large appropriate, but we alter the reader that a relativistic treatment, as
required, for example, for many systems in hadron physics, adds yet more facets
to the study of resonances.
Readers interested in this topic may find it useful to explore the recent review
by \citet{Mai:2022eur}.

We start by writing the differential cross section as
\begin{equation}
 \frac{\dd\sigma}{\dd\Omega} = \abs{f(\vec{k},\vec{k}')}^2 \,,
\label{eq:diff-xsec-f}
\end{equation}
with the scattering amplitude $f(\vec{k},\vec{k}')$ that depends on the initial
and final relative momenta $\vec{k}$, $\vec{k}'$.
Assuming rotational invariance and elastic scattering,
$\abs{\vec{k}}=\abs{\vec{k}'}=k$ and the on-shell energy is $E_k = k^2/(2\mu)$,
with $\mu$ denoting the reduced mass.
In this case, one has the familiar partial-wave expansion
\begin{equation}
 f(\vec{k},\vec{k}')
 = \sum_{\ell=0}^\infty (2\ell+1) \frac{\ee^{2\ii\delta_\ell(k)}-1}{2\ii k}
 P_\ell(\cos\theta)
 \equiv \sum_{\ell} (2\ell+1) f_{\ell}(k) P_\ell(\cos\theta) \,,
\label{eq:pw-f-delta}
\end{equation}
where $\theta$ is the angle between $\vec{k}$ and $\vec{k}'$, and the expansion
of $f$ induces an analogous expansion of the cross section.
Since there are no cross terms, for the \emph{total cross section} one ends up
with
\begin{equation}
 \sigma = \int \dd\Omega \frac{\dd\sigma}{\dd\Omega}
 = \sum_{\ell} \sigma_\ell \,.
\label{eq:sigma-sum}
\end{equation}
In the conventions we are using here, which are inspired by (but not identical
to) Sakurai's textbook~\citep{Sakurai:1994}, the partial-wave scattering
amplitude $f_{\ell}(k)$ is related to the $T$ matrix (in the same partial wave)
via
\begin{equation}
 f_{\ell}(k) = {-}\frac{\mu}{2\pi} T_{\ell}(E_k;k,k) \,,
\label{eq:f-T-ell}
\end{equation}
where we have written the latter in terms of three redundant argument ($E_k;k,k$)
in anticipation of generalizing it away from the on-shell case.
To conclude the basic definitions, we furthermore note that the partial-wave $S$
matrix is
\begin{equation}
 S_{\ell}(k) = \ee^{2\ii\delta_\ell(k)}
\label{eq:S-delta-ell}
\end{equation}
with $\delta_\ell(k)$ the scattering phase shift, and we can see that in the
partial-wave representation the relation between the $S$ and $T$ matrices,
stated generically in Eq.~\eqref{eq:S-T-E}, becomes
\begin{equation}
 S_{\ell}(k) = 1 - \ii \frac{\mu k}{\pi} T_{\ell}(E_k;k,k) \,.
\end{equation}

The $S$ and $T$ matrices are closely related to the wave functions
$\psi_{\ell}(k,r)$ that describe the scattering process at a given momentum $k$,
as a function of the particle separation $r$, in a fixed partial wave with
angular momentum $\ell$.
This formalism, which we encourage to reader to study with a dedicated textbook
such as the one by \citet{Taylor:1972}, will be most useful to understand the
influence of a resonance state.
For $r$ larger than the range of the interaction potential $V$,\footnote{The
only assumption we make here is that the potential $V$ is \emph{short-ranged},
i.e., falling off faster than any power law at large
distances.
This notably excludes the case of charged particles, subject to the long-range
Coulomb interaction, for the time being.}
the scattering wave function can be written as
\begin{spliteq}
 \psi_{\ell}(k,r)
 &\sim j_{\ell}(kr) + k f_{\ell}(k) h_{\ell}^+(kr) \\
 &\sim \frac{\ii}{2}\left[h_{\ell}^-(kr) + S_{\ell}(k) h_{\ell}^+(kr)\right] \,,
\label{eq:psi-f-s}
\end{spliteq}
where $j_{\ell}(z)$ and $h_{\ell}^\pm(z)$ are spherical Bessel and Hankel
functions, respectively.
As we are dealing with scattering wave functions, the overall normalization here
is not unique (even in magnitude, not just in phase), and from the previous
discussion it is already clear that all physics is encoded in just the
scattering phase shift $\delta_\ell(k)$.
What we do know, of course, is that $\psi_{\ell}(k,r)$ can be obtained as a
solution of the Lippmann-Schwinger equation (see \citep{Taylor:1972} for
details) and that it satisfies the radial Schrödinger equation
\begin{equation}
 \left[\frac{\dd^2}{\dd r^2} - \frac{\ell(\ell+1)}{r^2}
 - 2\mu V(r) + k^2\right] \psi_{\ell}(k,r) = 0 \,.
\label{eq:psi-radseq}
\end{equation}

As $h_{\ell}^\pm(z) \sim \exp(\pm\ii z)$ for large arguments ($\abs{z}\to\infty$),
one important aspect we can note from Eq.~\eqref{eq:psi-f-s} is that if one
considers all quantities as functions of a \emph{complex} momentum $k$, then a
pole in $S_{\ell}(k)$ corresponds to the $h_{\ell}^+(kr)$ term dominating.
Keeping in mind the arbitrariness in normalization, one can in fact divide the
second line of Eq.~\eqref{eq:psi-f-s} by the $S$ matrix factor and then finds
that the coefficient of $h_{\ell}^-(kr)$ \emph{vanishes} at this particular
momentum.
The latter is perhaps the more intuitive perspective, and consequently,
recalling the asymptotic form of the spherical Hankel functions, one refers to
these states as satisfying \emph{purely outgoing boundary conditions}.
For the special case $k = \ii\kappa$ with real $\kappa>0$, we merely recover
from this formalism the physics of bound states, and in that case $E =
k^2/(2\mu) < 0$.

If instead $k$ (and $E$) are in the fourth quadrant of the complex plane, the
state describes a resonance.\footnote{More rigorously, the $S$ matrix needs to
be considered as a multi-valued function of $E$, defined on Riemann sheets that
are connected by a cut along the positive real axis.
In this picture, resonance poles actually reside in the fourth quadrant of the
so-called \emph{second Riemann sheet} of the $S$ matrix.}
Noting that in such a case the particular partial wave in which the pole occurs
will dominate the total cross section, cf.~Eq.~\eqref{eq:sigma-sum}, we arrive
back at Eq.~\eqref{eq:sigma-BW}.
What we have gained from the present discussion is the realization that the
resonance is associated with a wave function that solves
Eq.~\eqref{eq:psi-radseq} for complex $k^2$ and with a purely outgoing boundary
condition, $\psi_{\ell}(k,r) \sim h_{\ell}^+(kr)$ for large $r$, imposed.

\subsection{Non-Hermitian quantum mechanics}
\label{sec:NHQM}

Traditional formulations of quantum mechanics rely on the assumption that all
observable quantities are represented by real-valued operators, which in turn
implies that the Hamiltonian must be Hermitian.
While this assumption is formally appropriate for closed (i.e., isolated)
quantum systems, in realistic scenarios one typically deals with systems that
are ``open'' in the sense that they are coupled to an external environment
and/or to a measurement apparatus.
For such open systems, the constraints imposed by a Hermitian description that
strictly conserves probability can become overly restrictive.

A simple illustrative example is provided by slow-neutron-induced fission of a
heavy nucleus, which results in the emission of several fragments.
From a formal standpoint, time-reversal symmetry enforces detailed balance
and thus guarantees the existence of the reversed process, in which all emitted
fragments recombine in a highly correlated manner at a common center,
reconstituting the original heavy nucleus and the incident slow neutron.
In practice, however, such a reverse process is never observed, as it is
statistically unrealizable.

Consequently, in the context of quantum scattering theory, one designates as
physical those processes that generate outgoing partial waves, yielding the
physically relevant solutions of the Schrödinger equation, denoted by $\Psi$
in the previous section (Sec.~\ref{sec:ScattTheory}).
Owing to the Hermiticity of the Hamiltonian, the complex-conjugate function
$\Psi^*$ is also a solution of the same Schrödinger equation, corresponding
to the time-reversed process.
For the reasons mentioned above, this formally possible solution is generally
regarded as unphysical.

For the same reason, for the description of resonances one is specifically
interested in eigenvalues of the Schrödinger equation that satisfy purely
outgoing boundary conditions and are associated with poles of the $S$ matrix
located in the fourth quadrant of the energy plane on the second Riemann sheet.
However, due to the unitarity of the $S$ matrix, there must also exist a
complex-conjugate pole associated with each such resonance, corresponding to an
incoming boundary condition and commonly referred to as an anti-resonant
state or capturing resonance.\footnote{
This scenario can be violated in situations where the $S$ matrix is deliberately
set up to violate unitarity, for example by employing a so-called ``optical
potential.''  We briefly get back to this in Sec.~\ref{sec:OP}.
}

\begin{figure}
\centering
\includegraphics[width=0.66\textwidth]{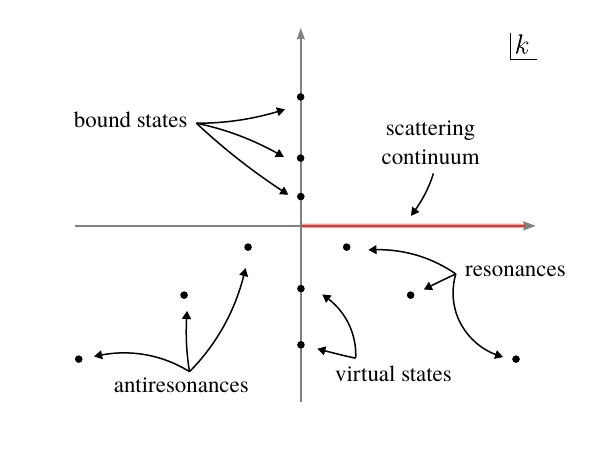}
\caption{The analytic structure of the $S$ matrix indicating its poles in the
complex $k$ plane (see text for details).
Adapted from original figure by N.~Yapa.}
\label{fig:s_matrix-k}
\end{figure}

In Fig.~\ref{fig:s_matrix-k}, the overall structure of the $S$ matrix in the
complex $k$ plane is shown as an illustration.
As mentioned in the previous section, bound states are located on the positive
imaginary axis, while resonances are located in the fourth quadrant, with their
anti-resonance counterparts as mirror images (upon reflection about the
imaginary axis) in the third quadrant.
As $E = k^2/(2\mu)$, the complex $k$ plane maps onto two copies of the complex
energy plane, which are precisely the two aforementioned Riemann sheets.
In this picture, the two sheets are connected by the ``scattering cut,''
located along the positive real axis.
Also indicated in the Fig.~\ref{fig:s_matrix-k} are so-called \emph{virtual
states}, which are discussed further in Sec.~\ref{sec:VirtualStates}.
For the time being, we merely note that these states, located on the negative
imaginary $k$ axis, map onto negative real energies, but they have to be
regarded as lying on the second Riemann sheet -- like resonances, and unlike
bound states, which have $E < 0$ on the first (physical) Riemann sheet.

As discussed in a previous section, resonance poles correspond to complex
eigenvalues of the Hamiltonian, i.e., they are associated with eigenstates
$\ket{\psi}$ that satisfy
\begin{equation}
 H \ket{\psi} = E \ket{\psi}
 \qquad
 E = E_{\text{res}} - \ii\Gamma/2 \,,
\label{eq:SEq-res}
\end{equation}
where, as introduced before, the real part $E_{\text{res}}$ denotes the
resonance position (located above the reaction threshold\footnote{
For the simple two-body scattering of two particles without internal structure,
this threshold would typically be located at energy $E=0$.
It may be at a different energy in situations where one or both fragments are
bound states, or if there is some external potential acting on them.}),
and $\Gamma$ is the total decay width of the state.
In writing Eq.~\eqref{eq:SEq-res} we have, of course, assumed that $H$ is
non-Hermitian, as otherwise only real eigenvalues $E$ would be possible.
We will comment on this further below.

For a time-independent Hamiltonian one can straightforwardly construct a
time-dependent solution of the resonant state that satisfies the time-dependent
Schrödinger equation:
\begin{equation}
 \psi(r,t) = \braket{r|U(t)|\psi}
 = \psi(r) \ee^{{-}\ii E_{\text{res}} t}\,\ee^{{-}\Gamma t/2} \,,
\end{equation}
where have used the well-known time evolution operator
$U(t) = \exp({-}\ii H t)$.
This then reproduces the probability density $P(t) \sim \ee^{{-}\Gamma t/\hbar}$
stated in Eq.~\eqref{eq:P-res}, and more generally we can see that the wave
function $\psi(r)$ at fixed $r$ decays with an exponential factor
$\ee^{{-}\Gamma t}$.
From this behavior one directly obtains the relation between the width
$\Gamma$ and the lifetime $\tau$ of the state:
\begin{equation}
 \tau = \frac{1}{\Gamma} \,.
\end{equation}

The resonant-state wave function thus describes particles that propagate to
infinity (decay) over time from any given point in coordinate space, thus
violating the usual unitarity assumption.
Formally, if $H$ is Hermitian, then $U(t)$ as defined above is unitary, but this
no-longer holds for non-Hermitian $H$.
This feature can be illustrated by considering a wave function describing the
decay of a radioactive nucleus, initially localized at the origin in coordinate
space.
After a time interval $\Delta t = t_{1} - t_{0}$, the probability density of
remaining nuclear fragments at the origin decreases by a factor $\ee^{{-}\Gamma
\Delta t}$.
At the same time, fragments of the nucleus that decayed at time $t_{0}$ will,
on average, have separated from each other by a distance
\begin{equation}
 \Delta r = \left(\frac{\operatorname{Re}(k)}{\mu}\right)\Delta t \,,
\end{equation}
where $\mu$ is the reduced mass and $k = \sqrt{2\mu E}$ is the complex
resonance momentum.
Consequently, the flux of fragments at $(r = \Delta r, t = t_{1})$ reflects
the decay that occurred at the origin $(r = 0, t = t_{0})$.
The overall probability balance is ensured by the asymptotic divergence of the
resonance wave function with $r$:
\begin{equation}
 \abs{\psi(r \to \infty)}^{2}
 \propto \exp\Big[{-}2\operatorname{Im}(k)\Delta r\Big]
 = \exp\left[{-}2\operatorname{Im}(k) \left(
  \frac{\operatorname{Re}(k)}{\mu}\right)\Delta t
 \right] \,,
\end{equation}
taking into account that
\begin{equation}
 \frac{\Gamma}{2} = {-}\operatorname{Im} E
 = {-}\operatorname{Im}{k^2}/(2\mu)
 = {-}\frac{\operatorname{Im}(k)\operatorname{Re}(k)}{\mu} \,.
\end{equation}

One might expect that a Hamiltonian of interacting particles constructed from
using real-valued potentials is Hermitian, i.e., that in coordinate
representation it satisfies\footnote{
This is assuming, for simplicity, a local potential:
$\braket{r|V|r'} = V(r)\delta(r-r')$.}
\begin{equation}
 \int_{0}^{\infty} f(r) H(r) g(r)\,\dd r
 =
 \int_{0}^{\infty} g(r) H^{\ast}(r) f(r) \dd r \,.
\end{equation}
By integrating the kinetic-energy term by parts one finds that this relation
holds provided the following boundary condition is satisfied by the functions
$f$ and $g$:
\begin{equation}
 \left.\bigg(f(r) \frac{\dd g(r)}{\dd r} - g(r)\frac{\dd f(r)}{\dd r}
  \bigg)\right|_{0}^{\infty} = 0 \,.
\label{eq:fg-BC}
\end{equation}
In other words, Hermiticity of the Hamiltonian depends crucially on
the boundary conditions imposed on the functions $f(r)$ and $g(r)$ on which it
acts.
Usually in quantum mechanics one considers the Hilbert space of states such that
in coordinate representation any two wave functions satisfy
Eq.~\eqref{eq:fg-BC}.
By relaxing this requirement, it is possible to find solutions of the
Schrödinger equation that belong to the non-Hermitian regime.
If one merely integrates the equation in differential form while imposing purely
outgoing boundary conditions matching a known resonance energy, this can be done
without formally changing $H$ -- it remains the sum of a derivative operator for
the kinetic part and a real-values multiplicative operator for the potential.
Alternatively, it is also possible to construct a basis of the extended vector
base in which $H$ maps directly onto a non-Hermitian matrix representation.
This technique is discussed further in Sec.~\ref{sec:Methods}, with
Sec.~\ref{sec:ComplexScaling} introducing a particularly simple way to achieve
this.

\section{Resonances in nuclear physics}

\subsection{What are the observables?}
\label{sec:Observables}

Our emphasis so far has been heavily on the formal aspects of resonances.
While conceptually it is very relevant and useful to understand these concepts,
it is equally important to note that complex-energy eigenstates are \emph{not}
what is being measured directly in experiments.
Instead, what an experimentalist observes, no matter how intricate the setup in
a particular case may be, are ultimately count rates in detectors, from which
they then infer scattering cross sections.
While we mentioned this in Secs.~\ref{sec:QuantumResonances}
and~\ref{sec:ScattTheory} to motivate the non-Hermitian/complex-energy
formalism, we now focus on the important phenomenological aspects before we
return in Sec.~\ref{sec:Methods} to the -- equally important -- question of how
theorists can perform calculations of resonances \emph{in practice}, i.e., how
to actually implement the non-Hermitian formalism introduced in the previous
section.

Cross sections are the most commonly employed quantity to describe collider
experiments, and we assume that the reader is familiar with the basic idea of
how these are defined in terms of the ratio of detected outgoing particles --
either overall (total cross section $\sigma$) or in a given angular region
(differential cross section $\dd\sigma/\dd\Omega$) -- compared to the incoming
particle flux.\footnote{Cross sections are frequently
considered as differential with respect to kinematical variable beyond a single
solid angle, such as energies/momenta or angles of multiple decay products.
We are neglecting this here to keep the discussion focused and simple.}
There are countless different ways in which a nuclear collider experiment can be
set up, far beyond what we can comprehensively discuss in this context, so we
consider, for a generic illustration, a target nucleus with $A$ nucleons ($Z$
protons and $N$ neutrons), on which another particle is incident from a
collider, with some tunable kinetic energy $E$.
This beam of incident particles can be mostly anything -- electrons, nucleons,
or other nuclei -- but let us pick the case of a proton beam.
As a proton interacts with the target nucleus upon approach, a number of things
can happen, ranging from simple elastic scattering to the proton being absorbed
by the target.
Among these, we highlight here a few important cases in which resonances enter
the picture.
\begin{enumerate}
\item The proton may scatter off the target nucleus, but transfer some amount of
 energy to it, just enough to lift it from its ground state into an excited
 state.
 How much energy exactly was transferred can be determined by measuring the
 scattered proton.
 If this excited state is another bound level of the target nucleus (with
 respect to the strong interaction), it will
 subsequently transition back to the ground state (via electromagnetic
 interactions), emitting one or more gamma rays.
 If, on the other hand, enough energy was transferred to lift the target nucleus
 into the \emph{continuum}, i.e., above the energy of the least bound state for
 the given configuration of $Z$ protons and $N$ neutrons,
 the excited state is a resonance.
 This resonance will typically then decay by particle emission, and these
 products can also be measured.
 Of course, another possibility is that the transferred energy may immediately
 break up the target nucleus.
 What makes the resonance scenario interesting and relevant is that one can
 analyzes the breakup process as a function of the transferred energy.
 What one then finds is that if the energy is close to the real part of the
 complex resonance energy, the cross section for breaking up the nucleus will be
 significantly enhanced, which is precisely the manifestation of the
 Breit-Wigner peak discussed earlier.
\item While the scenario discussed in the previous item can reveal resonances in
 the $A$-body nucleus, another possibility is the population of states in the
 $(A+1)$-body system, namely by capturing the incident proton.
 The likelihood for this to happen will in general be very low (it depends on the
 energy), but if there exists a resonance state of $Z+1$ protons and $N$
 neutrons, then, when the beam energy is tuned just right, the Breit-Wigner
 peak in the cross section may result again in significant enhancement of the
 capture process.
 Of course, conservation of both energy and momentum requires that a photon is
 created and radiated off as the target captures the proton, and the key
 signature for the process to have happened is the detection of this photon.
 The resonance energy in this scenario is determined by the tuning of the beam.
\item One more possibility is related to the situation in the first item above:
 the proton hitting the target nucleus may break it up immediately without first
 exciting an $A$-body resonance state, but any of the \emph{fragments} may be in
 such a state and subsequently decay further.
 In this situation, in order to observe the characteristic resonant enhancement
 in the cross section, one needs to measure all the fragments and sub-fragments
 that eventually hit the detector, select the ones that have emerged from the
 secondary decay vertex, and use their kinematic information in order to
 reconstruct the \emph{invariant mass} of the decaying original fragment.
 From this process, although very intricate, one can indeed obtain interesting
 information, in particular about resonance states that are difficult to produce
 via direct excitation or capture.
\end{enumerate}

\medskip
In the following, we will discuss how different types of resonances, from the
theoretical point of view, can be classified and how they related to the
exemplary experimental situations listed above.

\subsection{Different types of resonances}
\label{sec:ResonanceTypes}

Nuclear resonances cover a broad range of complex dynamical phenomena, but
nevertheless quite often -- not always and not necessarily without ambiguity --
a dominant mechanism can be associated with a particular state or system.
These mechanisms include:
\begin{enumerate}
 \item Shape resonances
 \item Virtual states
 \item Feshbach resonances
 \item Near-threshold few-body resonances
 \item Efimov three-body resonances
 \item Giant resonances
\end{enumerate}
In the following, we briefly discuss each of these from a high-level
phenomenological perspective.
This overview is followed by more specific physical considerations, while in
Sec.~\ref{sec:Methods} we focus in more detail on theory methods for concrete
calculations of resonances.

\subsubsection{Shape resonances}

Shape resonances are the simplest manifestation of the resonance phenomenon.
Such a resonance appears in particle scattering from a potential that
contains a repulsive barrier in addition to an attractive part.
For example, this scenario arises when a strong short-range attractive
interaction is combined with a long-range repulsive potential (e.g., a repulsive
Coulomb force).
For states with non-zero angular momentum, the repulsive part can also be merely
the effective potential generated by the centrifugal term.
With specifically a \emph{local} potential -- $V(r)$, with $r$ the coordinate of
the particle -- in mind, it is this overall form of the potential graph that
gives rise to the name ``shape resonance.''

If a sufficiently deep pocket exists in the resulting overall potential,
the particle may be temporarily trapped inside it, forming a long-lived
(quasi-stationary) bound-state-like structure at positive energy.
Eventually, however, the particle escapes from the inner region by tunneling
through the repulsive barrier, so that the resonance decays.

Shape resonances are ubiquitous in nuclear physics.
Well-known examples include neutron strength-function resonances in the optical
model of slow-neutron scattering by nuclei and single-particle resonances in the
mutual scattering of light nuclei such as $^4$He + $^4$He, $^4$He + $^3$He,
$^3$He + n, etc.
Generally, such resonant states become manifest in nuclear collisions with a
single open reaction channel, and they can be identified by the presence of a
peak in the corresponding scattering cross section or in the cross section for
an associated radiative capture/decay process, as discussed in
Sec.~\ref{sec:QuantumResonances}.
Neutron shape resonances are typically relatively broad, with a characteristic
width of $\Gamma \sim 1$ MeV.
However, shape resonances may also become very narrow in the case of charged
clusters with a strong Coulomb barrier.
Notably, the theoretical framework for describing this type of resonant states
was first formulated by \citet{Gamow:1928zz} in his investigation of
$\alpha$-decaying nuclei, where he analyzed the quantum tunneling of a particle
through a potential barrier.

\subsubsection{Virtual states}
\label{sec:VirtualStates}

Virtual states are single-particle phenomena that are closely related to shape
resonances.
As discussed in Secs.~\ref{sec:Intro} and~\ref{sec:States}, resonances
correspond to poles of the $S$ matrix located in the fourth quadrant of the
complex momentum plane (with a positive real part and a negative imaginary
part).
A virtual state, in contrast, is linked to an $S$-matrix pole whose real part
vanishes and whose imaginary part is negative, i.e., a pole situated on the
negative imaginary momentum axis.
Because bound states in this framework are represented by poles on the positive
momentum imaginary axis, virtual states are sometimes also called ``antibound
states.''
In the energy plane, both bound and virtual states map onto the negative real
axis, but on different Riemann sheets (the first and second, respectively).

In contrast to shape resonances, which, as mentioned above, arise from a balance
between attractive and repulsive parts of the potential, virtual states  appear
when the potential does not contain a repulsive barrier.
They are associated with orbital momentum $\ell=0$ systems -- not exhibiting any
effective repulsion from the centrifugal term -- for which the interaction is
attractive, but not attractive enough to generate a bound state. In the complex
momentum plane, the pole associated with such a state would move from the
negative to the positive imaginary axis if the interaction is made more
attractive.

Like resonances, virtual states leave observable signatures in scattering
experiments.
When such a state lies close to threshold (i.e., near zero energy or momentum),
it manifests as an enhancement of the scattering cross section.
This effect can be seen, for instance, in $S$-wave neutron-nucleus scattering,
where a sharp rise of the cross section as the neutron energy approaches zero
signals the presence of such a state.
A shallow bound state near threshold would generate a similar enhancement, but
if experimental evidence rules out the existence of a bound state (for example,
because it could otherwise be produced directly), one can infer the presence of
a virtual state instead.

Perhaps the most prominent example of this phenomenon is found in neutron-proton
scattering.
In the spin-triplet channel, the interaction gives rise to the well-known
deuteron bound state, whereas no bound state exists in the spin-singlet channel.
Nevertheless, the interaction in the singlet channel remains strong, and the
observed enhancement of the scattering cross section near threshold reveals a
virtual state.
This state can be regarded as the spin-singlet partner of the deuteron.

Similarly, a virtual state exists also in the two-neutron system, and if it were
not for the electromagnetic interaction, the situation would be the same for two
protons.
As first shown by \citet{kok80_1994}, what happens instead in the two-proton
$S$-wave configuration is that the Coulomb repulsion between the two protons
generates an effective barrier that pushes the pole off the imaginary axis into
the fourth quadrant of the complex momentum plane, with imaginary part larger
than the real part.
Because momentum poles in this region (below the 45-degree line in the fourth
quadrant of the momentum plane) map to energies with negative real part (and
negative imaginary part), this phenomenon is sometimes referred to as a
``subthreshold resonance.''
Alternatively, one may think of this as merely a very broad shape resonance.

\subsubsection{Feshbach resonances}
\label{sec:Feshbach}

Feshbach resonances can be described as a bound state
embedded in the continuum of an unperturbed Hamiltonian that becomes metastable
due to its coupling to the surrounding continuum.
Although Feshbach resonances originate as a many-body phenomenon, their
essential features can be illustrated using a much simpler model, which is so
important that we discuss it in more detail in Sec.~\ref{sec:Feshbach-Details}.
The concept was introduced by Herman Feshbach~\citep{feshbach1962unified} within
the framework he developed for a unified theory of nuclear
reactions, characterizing the coupling between open and closed
channels in scattering processes involving the formation of a compound nucleus,
thus providing the flexibility to model much of what we described in
Sec.~\ref{sec:Observables}.

The simplest realization of a Feshbach resonance can be found in a
one-dimensional coupled system with two scattering channels, characterized by
thresholds at different energies $E_2>E_1$.
Suppose that in the absence of coupling between the two channels each channel is
governed by its own Hamiltonian, $H_1$ and $H_2$, respectively, and assume
further that $H_2$ supports a bound state with energy $E$ that satisfies
$E_2>E>E_1$.
When a weak coupling between the two channels is introduced, this bound state
becomes unstable with respect to decay into the lower-energy channel.
If the coupling is sufficiently weak, the lifetime of such a metastable state
can be very long, leading to a narrow resonance in the vicinity of energy $E$
for the overall coupled system.
In Sec.~\ref{sec:Feshbach-Details} we explain exactly how this mechanism works.

In nuclear physics, the underlying interactions are typically strong, but
many-body dynamics can nevertheless give rise to an \emph{effective} interaction
that weakly couples different channels.
Representative examples include slow-neutron scattering on heavy nuclei, such as
n+\isotope[238]{U} or n+\isotope[232]{Th} and various other processes in
which an intermediate compound nucleus is formed.
In heavy nuclei, the number of available nucleon configurations can be extremely
large, so that an incident nucleon may be captured and have its energy rapidly
redistributed across the nucleons in the target.
The resulting compound state can have a lifetime much longer than the time
required for a wave packet to traverse the nucleus, due to its weak overlap with
the final decay channels.
At low excitation energies, the resulting compound nuclear states appear as
discrete Feshbach resonances.
At higher energies, however, the density of states increases dramatically, so
much so that individual resonances overlap to produce continuous cross section
with a broad enhanced structure instead of a clear resonance peak.
This picture is consistent with the compound nucleus model proposed by
\citet{bohr1936neutroneneinfang}, in which the incident particle loses its
individual identity and shares its energy with all nucleons in the target
nucleus, forming a fully equilibrated, highly excited system.

\subsubsection{Near-threshold few-body resonances}

\citet{baz1967} predicted another general class of multichannel long-lived
states arising from a near-threshold singularity in the $S$-matrix.
In this framework, summarized for example in the monograph by
\citet{Kukulin:1989}, one assumes that in the absence of inter-channel coupling
there exists a weakly bound compound state with a very small binding energy and,
consequently, a wave function with a large spatial extent.
This extended spatial structure implies a strongly suppressed overlap between
the compound-state wave function and the wave functions of the open-channel
states.\footnote{%
For a detailed discussion of the threshold behavior, see also
the monograph by \citet{baz69_b3}.}
As a result, the effective coupling between channels is weak, and the compound
state manifests itself as a long-lived Feshbach resonance.

\subsubsection{Efimov three-body resonances}

A particularly striking quantum three-body effect was discovered by
\citet{efimov1970energy}.
Consider a system of three identical bosons interacting via a two-body potential
that supports a very weakly bound dimer state (i.e., a two-body bound state with
almost zero binding energy).
Efimov predicted the emergence of an infinite series of three-body bound states
exhibiting a discrete scaling symmetry, now known as Efimov states.
If the
interaction is tuned such that the two-body bound state is exactly at zero
binding energy -- the \emph{unitarity (or unitary) limit}, equivalently
characterized by an infinite two-body $S$-wave scattering length -- these Efimov
states are bound and their energies accumulate at zero three-body binding
energy.
When the two-body interaction is tuned away from unitarity in a specific way (in
particular, when the near-threshold bound state crosses into the continuum and
becomes a virtual state), these three-body bound states can evolve into resonant
states.

Although the theoretical formulation is most transparent for identical, spinless
bosons, the inclusion of spin and isospin degrees of freedom allows for
analogous phenomena in nuclear systems.
To date, no genuine nuclear Efimov \emph{resonance} has been conclusively
observed experimentally.
Nonetheless, the triton (\isotope[3]{H}) is widely regarded as an Efimov-like
three-body bound state.
Moreover, despite the absence of an unambiguous observation of Efimov resonances
so far, the search for such effects remains an active field of research,
particularly in the context of halo nuclei, i.e., nuclear states with one or
more very weakly bound nucleons that can be effectively described within a
few-body framework.

\subsubsection{Giant resonances}

Giant resonances are highly collective modes of nuclear excitation.
The most prominent giant nuclear resonance, called the ``giant dipole resonance
(GDR)'' is commonly interpreted as a correlated oscillation of the protons in
a nucleus relative to the neutrons.
It was first observed experimentally by \citet{baldwin1947photo}.
These resonances are commonly observed in photonuclear reactions, induced by
energetic gamma rays, but they may also be excited by high-energy electrons
(interacting with the nucleus via virtual-photon exchange), or through inelastic
hadronic collisions.
It is generally observed that the peak energy of a giant resonance for a given
nucleus is anticorrelated with its charge radius, while at the same time the
width decreases.

\subsection{Decay modes and competition}
\label{sec_competition}

As we discussed, a key feature of a resonance is its decay over time, which the
context of nuclear physics generally refers to a given unstable nucleus
disintegrating into fragments.
While the simplest and most common process for this is decay by emission of a
single proton or neutron, in actual nuclei, depending on the condition, other
decay modes (also referred to as channels) can compete with single-nucleon
emission.
These conditions are determined by the relative $Q$ values (characterizing the
amount of energy released in the decay) of the different channels, as well as
their relative decay rates.
The relevant concept here is that of
so-called \emph{partial widths} (or branching ratio) $\Gamma_i$, which sum up to
the total width of the resonance: $\Gamma = \sum_i \Gamma_i$.
If two or more decay modes have a positive $Q$ value, barring any other
considerations such as angular momentum and parity selection rules, competition
between different decay modes is possible.
However, if the various possible decay modes involved have vastly different
lifetimes ($\tau_i = 1/\Gamma_i$), a positive $Q$ value might not be enough
to observe a particular decay in practice, as the channel with the shortest
lifetime will dominate.

One common condition encountered in neutron- and proton-rich nuclei is
$\beta$-decay competition, which can be understood intuitively as follows:
Away from the valley of stability, the $N/Z$ imbalance in a nucleus makes either
the neutron or the proton mean-field potential deeper than its isospin
counterpart.
In neutron-rich nuclei, $Q_{\beta^-}$ is usually significantly larger than $Q_n$
(assuming both are positive), but because $\beta$-decay is considerably slower
than neutron decay ($10^{-3}-10^{2}$~s versus $10^{-15}-10^{-9}$~s), when $Q_n$
is very small, both branching ratios can be comparable.
This situation is even more common on the proton-rich side where not only the
condition $Q_{\beta^+} \gg Q_p > 0$ can happen, but where also the Coulomb
barrier suppresses the proton-decay rate.
Electron capture is also possible when $\beta^+$-decay is likely.

We note in passing that $\beta$-decay in neutron- and proton-rich nuclei has an
interesting consequence in some cases.
The conversion of an excess neutron or proton into its isospin counterpart often
populates a daughter nucleus in a highly excited state above the neutron or
proton threshold.
The subsequent decay of this daughter nucleus via neutron- or proton-emission is
what is called $\beta$-delayed neutron or proton emission ($\beta n$ or $\beta
p$).

Another type of competition with one-nucleon decay is $\gamma$-decay
competition.
It is fairly rare, and especially so in neutron-rich nuclei.
This might seem surprising given that $\gamma$-emission, governed by
the electromagnetic interaction, is considerably faster than the
weak-interaction process of $\beta$-emission ($10^{-14}-10^{-12}$~s versus
$10^{-3}-10^{2}$s).
However, for $\gamma$-decay there is no inherent enhancement from $N/Z$
imbalance, and thus the condition $Q_{\gamma} \gg Q_{n/p} > 0$ is rarely
satisfied.
In certain heavy proton-rich nuclei, the Coulomb barrier can be high enough to
allow for $\gamma$-decay competition, but a large number of protons also has the
tendency of increasing the $Q$ value of $\alpha$-decay, which is yet another
possible decay mode to compete against proton decay.
In fact, $\alpha$-decay competition is very common in proton-rich nuclei due to
the combined effect of Coulomb repulsion, nuclear forces and antisymmetry
favoring the formation of $\alpha$ clusters, and the Coulomb barrier suppressing
proton decay.
In superheavy nuclei, spontaneous fission into other complex nuclei eventually
becomes the dominant decay mode, but this is beyond the scope of this survey.

Finally, it is also possible to have competition with unusual decay modes
such as two-neutron or two-proton emission.
These are new forms of radioactivity (beyond the well-known processes covered in
textbooks) and are considered ``exotic'' decay modes; they become possible
when $Q_{2n} > 0$ and $Q_{2p} > 0$, respectively.\footnote{
Of course, to have decay \emph{competition}, one must also have $Q_{n} > 0$ and
$Q_{p} > 0$, respectively.}
This usually happens near the drip lines where pairing correlations enhance
$Q_{2n/2p}$ relative to $Q_{n/p}$, as well as in certain excited states due to
local circumstances.

What makes this type of decay competition unique is the variety of decay
dynamics that can emerge depending on the widths of the emitting state of the
$A$-body system, and that of the intermediate states of the $(A-1)$ nucleus, as
well as the relative $Q$ values involved in the $A$, $A-1$, and $A-2$ systems
involved.
In general, the tunneling rate of two neutrons or protons is significantly lower
than that of only one nucleon, and thus competition is often limited to
situations in which $Q_{2n} > Q_n > 0$ and $Q_{2p} > Q_p > 0$, but this is not
always the case if an intermediate state in the $A-1$ system can act as a
doorway state.
The decay dynamics in which competition is possible are the sequential and
democratic ones.
Of course, even more complex exotic decay modes involving more nucleons exist,
which are discussed in Sec.~\ref{sec_drip_lines}.


\subsection{Narrow vs.\ broad resonances}

As already indicated in the previous section, one can in general estimate a
characteristic timescale associated with the interaction responsible for the
formation of a resonance state.
We denote this by $\tau_\text{int}$ and note that it is then natural to compare
the resonance lifetime $\tau \sim 1/\Gamma$ with $\tau_\text{int}$.
When there is a clear separation of timescales, $\tau \gg \tau_\text{int}$, the
resonance is classified as a ``long-lived'' state, for which it is meaningful to
describe the reaction as a two-step process in which a resonant state is first
formed and subsequently decays.
Resonances of this type, characterized by a small decay width $\Gamma$, are
commonly referred to as \emph{narrow resonances}.
For such states, the decay dynamics is predominantly governed by the many-body
structure of the system, which constitutes the basic assumption underlying the
quasi-stationary formalism introduced in the preceding section.
In contrast, when the relevant timescales are comparable, $\tau \sim
\tau_\text{int}$, the system effectively disintegrates on the same timescale as
its formation.
In a quantum-mechanical description, it is then no longer conceptually
meaningful to regard formation and decay as distinct, sequential processes.
Consequently, the decay dynamics of a short-lived \emph{broad} resonance is
determined not only by the internal structure of the system but also by the
details of the reaction mechanism (or, potentially, the primary decay of another
state) that populates it.

In the limiting case, one may consider reactions in which the projectile and
target interact only transiently near the nuclear surface for a duration $t \sim
\tau_\text{int}$ and then promptly separate.
In such a scenario, it becomes questionable whether one can properly speak of a
compound projectile–target system, because by construction not all nucleons
originally belonging to the projectile and target have sufficient time to
mutually interact before separation occurs.
A broad resonance can thus be regarded as a metastable compound configuration
only to the extent that a genuine many-body system is formed, which in turn
defines a lower bound on its existence time.
In practice, this bound must exceed $\tau_\text{int}$ and is system dependent.
This requirement can also be interpreted as a causality condition: decay cannot
occur unless a well-defined composite system has first been established.

A rough estimate of the minimum existence time of a given nucleus can be
obtained by assuming that the system is spherical, that nucleons move with the
Fermi velocity $v_F \approx 0.27c$, and that during the interval in which a
single nucleon traverses the nucleus, all other nucleons have sufficient time to
interact with one another.
For nuclei with mass numbers $A = 10 \cdots 40$, and adopting an interaction
time $\tau_\text{int} \approx 10^{-24}$~s, this argument yields lower
limits on the existence time of order $10^{{-}23}$ - $10^{{-}22}$~s.
Experimentally, the lower limit is conventionally taken to be
$10^{-22}$~s~\citep{thoennessen04_1165}, although it may be smaller for
light nuclei.

Given that the width of a resonance is inversely proportional to its half-life
$t_{1/2} = \tau \ln(2) = (\hbar \ln(2))/\Gamma$, we see that long- and
short-lived resonances correspond to narrow and broad energy peaks in cross
sections, respectively.
The limit of existence implies a maximal width that a nuclear resonance can
have.
Using the experimental limit, we obtain $\Gamma \approx 4.5$~MeV, but,
in line with the statement above, it could be larger in light nuclei.
To make the connection with the formal picture of a resonance as an $S$-matrix
pole, we note that for a broad resonance this pole is located relatively far
from the real axis, and if it is too far away (corresponding to large $\Gamma$
and consequently small $\tau$), such a pole may have no noticeable impact on the
magnitude or shape of the scattering cross section.

Direct measurements of nuclear half-lives are experimentally feasible down to
approximately  $t_{1/2} \approx 10^{{-}12}$~s.
For states with shorter half-lives, the corresponding decay widths are inferred
indirectly from fits to the resonance lineshape, i.e., from the energy width of
resonances in cross sections, typically employing Breit–Wigner
parameterizations.
Such an extraction is straightforward for isolated, narrow resonances that
display a symmetric Breit–Wigner profile.
However, the situation becomes considerably more intricate for resonances that
interfere with a non-resonant background (leading to a Fano-type lineshape), for
overlapping resonances (requiring, e.g., $R$-matrix theory), and for broad
resonances.
As discussed above, broad resonances are expected to manifest differently
depending on the specific production mechanism (population channel) employed.
Moreover, detailed properties of the reaction dynamics—such as the
center-of-mass energy, scattering angle, or impact parameter—can significantly
influence the observed resonance profile.
In addition, broad resonances are frequently affected by threshold phenomena
arising from their extended lineshape.

From an intuitive standpoint, a resonance formed just above a reaction threshold
possesses barely sufficient energy to decay.
This effectively reduces the available phase space and drives the lifetime of
the resonance to zero exactly at threshold.
This behavior is a direct consequence of the unitarity of the $S$ matrix, which
enforces conservation of flux and thus of total probability.
In the vicinity of the threshold, the resonance lineshape must be distorted to
comply with this unitarity constraint.
It can be demonstrated that, under these conditions, the cross section can no
longer be accurately described with a Breit–Wigner form characterized by a
constant resonance energy $E_R$ and a constant width $\Gamma$.
Instead, the lineshape of a near-threshold resonance must be parametrized using
a Breit-Wigner distribution with an explicitly energy-dependent width
$\Gamma(E)$.\footnote{In this context, we also refer the reader to the
discussion in Sec.~\ref{sec:DampedDrivenOsc}.}
Because the description of a
broad resonance lies at the boundary of validity of the quasi-stationary
formalism, its position $E_\text{peak}$, as inferred from the maximum of the
cross section, generally does not coincide with the real part $E_R$ of the pole
energy $\tilde{E} = E_R - \ii \Gamma/2$.
Rather, the peak position is more closely related to the absolute value of the
complex pole energy, $E_\text{peak} = |\tilde{E}|$.
Furthermore, extracting the energy-dependent width at the peak typically yields
an effective width $\Gamma(E_\text{peak})$ that exceeds the nominal width,
$\Gamma(E_\text{peak}) > \Gamma$.
Such discrepancies complicate a direct, quantitative comparison between
theoretical predictions and experimental observables.

Given that establishing a precise correspondence between experimentally
accessible cross sections and the poles of the $S$ matrix is a nontrivial
problem, it is often advantageous for theorists to concentrate primarily on the
pole structure when aiming at high-precision comparisons.
Although $S$-matrix poles are not observables in themselves, they are intrinsic
features of its analytic structure and, for a fixed Hamiltonian, are independent
of the particular computational method used to solve the problem.
A further important consideration is that the calculation of reaction
observables (such as differential or total cross sections) can be technically
demanding in many-body systems, whereas the determination of resonance poles is
frequently more tractable within the quasi-stationary formalism.

In the context of  the scattering matrix $S$ poles, a broad resonance can be
defined as a resonance characterized by $E_R > 0$ and $E_R \lesssim \Gamma/2$.
If we associate this resonance with a complex linear momentum
\begin{equation}
 k = \sqrt{2\mu E} \,,
\end{equation}
then broad resonances correspond to $S$-matrix poles located in the fourth
quadrant of the complex momentum plane, between the rays with arguments
${-}\pi/8$ and ${-}\pi/4$.

Indeed, writing
\begin{equation}
 k = \sqrt{2\mu E}
 = \sqrt{2\mu E_R\left(1 - \ii \frac{\Gamma}{2E_R}\right)},
\end{equation}
makes explicit that broad resonances are distinguished from narrow ones by the
condition $E_R \lesssim \Gamma/2$.
Below the ${-}\pi/4$ ray, one enters the region of subthreshold resonances with
$E_R < 0$ and $\Gamma > 0$, while along ${-}\pi/2$ (the negative imaginary axis)
one encounters antibound (virtual) states with $E_R < 0$ and $\Gamma = 0$ (see
Sec.~\ref{sec:Virtual} for a discussion of both).

To conclude this section, we emphasize that, in quantum systems, it is essential
to distinguish between the experimentally observable resonance peak position and
width in a cross section and the corresponding pole parameters of the $S$
matrix.
In order to establishing rigorous relation between these quantities -- a
reaction-theory framework is required. In many-body systems, the pole parameters
can be determined using the quasi-stationary formalism with a non-Hermitian
extension, as discussed in Sec.~\ref{sec:NHQM}.
Several concrete computational schemes for implementing this program are
presented in Sec.~\ref{sec:Methods}.
Once the pole parameters are known, one can unambiguously classify a resonance
as narrow or broad by inspecting the ratio $\Gamma/(2E_R)$; this ratio provides
crucial information for ensuring a meaningful and consistent comparison with
experimental observables.

\subsection{Structure vs.\ reaction, dynamical processes}
\label{sec_struct_react}

The distinction between the observable cross section and the complex-energy
parameters of a resonance state points to a deeper question: what really is the
proper theoretical framework for describing a many-body system decaying into a
continuum of final states?
For well-bound nuclei sitting far below particle-emission thresholds, the
non-resonant continuum contributes only as a perturbation, and it is justified
to compute structure observables (masses, radii, transition matrix elements) and
reaction observables (cross sections, phase shifts) within largely independent
frameworks.
Near the drip lines, however, this separation breaks down as particle-emission
thresholds are close to the low-lying spectrum and the interaction effectively
couples bound, resonance, and scattering states.
As mentioned already in Sec.~\ref{sec:NHQM}, nuclei in this situation must be
described as \emph{open quantum systems}
(OQSs)~\citep{dobaczewski07_17,michel09_2,rotter15_2464,michel21_b260}.
In this framework, a finite ``system'' subspace, spanning the configurations
relevant for a localized many-body state, is coupled to an infinite
``environment'' of continuum states and decay channels through an effective,
non-Hermitian Hamiltonian either derived from the Feshbach projection
formalism~\citep{Feshbach:1992} (introduced in Sec.~\ref{sec:Feshbach} and
discussed in more detail in Sec.~\ref{sec:Feshbach-Details}), or from the
quasi-stationary formalism based on the pole expansion of the
resolvent~\citep{baz69_b3}.

A consequence of treating nuclei as OQSs is that nuclear structure and reactions
cease to be distinct problems as both derive from the same underlying
$S$ matrix.
Its poles yield the positions and widths of resonant states, its phase shifts
and branch cuts encode reaction cross sections, and the residues of its poles
encode properties such as partial widths and asymptotic normalization
coefficients.
There exist several concrete many-body implementations of this idea.
Historically, the ``continuum shell-model''~\citep{rotter91_448,volya06_94} and
the ``shell model embedded in the
continuum''~\citep{bennaceur99_46,bennaceur00_44,okolowicz03_21}, both based on
the Feshbach projection formalism (see Sec.~\ref{sec:Feshbach-Details}, were
formulated first.
Later, structure approaches based on the quasi-stationary formalism were
introduced such as the Gamow Shell Model~\citep{michel09_2,michel21_b260} and
the Gamow Density Matrix Renormalization
Group~\citep{rotureau06_15,rotureau09_140}
method based on the Berggren basis~\citep{berggren68_32,berggren93_481}.
More details about these techniques are discussed in Secs.~\ref{sec_GSM}
and~\ref{sec_GDMRG}.
In parallel, other approaches such as the
Faddeev-Yakubowsky~\citep{lazauskas18_2032,lazauskas19_2363} equations and the
Finite-Volume Discrete Variable Representation~\citep{yu24_3045} were extended
using the uniform complex-scaling (see Sec.~\ref{sec:ComplexScaling}.
From the reaction side of the problem, an alternative to these methods, closer
in spirit to the continuum shell model, is given by approaches based on the
Resonating Group Method (RGM)~\citep{wheeler37_2315,fliessbach82_638}, notably
the No-Core Shell Model with Continuum (NCSMC)~\citep{navratil16_1956}.
The NCSM has also been recently extended to implement uniform complex
scaling~\citep{Yaghi:2025ftv}.

Within a single calculation, they can all give access to energies and widths, as
well as spectroscopic and reaction observables in some cases.
This unified description naturally accommodates a number of dynamical processes
that are otherwise difficult to capture consistently, in particular the
reorganization of the wave function due to continuum couplings and, conversely,
the modification of the continuum due to the structure.
Such feedback between structure and reactions is a key feature of describing
nuclei as OQSs.

\subsection{Exotic states, drip lines, nuclear forces}
\label{sec_drip_lines}

The resonance phenomenon plays a central role in nuclear physics, starting with
low-energy few-nucleon systems.
The proton-neutron (np) system admits only one weakly bound state
(\isotope[2]{H}) with a binding energy of about $2.2$ MeV, which is
considerably lower than the average $8.0$ MeV per nucleon in heavier nuclei.
This bound state occurs in the spin-triplet isospin-singlet channel ($^3S_1$),
while the spin-singlet isospin-triplet channel ($^1S_0$) only supports a virtual
state.
As mentioned in Sec.~\ref{sec:Virtual}), the same is true for the other two
isospin-triplet configurations (2n and 2p), only that the Coulomb repulsion for
between protons actually moves the pole off the imaginary to become a
subthreshold resonance~\citep{kok80_1994}.

The virtual state in the 2n system is equivalently characterized by a large
negative $S$-wave scattering length ($-23.7$ fm, much larger than the typical
length scale of the nuclear force set by one-pion exchange, $1/m_\pi \sim 1.4$
fm.
This remarkable fact has important consequences for neutron-rich nuclei.
The 2n system is so close to forming a bound state that significant efforts,
both experimental and theoretical, have been devoted to testing whether or not a
four-neutron (4n) system could form.
While it was initially claimed that the 4n system might
exist~\citep{marques02_1460,kisamori16_1463,pieper03_1461,%
shirokov16_1791,li19_2634}, there is now considerable evidence that it does
not~\citep{hiyama16_1624,fossez17_1916,deltuva18_2079,higgins20_2495} and the
peak observed~\citep{duer22_2494} corresponds to four correlated neutrons in
\isotope[8]{He}~\citep{lazauskas23_2744}.

Similarly, somewhat longer ago there have been investigations of possible
three-neutron (3n) states~\citep{Glockle:1978zz,Offermann:1979wbx}, but the
general consensus to today is that neither 3n nor 3p systems exist.
However, the \isotope[3]{H} and \isotope[3]{He} isotopes each support one bound
state at $E \sim {-}8.0$ MeV.
Adding one more nucleon, the \isotope[4]{H} and \isotope[4]{Li} isotopes are
unbound and exhibit several resonances, while the \isotope[4]{He} isotope is
strongly bound (${-}28.3$~MeV) and supports many resonances at an excitation
energy of about $20$~MeV and higher.

We know that three-body and higher-body forces play an important role in nuclei,
but the fact remains that two-body forces dominate.\footnote{%
This hierarchy of forces arises forces is naturally in the context of
effective field theories~\citep{Hammer:2019poc}}.
For that reason, and given the nature of few-body systems, as we build up larger
systems of nucleons it is evident that nuclei must eventually become unbound
when either more protons are added along an isotonic chain, or more neutrons are
added along an isotopic chain.
These limits of nuclear binding are called the \emph{drip lines} and, evidently,
are determined to an extent by the resonant nature of the 2n and 2p systems.
We note in passing that the problem of how large can a nucleus be (when adding
both protons and neutrons) is beyond the scope of this survey, and still
unsolved.

The large 2n and 2p scattering lengths lead to strong dineutron and diproton
correlations in nuclei, respectively.
One way to learn about nuclear forces from these correlations is to perform
precision studies of the two-neutron and two-proton exotic decay modes mentioned
in Sec.~\ref{sec_competition}.
Pairing in the $S$-wave channels naturally favors a low relative momentum
between the two nucleons, which tends to localize the pair spatially.
This compact ``dinucleon'' configuration competes against the anti-correlated
``cigar'' configuration in which nucleons sits on opposite sides of the
effective core.
The relative weight of these two configurations is primarily controlled by the
interference between positive-parity and negative-parity continuum partial
waves~\citep{catara84_1880,pillet07_1879}, making the imprint of the
correlations on the angle-energy distributions of the pair very sensitive to
details of nuclear forces, and most notably so for high-$\ell$
components~\citep{matsuo05_3139}.
A dinucleon configuration produces a forward peak in the differential cross
section at small opening angle, while a cigar configuration yields a
back-to-back pattern~\citep{pfutzner23_2859}.
Such correlations are not limited to exotic decay modes and also affect halo
structures and their properties, notably soft-dipole strengths and
Coulomb-dissociation cross sections.

Beyond 2n and 2p decay modes, experimental studies have revealed the existence
of multi-neutron and multi-proton resonances.
Notable examples include the sequential 4n decay of
\isotope[28]{O}~\citep{kondo23_2923} and the sequential 5p decay of
\isotope[9]{N}~\citep{charity23_2960}.
These subtle structures impose extreme challenges on theory for the simple
reason that decay widths depend exponentially on separation energies, and thus
describing, for example, a 5p emitter requires obtaining five separation
energies along an isotonic chain both precisely and accurately.
This makes the exploration of the drip lines, and thus the proper description of
exotic nuclei, critical for the advancement of nuclear science.

In a similar vein, antibound (virtual) states and subthreshold resonances in
many-body systems, such as (possibly) the $J^\pi = {1/2}^+$ state of
\isotope[9]{He}~\citep{kalanee13_1909} and the ground state of
\isotope[10]{Li}~\citep{chartier01_2576}, provide sensitive probes of nuclear
forces.
While the properties of such states are determined by universal low-energy
$S$-wave physics, it is nuclear forces that shape the subtle conditions in the
many-body system for such states to emerge.
The poles of the $S$ matrix associated with these states are located very close
to thresholds, and their positions depend exponentially on small variations of
the underlying interaction.
Exotic nuclei near the drip lines therefore provide an opportunity to test
chiral effective field theory
interactions~\citep{epelbaum09_866,machleidt11_414,Hammer:2019poc} in extreme
isospin conditions.

The presence of resonances and weakly bound states can lead to various kinds of
interesting interplay with traditional emergent phenomena in nuclei, such as
deformation, clustering, superfluidity, and shell evolution.
In the shell-model picture, continuum couplings favor the occupation of
low-$\ell$ orbitals, leading to a reorganization of the shell structure that
can, in turn, favor or disfavor the emergence of certain phenomena.
For example, continuum couplings appear to induce deformation near the $N=20$
island of inversion by promoting $P$-wave occupation -- which, in turns, enables
couplings between $P$ and $F$ waves ($\Delta \ell = 2$) and therefore the
emergence of quadrupole deformation~\citep{fossez22_2540,wang26_3366}.

Proximity to a particle-emission threshold can also lead to a phenomenon called
``alignment of the wave function''~\citep{Feshbach:1992}.
This refers to a scenario in which the structure of the state considered
reorganizes to match the corresponding decay channel.
This can lead to trapped resonances, where the wave function aligns with a
closed channel and sees its decay width reduced, as well as near-threshold
clustering near cluster thresholds (see Sec.~\ref{sec:Feshbach}
and~\ref{sec:Feshbach-Details} for the general formalism).
The same phenomenon can also be interpreted as the origin of so-called
\emph{superradiant states}, where two or more states of same spin and parity
couple in the continuum so that one state ``absorbs'' all the width while the
others become narrow resonances~\citep{auerbach11_879,zelevinsky23_b337}.

Near-threshold resonances also have important consequences for nuclear
astrophysics, most notably in stellar environments where the thermal energy is
usually below 1.0 MeV.
For charged particles, the capture probability is given by the overlap between
the thermal energy distribution and the probability of tunneling through the
Coulomb barrier. At low energy, this capture probability
is highest in a narrow energy range that defines the Gamow window.
For neutral particles, it is the density of states at low energy that matters.
In particular, proton and neutron radiative-capture reaction rates are typically
exponentially sensitive to the energy position of a handful of low-lying
resonances.
When used as input for astrophysical simulations, an error of a few hundred keV
on the position of a single near-threshold resonance can translate into orders
of magnitude uncertainty on the predicted observables.
This sensitivity strongly affects the rapid neutron-capture ($r$-) process,
which is responsible for the synthesis of about half of the elements heavier
than iron.
In the nuclear chart, the path of this process runs very close to, and in some
conditions along, the neutron drip line~\citep{mumpower16_2386,martin16_2292}.

In summary of this section, resonances play a critical role at the level of
nuclear forces and few-body systems, which in turn shapes the drip lines.
Understanding how OQS physics and emergent phenomena affect each other is
critical to determine the position of the drip lines and understand properties
of exotic nuclei.
In many-body nuclei, resonances provide a unique way to test nuclear forces and
affect properties of exotic nuclei.
Finally, resonances in exotic nuclei control critical astrophysical processes
such as the $r$-process.

\section{Theory methods}
\label{sec:Methods}

\subsection{Feshbach's projection formalism}
\label{sec:Feshbach-Details}

A phenomenologically very useful formalism for the theoretical modeling of
resonances has been developed by Feshbach and is nicely detailed in his book on
nuclear reactions~\citep{Feshbach:1992}.
While originally developed for nuclear systems, a version of this formalism has
found widespread application also in atomic physics.
While conceptually we introduced Feshbach resonances already in
Sec.~\ref{sec:Feshbach}, we provide here a summary of the concrete theoretical
framework used to describe them.
For a more comprehensive discussion, we refer the reader to
Feshbach's book (cited above).
To be concrete, let us imagine a situation such as the scattering process
described as the first scenario in Sec.~\ref{sec:Observables}.
That is, we consider a scattering system comprised of a proton incident on a
nucleus with mass number $A$.

The overall $p+A$ system is described by states in a Hilbert space
$\mathcal{H}$ that includes all the various physical possibilities for this
setup.
Exactly which states in $\mathcal{H}$ are accessible depends on the energy $E$
of the incident proton.
Following \citet{Feshbach:1992}, we separate $\mathcal{H}$ into a subspace
$\mathcal{P}$ of these states (which we call the ``open channels'') and its
orthogonal complement $\mathcal{Q}$ (containing the  ``closed channels'').
We define projection operators $P$ and $Q$ onto these spaces, satisfying
\begin{equation}
 P + Q = \Id \,, \quad
 P^2 = P \,, \quad
 Q^2 = Q \,,
\end{equation}
along with $P = P^\dagger$ and $Q = Q^\dagger$.
The Schrödinger equation
\begin{equation}
 (E - H)\ket{\Psi} = 0
\end{equation}
for a state $\ket{\Psi} \in \mathcal{H}$, where $H$ is the Hamiltonian for the
$p+A$ system, can be decomposed with the help of the projection operators.
Specifically, we can write
\begin{equation}
 \ket{\Psi} = P\ket{\Psi} + Q\ket{\Psi} \equiv \ket{\Psi_P} + \ket{\Psi_Q}
\end{equation}
and use formal manipulations to find an effective equation for $\ket{\Psi_P}$
alone.
This is achieved by defining
\begin{equation}
 H_{PP} = PHP \,, \quad
 H_{PQ} = PHQ \,, \quad
 H_{QP} = QHP \,, \quad
 H_{QQ} = QHQ \,,
\end{equation}
where the off-diagonal terms arise only from the interaction $V$ if we split
$H=H_0+V$ into kinetic and interactions terms, in the usual way.
That is, we can write $H_{PQ} = V_{PQ}$ and $H_{QP} = V_{QP}$ in the following
and therefore make explicit that it is the interaction among nucleons that is
responsible for coupling the $\mathcal{P}$ and $\mathcal{Q}$ spaces.
With these definitions, we find the coupled equations
\begin{subalign}
 (E-H_{PP})\ket{\Psi_P} &= H_{PQ} \ket{\Psi_Q} \,, \\
 (E-H_{QQ})\ket{\Psi_Q} &= H_{QP} \ket{\Psi_P} \,,
\end{subalign}
and we can formally solve the second of these by writing
\begin{equation}
 \ket{\Psi_Q} = \left(E + \ii\epsilon - H_{QQ}\right)^{{-}1} H_{QP}
 \ket{\Psi_P} \,,
\end{equation}
where a small imaginary part (with implied limit $\epsilon\to 0$) has been
included in the usual manner in order to impose the appropriate boundary
condition.
This formal solution can be substituted back into the first of the coupled
Schrödinger equations to arrive at
\begin{equation}
 (E - H_{\text{eff}})\ket{\Psi_P} = 0
\end{equation}
with
\begin{equation}
 H_{\text{eff}} = H_{PP}
 + H_{PQ} \left(E + \ii\epsilon - H_{QQ}\right)^{{-}1} H_{QP} \,.
\label{eq:H-eff}
\end{equation}
This is the effective Hamiltonian we have been looking for, and we can see
immediately that due to the presence of the Green's function
\begin{equation}
 G_{QQ}(E) = \left(E + \ii\epsilon - H_{QQ}\right)^{{-}1} \,,
\end{equation}
it is energy-dependent, complex (due to the $\ii\epsilon$ term),
and non-Hermitian.

Given these properties, it is certainly possible that it allows for
$\ket{\Psi_P}$ solutions that describe complex-energy resonance states, and
there is also an elegant physical explanation for when they occur.
To see that, consider an energy $E$ that is close to the energy of a bound state
of the target nucleus, which in Feshbach's formalism are described as bound
eigenstates of $H_{QQ}$, labeled by $\ket{\Phi_n}$ with for eigenvalue $E_n$ in
the following.
By the spectral representation of $G_{QQ}(E)$,
\begin{equation}
 G_{QQ}(E) = \sum_{n} \frac{H_{PQ}\ket{\Phi_n}\bra{\Phi_n}H_{QP}}{E-E_n}
 + \int\dd E' \frac{H_{PQ}\ket{\Phi(E')}\bra{\Phi(E')}H_{QP}}{E-E'} \,,
\label{eq:G-QQ}
\end{equation}
such a bound state will give rise
to a pole at $E=E_n$ for some particular $n$, and for $E$ near $E_n$, one may
approximate all energy dependence as arising from this single pole term, setting
$E$ equal to $E_n$ for all other terms.
The effective Hamiltonian consequently becomes
\begin{equation}
 H_{\text{eff}} \approx \bar{H}_{PP} +
 \frac{H_{PQ}\ket{\Phi_n}\bra{\Phi_n}H_{QP}}{E-E_n} \,,
\label{eq:H-eff-pole}
\end{equation}
where $\bar{H}_{PP,n}$ is the sum of the original $H_{PP}$ and all other terms
from Eq.~\eqref{eq:G-QQ}, evaluated at $E = E_n$.
Referring again to \citet{Feshbach:1992} for the details, the final step is to
formally solve the Lippmann-Schwinger equation in the $P$ space with
$H_{\text{eff}}$ as in Eq.~\eqref{eq:H-eff-pole}.
If $\ket{\chi^{+}}$ and $\ket{\chi^{-}}$ label scattering eigenstates (at energy
$E$) of $\bar{H}_{PP}$ with outgoing (superscript $+$) and incoming (superscript
$-$) boundary conditions, one finds that the effective $T$ matrix for this
scenario can be written as
\begin{equation}
 T(E) = \braket{\chi^{-}|\chi^{+}}
 + \frac{\braket{\chi^{-}|H_{PQ}|\Phi_n}\braket{\Phi_n|H_{QP}|\chi^{+}}}
 {E - E_n + \braket{\Phi_n|W_{QQ}|\Phi_n}} \,,
\label{eq:T-eff}
\end{equation}
where
\begin{equation}
 W_{QQ} = H_{QP} \left(E + \ii\epsilon - \bar{H}_{PP}\right)^{{-}1} H_{PQ} \,.
\end{equation}
Separating the matrix element in the denominator of Eq.~\eqref{eq:T-eff} into
real and imaginary parts by defining $\braket{\Phi_n|W_{QQ}|\Phi_n} =
\Delta_n(E) - \ii\Gamma_n(E)/2$.
Note that the imaginary part here arises specifically from the presence of the
$\ii\epsilon$ term, treated via the standard principal-value prescription in the
implied limit $\epsilon\to0$.
Schematically, this can be written as
\begin{equation}
 \lim_{\epsilon\to0} \frac{1}{E + \ii\epsilon - \bar{H}_{PP}}
 = \text{PV} \frac{1}{E - \bar{H}_{PP}} + \ii\pi\delta(E - \bar{H}_{PP}) \,,
\end{equation}
and using the spectral representation as in Eq.~\eqref{eq:G-QQ}, it can be
expressed concretely in terms of energy eigenvalues and eigenstates.
Assuming that the energy dependence of $\Delta_n$ and $\Gamma_n$ is weak
compared to the explicit linear term $E$, one finally arrives at
\begin{equation}
 T(E) = \braket{\chi^{-}|\chi^{+}}
 + \frac{\braket{\chi^{-}|H_{PQ}|\Phi_n}\braket{\Phi_n|H_{QP}|\chi^{+}}}
 {E - E_R + \ii\Gamma/2} \,,
\end{equation}
which gives rise to a Breit-Wigner distribution (with $E_R = E_n -
\Delta_n(E)$), plus a background term describing elastic scattering via
$\bar{H}_{PP}$.

From the various approximations and formal manipulations involved in getting to
this final form, it should be clear that, from a practical perspective, the
formalism described above is useful primarily for constructing relatively
simple phenomenological models.
For example, so-called ``Feshbach resonances'' have become a broadly used
experimental tool that can conveniently be modeled theoretically based on the
ideas summarized here.
For more details about this, as well as broader historical context that includes
also the work of other authors, the reader is referred to the review by
\citet{Chin:2010crf}.
Apart from that, the formalism provides a very nice and strikingly simple
conceptual explanation for how resonances arise naturally out of otherwise
highly complex compound reactions.

\subsection{Complex absorbing potentials}
\label{sec:CAP}

As discussed in Sec.~\ref{sec:ScattTheory}, resonance wave functions satisfy
purely outgoing boundary conditions in the asymptotic region and therefore
diverge $\sim \exp(\operatorname{Im}(k)r)$, where $k = \sqrt{2\mu E}$ is the
wave vector (i.e., the momentum scale, in our units with $\hbar=1$)
for a resonance with energy $E$.\footnote{%
 We are assuming here still a simple two-body system with reduced mass $\mu$ for
illustration.}
The \emph{complex absorbing potential (CAP)}
method~\citep{Neuhauser_ABP} aims to  modify the underlying Schrödinger equation
such that the effective wave vector is altered beyond the physical interaction
region (for $r>R$, with $R$ range of the potential) according to
\begin{equation}
 k \rightarrow k + \ii \epsilon(r) \,,
\end{equation}
in a manner that ensures $\operatorname{Im}(k + \ii\epsilon(r)) \gg 0$.
Under this condition, the wave function is strongly damped in the asymptotic
region while remaining essentially unchanged within the physical interaction
domain.

This modification is typically implemented by adding to the Hamiltonian a
complex absorbing potential $V_{\text{CAP}}$ of the form
\begin{equation}
 V_{\text{CAP}}(r) = \left\{%
 \begin{array}{ll}
  0 \;; & r<R \,, \\
  {-}\ii w(r) , \quad w(r) > 0 \;; & r>R \,.
 \end{array}\right.
\end{equation}

Although the method is conceptually straightforward to implement, constructing
an absorbing potential that is effectively reflection-less (to ensure that it
does not perturb the resonance wave function in the internal region and thereby
avoids shifting the computed eigenvalues) constitutes a nontrivial problem.
For one-dimensional systems, several efficient strategies are available for
designing such reflectionless absorbing potentials~\citep{Manolopoulos_ABP};
however, analogous constructions are considerably more challenging for
higher-dimensional or otherwise more complex systems.

\subsection{Optical potentials}
\label{sec:OP}

While the CAP method discussed in the previous subsection introduces an
imaginary term the potential purely as a computational device, it should be
noted that in phenomenological nuclear calculations it is also common to
construct so-called \emph{optical potentials} (also referred to as
\emph{effective interactions}) that include an imaginary term in the potential
as a means to effectively parameterize physics effects that are not explicitly
accounted for.
This is used in particular in calculations of nuclear reactions, to account for
the flux of probability into channels not resolved in the calculation.

A comprehensive recent overview of how optical potentials are constructed and
used has been given \citet{Hebborn:2022vzm}, and we refer the interested reader
to this review article for more details.
One thing to point here is, however, that in the construction of optical
potentials, the unitarity of the $S$ matrix is intentionally violated in a
controlled manner so that $\mathrm{Tr}(S^\dagger S) < 1$.
With regard to the analytical structure, one aspect to note is that  in these
non-unitary descriptions, the $S$ matrix may exhibit poles in the fourth energy
quadrant without the corresponding complex-conjugate partner poles, thereby
modeling absorption and effectively discarding the time-reversed, anti-resonant
counterparts, thereby deviating from the general scenario otherwise discussed in
Sec.~\ref{sec:NHQM}.

\subsection{Complex scaling}
\label{sec:ComplexScaling}

The technique known as ``complex scaling'' provides a particularly simple and
transparent procedure for transforming an initially Hermitian Hamiltonian into a
non‑Hermitian one, in such a way that essential physical properties are
preserved while resonant states are exposed as complex energy eigenvalues of the
transformed Hamiltonian.
Before we substantiate these statements, we introduce complex scaling in its
simplest form, commonly referred to as ``uniform complex scaling.''

For a two‑body system with relative coordinate $\vec{r}$, uniform complex
scaling consists of multiplying the radial separation $r = |\vec{r}|$ by a
complex phase,
\begin{equation}
 r \to r \ee^{\ii \theta} \,,
 \label{eq:r-scaled}
\end{equation}
while leaving the angular dependence of $\vec{r}$ unchanged.
The rotation angle $\theta$ is a continuous parameter the value of which may be
chosen freely, provided it exceeds a certain minimum that depends on the
resonance under investigation.
Specifically, for a resonance with complex energy $E$, one must ensure that
$\theta > {-}\frac{\arg{E}}{2}$ (as will be discussed in detail below, it is in
addition generally required that $\theta < \pi/2$).
For a Hamiltonian of the form $H = H_0 + V$, with kinetic energy $H_0$ and a
spherically symmetric local potential $V$, the complex scaling transformation
leads (in coordinate representation) to
\begin{equation}
 H \rightarrow H_\theta = \cphase{{-}2} \frac{1}{2 \mu}\frac{\dd^2}{\dd r^2}
 + V(r \cphase{})
\end{equation}
for a two‑body system with reduced mass $\mu$.
This expression makes it evident that a key prerequisite of the method is that
$V$ be specified in a form that admits an efficient and accurate analytic
continuation to complex arguments.
This is most straightforward when $V$ is available as a closed‑form analytical
function.
We emphasize that the method is not restricted to local central interactions; it
can be applied equally well to more general classes of potentials, such as
non‑local interactions or those containing tensor components, and it can be
extended in a natural way to systems with more than two particles.
Furthermore, if $\vec{p}$ denotes the momentum conjugate to $\vec{r}$, one may
equivalently formulate complex scaling in momentum space by implementing the
transformation
\begin{equation}
 p \to p \ee^{{-}\ii \theta}
\label{eq:p-scaled}
\end{equation}
on the magnitude of the momentum.

\begin{figure}[htbp]
 \centering
 \includegraphics[width=0.55\textwidth]{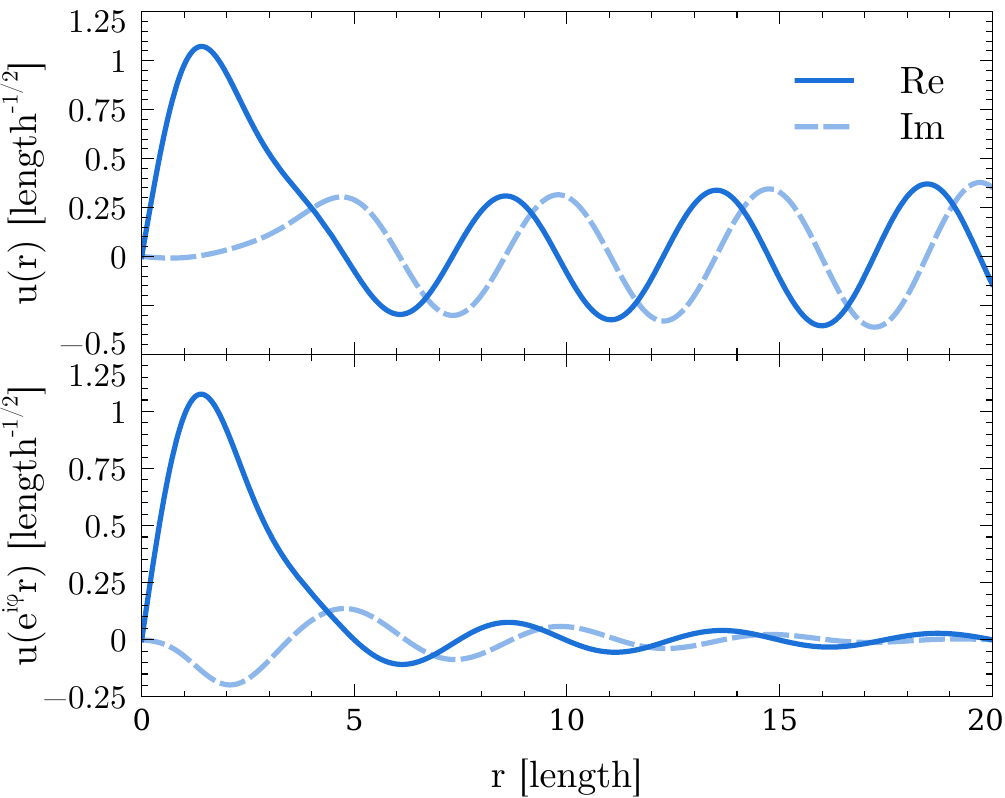}
 \caption{Radial resonance wave function calculated without (top panel) and
  with (bottom panel) complex scaling.  Calculated wave function courtesy of
  N.~Yapa.
  \label{fig:ResonanceWF}
 }
\end{figure}

Coming back to our specific simple two-body example, where we assume that $V$ is
short-ranged (falling off faster than any power law for large separation $r$
between the particles), then outside the interaction range the reduced radial
wave function in coordinate representation becomes
\begin{equation}
 \label{eq:asymptotic_wf}
 u_{l,k}(r) \sim \hat{h}^+_l(kr)
\end{equation}
for a state with angular momentum $l$.
The Riccati-Hankel function $\hat{h}^+_l(kr)$, in turn, behaves like an
exponential $\ee^{\ii kr}$ for large $r$.
For $E$ and $k=\sqrt{2\mu E}$ located in the fourth quadrant of their respective
complex planes, this form leads to an exponentially growing amplitude of the
asymptotic wave function.
Complex scaling works because the rotation of $r$ with an angle $\phi > \arg E$
counteracts this exponential growth and renders the wave function normalizable
when expressed along the rotated axis; this is illustrated in
Fig.~\ref{fig:ResonanceWF}.

While for the simple two-body system we considered the intuitive explanation is
immediately clear, it should be noted that complex scaling can readily be
generalized to different coordinate systems and to more particles.
In that case, the asymptotic form of the wave function becomes more involved,
but one should keep in mind that one can always consider the component of the
resonance wave function that is associated with, via separating the overall
system into clusters, a given breakup channel.
For two-body breakup, the relative asymptotic wave function between clusters
will again have an exponential form, and complex scaling amounts to a scaling of
the relative coordinate describing this channel (assuming that each individual
coordinate is rotated equally) because the coordinate transformation is a linear
operation.
Systems of charged particles follow the same logic, only replacing the
Riccati-Hankel function with an appropriate pure Coulomb wave function.

Moreover, the method can be expressed using an abstract operator formalism, and,
reassuringly, there exists a rigorous underpinning of what we discussed above on
intuitive grounds: the Aguilar-Balslev-Combes (ABC)
theorem~\citep{Aguilar:1971ve,Balslev:1971vb} establishes that bound-state
energies remain unchanged under complex scaling and that the continuum branch
cut (describing scattering states) that would normally be located along the
positive real axis is rotated by an angle ${-}2\theta$ into the lower half
plane of the energy, thus revealing resonance poles on the second Riemann sheet,
cf.~Eq.~\eqref{eq:p-scaled}.

Variants such as \emph{exterior complex scaling} exist to deal with potentials
that have an analytic structure precluding a simple uniform coordinate rotation
-- or which are only known numerically and can therefore not easily be
analytically continued.
The general idea of this version of complex scaling is that in order to render
wave functions square integrable, it is really only necessary to tame the
exponentially growing tail.
The ``interior part'' of the wave function, corresponding to the domain over
which the potential cannot be neglected, can still be expressed along the
original, unrotated, coordinate.
For more details about these topics, we refer the interested reader to the
review by \citet{Myo:2020rni} and to the textbook of \citet{Moiseyev:2011}.

\subsection{Berggren basis and Gamow Shell Model}
\label{sec_GSM}

In the so-called Continuum Shell Model, the Feshbach projection
formalism discussed in Sec.~\ref{sec:Feshbach-Details} is used to couple the
space of bound states with the continuum of scattering states.
On the other hand, in the uniform complex scaling approach, introduced in
Sec.~\ref{sec:ComplexScaling}, the real-momentum continuum of scattering states
is rotated into the fourth quadrant of the complex momentum plane to
reveal low-lying Gamow states.
This begs the question: could we directly express the Hamiltonian in a basis
that includes not only bound and scattering states, but also resonances as Gamow
states?

The \emph{Berggren basis}~\citep{berggren68_32,berggren93_481} is a scheme
that implements this idea based on a completeness relation at the
single-particle level.
Specifically, it is built upon the Newton completeness
relation~\citep{newton82_b6} between bound and scattering states, and Cauchy's
integral theorem.
For a system that supports only a set of discrete bound states
(labeled by an integer index $n$) in addition to the scattering continuum,
labeled by a momentum (or wave number) $k$, we assume that the reader is
familiar with the resolution of the identity written as follows:
\begin{equation}
 \sum_{n} \ket{u_{n}}\bra{u_{n}}
 + \int_0^{\infty} \dd k \, \ket{u_{k}}\bra{u_{k}} = \Id \,,
\label{eq_Newton_basis}
\end{equation}
where we use $u$ with a reduced radial wave function in mind.\footnote{
Specifically,for a spherically symmetric potential, we generally assume that
the system is decomposed in partial waves and that wave functions in coordinate
space are expressed as $\psi(r,\theta,\phi) = u(r)/r Y_{l}^{m}(\theta,\phi)$.}
In practice, for a many-body setup, one can arbitrarily pick a Hamiltonian with,
for example, a Wood-Saxon potential acting on each particle individually, to
generate a one-body basis that satisfies Eq.~\eqref{eq_Newton_basis}.
The basic idea of the Berggren basis is to consider a non-Hermitian extension of
this one-body setup so that resonance states can be included in the discrete sum
over states.

To achieve this, following an approach that is closely related to complex
scaling, the integral must be first extended over a closed contour going into
the first, second, and third quadrant (in this order) of the complex-momentum
plane.
Then, it must be deformed in the fourth quadrant to go around any physically
relevant poles of the single-particle $S$ matrix, and the contribution from
those poles must be included in the sum in Eq.~\eqref{eq_Newton_basis}.
The difficult part in carrying out this construction is to prove that only the
states located in the fourth quadrant contribute to the completeness (keeping in
mind that there are additional poles, stemming from the anti-resonance partners,
in the third quadrant).
This has been accomplished by \citet{berggren68_32,berggren93_481}, thus giving
the method its name.
Ultimately, for a given partial wave, the Berggren basis completeness
relation can be written as
\begin{equation}
 \sum_{n} \ket{u(k_n)}\bra{\tilde{u}(k_n)}
 + \int_{\mathcal{L}^+} \dd k \, \ket{u(k)}\bra{\tilde{u}(k)} = \Id \,,
 \label{eq_BB}
\end{equation}
where the label $n$ now runs over both bound and resonant states.
$\mathcal{L}^+$ denotes the contour that defines the continuum of
complex-momentum scattering states, and the tilde indicates time reversal
applied to a state.\footnote{This is a subtle but important point, closely
related to the use of the so-called $c$-product (or bi-orthogonal product)
in the context of complex scaling.
We refer the reader to the references cited in Sec.~\ref{sec:ComplexScaling}
and in particular to \citet{Moiseyev:2011}'s book for more details about this.}

The Berggren basis was originally introduced to solve a practical problem, which
is that when a pole is close to the continuum, the density of scattering states
is very high, which in practice requires discretizing the integral entering
Eq.~\eqref{eq_Newton_basis} with a considerable number of points, making the use
of this basis computationally inefficient.
In the Berggren basis, this problem is largely avoided by keeping the deformed
contour $\mathcal{L}^+$ away from any pole.
We emphasize here that working with Gamow states implies that we use outgoing
boundary conditions, and that therefore any operator associated with an
observable expressed in this basis is, by definition, represented by a
complex-symmetric (rather than Hermitian) matrix.

In the many-body context, the discretization of the Berggren basis makes it
possible to build Slater determinants that includes bound, scattering, and
resonance orbitals on an equal footing.
Doing so with the intent to solve the configuration-interaction (CI) problem
effectively generalizes the shell model in the complex-energy plane, which is
precisely the definition of the \emph{Gamow Shell Model
(GSM)}~\citep{michel09_2,michel21_b260}.
This approach makes it possible to solve the many-body time-independent
Schr\"odinger equation $H \ket{\Psi} = E \ket{\Psi}$ with outgoing
boundary conditions while maintaining the conceptual and computational aspects
of the shell model.

One of the most interesting aspects of this approach is that, in principle, one
can reconstruct any many-body asymptotic from the linear combination of Slater
determinants, provided the underlying Berggren basis in each partial wave is
properly defined.
In other words, just like with the uniform complex-scaling method, continuum
couplings scale automatically with the number of particles and there is no need
to define explicitly the decay channels that one wants to include.

Despite these desirable features, there are still challenges in practice.
The first difficulty encountered in the GSM -- as in any other approach rooted
in the quasi-stationary formalism, for that matter -- is the identification of
physical states.
Indeed, as the Hamiltonian matrix becomes complex symmetric, it can yield many
complex-energy solutions, most of which are not associated with actual many-body
resonance states, but rather arise from the rotated one-body continuum levels.
The variational principle becomes obviously inapplicable, and although there is
a stationary principle for complex-energy Gamow
states~\citep{moiseyev98_92,rotureau09_140}, this cannot be easily used in this
context.

By definition, physical states are those with a well-defined structure that can
be observed and that are associated with poles of the many-body $S$ matrix.
This means that physical states should correspond to solutions that are
essentially invariant under basis change, similar to what is ensured by the ABC
theorem for complex scaling.
However, repeatedly diagonalizing a large, dense, complex-symmetric matrix to
identify these invariant solutions is computationally costly.
An elegant solution to this problem is provided by the so-called overlap
method~\citep{michel02_8,michel03_10}.
The basic idea is to recognize that if ones uses a truncated Berggren basis
including only discrete states that desribe poles of the
single-particle $S$ matrix and contribute to the structure (``pole space''), all
many-body solutions obtained must be approximations of discrete many-body states
associated with poles of the many-body $S$ matrix.
The goal is then to extract the solution for the full problem (including all
scattering states) that has maximal overlap with the approximation of the
physical state selected in just the pole space.
In practice, the overlap method works well, even for fairly broad resonances,
where the concept of structure starts to dissolve.

A major limitation of the GSM is its computational cost.
Indeed, the discretization of the Berggren basis usually requires about 30-45
scattering states per partial wave.
This means that in practice truncations must be applied.
This can have adverse consequences in particular for reproducing the width of
resonances, which are more sensitive to the completeness of the many-body basis
than energies.
This problem is addressed in Sec.~\ref{sec_GDMRG}.

Another limitation is related to the extraction of reaction observables.
The asymptotic behavior of many-body resonances is obtained via configuration
mixing instead of being expressed via weights of distinct decay channels.
For that reason, the GSM was extended using the Resonating Group Method (RGM) to
allow for explicit channel couplings, constructed with the expected physical
structure in mind.
The method is constructed similarly to the No-Core Shell Model with Continuum
(NCSMC) approach, where the states of a target, described microscopically within
the model, are coupled to the states of a projectile.
The projectile can either be a single particle or a small cluster, itself
described within the model, just like the target.
Such a formulation is quite involved technically, but it allows for a
realization of the unification of nuclear structure and reactions that naturally
connects reaction observables to the structure, and vice versa.

While we have covered here the basic idea and features of the GSM, we refer
the interested reader to the thorough review by \citet{michel21_b260} for
further details.

\subsection{Gamow Density Matrix Renormalization Group}
\label{sec_GDMRG}

The DMRG method was originally introduced in condensed matter physics to
describe strongly correlated one-dimensional lattice
systems~\citep{white92_488,white93_491}, but it turned out to be more generally
useful.
Its central idea is that, when a many-body system is partitioned into two
complementary subsystems, the eigenvalues of the reduced density matrix of one
subsystem quantify the entanglement between the two subsystems and therefore
rank the many-body configurations by their importance in reconstructing the
target state.
By truncating these eigenvalues to retain only the most important
configurations, it is possible to build a low-dimensional renormalized
Hamiltonian that represent the target state optimally, with a controlled error.
Modern reformulations of the DMRG in terms of matrix product
states~\citep{rommer97_3246} make the underlying tensor-network structure
manifest and have turned the method into a standard tool in quantum chemistry
and condensed matter
physics~\citep{peschel99_b245,schollwock05_479,chan11_3194}.

In nuclear physics, the DMRG method was first applied to solve the shell model
problem~\citep{dukelsky99_2004,dukelsky01_2003,pittel01_2008,dukelsky02_1568},
but it was realized that the convergence of the DMRG method in nuclei tends to
be slow due to strong, highly non-perturbative correlations, and possibly also
due to a volume-law scaling of entanglement in
nuclei~\citep{pazy23_3230,gu23_3201}.

Nevertheless, the DMRG method emerged as a natural choice to overcome the
problem of the computational cost due to the large continuum space encountered
in the GSM (see Sec.~\ref{sec_GSM}).
In fact, the DMRG method is particularly well suited to the specific structure
of the GSM problem.
In the Berggren basis, the natural bi-partition of the single-particle space is
not between bound and scattering states, as in the continuum shell model, but
between, on the one hand, \emph{general} discrete states (bound and resonant)
and, on the other hand, scattering states.
Discrete states control the overall structure of a many-body state, while
individual scattering states act more like a correction.
This observation led to the proposition of the Gamow DMRG (G-DMRG) method, as a
generalization of the DMRG method in the Berggren basis that exploits the low
entanglement between the spaces of discrete and scattering
states~\citep{rotureau06_15,rotureau09_140}.

In the original Wilsonian formulation of the G-DMRG, the single-particle
orbitals are split into a ``reference space'' $\mathcal{H}_A$, built exclusively
from the discrete (pole) orbitals selected when constructing the Berggren basis,
and an ``environment'' $\mathcal{H}_B$ containing the discretized scattering
states along the contour $\mathcal{L}^+$.
The Hamiltonian is first diagonalized in $\mathcal{H}_A$ alone to produce a
reference state $\ket{\Psi_0}$, which plays the role of a zeroth-order
approximation to the targeted many-body Gamow state.
At each subsequent iteration, one scattering orbital is moved from
$\mathcal{H}_B$ into the active space, all possible many-body configurations
with well-defined total angular momentum and parity are constructed, and the
Hamiltonian is diagonalized in this enlarged space.
The key step is then to build the reduced density matrix $\rho_B$ of the newly
added subspace by tracing the resulting many-body state over the degrees of
freedom of $\mathcal{H}_A$, schematically
\begin{equation}
 \rho_B = \mathrm{Tr}_A \ket{\Psi}\bra{\tilde\Psi} \,.
\label{eq_rhoB}
\end{equation}
Diagonalizing $\rho_B$ yields eigenvalues $\omega_\alpha$ and eigenvectors that
provide an optimal representation of the environment subspace for the targeted
state, and only the most significant eigenvectors are retained to form the input
for the next iteration.

This iterative sweep over the orbitals of the environment constitutes the
``warm-up'' phase, and is typically followed by one or more ``sweep'' phases in
which the procedure is repeated in alternating directions until convergence is
reached for the energy and width of the target state.
In practice, it is more efficient to construct natural orbitals, defined as
eigenvectors of the one-body density
matrix~\citep{brillouin33,lowdin55_2499,lowdin56_2498}, and to perform the
warm-up phase again in this basis.
Natural orbitals capture most of the short-range correlations at the
single-particle level, so that the DMRG renormalization only has to handle the
remaining longer-range correlations, often making it possible to bypass the
sweep phase altogether~\citep{fossez17_1916,fossez22_2540}.

In the G-DMRG method, the problem of the identification of physical states is
solved using the overlap method of the GSM at each iteration, and more precisely
between the target state and the state retained at the previous iteration.
The main advantage of the G-DMRG method is that the rank of the largest reduced
density matrix to be diagonalized is nearly independent of the number of
scattering orbitals, so that calculations converge exponentially with the number
of retained states and the ratio to the full GSM dimension decreases rapidly as
the model space is enlarged.
For this reason, the G-DMRG has become a tool of choice for the description of
multi-nucleon resonances.

Of course, this approach also has limitations, most notably the difficulty to
calculate observables other than the energy.
The calculation of reaction observables is also currently an unsolved problem,
which would probably require an adaption of the RGM (as discussed in the context
of the Berggren basis in Sec.~\ref{sec_GSM}).

\subsection{Analytic Continuation in the Coupling Constant}
\label{sec:ACCC}

As discussed in Section \ref{sec:ResonanceTypes}, resonant states exhibit a
strong similarity to bound ones.
Moreover, many types of resonant states -- particularly shape resonances and
virtual states -- can be transformed into bound states by increasing the
attractive component of the Hamiltonian.
Building on this observation, \citet{kukulin1977description} introduced the
method of \emph{analytic continuation in the coupling constant (ACCC)} to
determine the trajectory of a resonant state as a function of the coupling
constant (strength parameter), controlled by introducing a strength
parameter $\lambda$ that governs the attractive interaction.

Specifically, for a given initial interaction $V$, ACCC introduces the
transformation $V \to \lambda V$ and computes wave numbers $k(\lambda)$ of a
given initial bound state as $\lambda$ is varied.
Near the threshold where the bound state turns into a resonance, the wave number
behaves as
\begin{equation}
 k = \sqrt{{-}2\mu E} \sim x \equiv \left\{
  \begin{array}{lr}
   \lambda -\lambda _{0} & \text{for a virtual state} \,, \\
   \sqrt{\lambda -\lambda _{0}} & \text{for a resonant state} \,,
  \end{array}%
 \right.
\label{ACCC_ff}
\end{equation}
when it is measured relative to this threshold (characterized by the critical
strength parameter $\lambda_0$).
This behavior permits one to treat $k$ as an analytic function of $x$, which can
be continued from the bound-state region ($\lambda > \lambda_0$) into the
resonance region ($\lambda < \lambda_0$).

In practical applications, the method requires as input only a set of
bound-state energies $E(\lambda)$ together with the critical coupling strength
$\lambda_0$ in order to determine the resonance position by extrapolating
former functional to the physical value of the coupling constant.
Although the procedure is straightforward to implement, the reliability of the
extracted resonance parameters depends sensitively on the numerical accuracy of
the bound-state energies and, in particular, on the precise determination of
$\lambda_0$, which is often the most delicate aspect of the calculation.
In the vicinity of the critical point, numerical challenges arise because the
wave function becomes increasingly spatially extended, rendering an accurate
determination of $\lambda_0$ difficult.\footnote{%
Regarding the general challenge of extrapolating from bound states to
resonances, we note that recent work~\citep{Yapa:2023xyf} has shown how it can
be achieved with an extension of the technique that has become known as
``eigenvector continuation''~\citep{Duguet:2023wuh}.}
The overall performance of the method is also influenced by the choice of the
auxiliary potential employed to generate the bound-state input and by the
resulting smoothness of the associated S-matrix pole trajectory.

Extending the ACCC method beyond two-body systems is challenging.
A naive global scaling of the two-body interaction is inadequate, as it modifies
the thresholds associated with bound subsystems.
This can, in turn, lead to the emergence of additional thresholds, thereby
invalidating the threshold behavior assumed in Eq.~\eqref{ACCC_ff}.
To circumvent these complications, one may introduce an additional attractive
many-body interaction that does not act within the subsystems.
However, this auxiliary interaction must be designed with particular care
closely matching the original Hamiltonian and preserving the structure of the
resonant-state wave function.

\subsection{Effective field theory and complex amplitudes}

As discussed in Sec.~\ref{sec:QuantumResonances},
resonance poles in the $S$ matrix, arising from the interaction among particles,
are really poles of the $T$ matrix, $T(E)$, or equivalently the scattering
amplitude.
For short-range interactions and $E = E_k = k^2/(2\mu)$ as defined in
Sec.~\ref{sec:ScattTheory}, the $T$ matrix, as a function of $k$, can famously
be written as
\begin{equation}
 T_{\ell}(k) \equiv T_{\ell}(E_k;k,k)
 = \frac{k^{2\ell}}{k^{2\ell+1}\left[\cot\delta_\ell(k) - \ii\right]} \,,
\label{eq:T-K}
\end{equation}
with the \emph{effective range expansion (ERE)}
\begin{equation}
 k^{2\ell+1}\cot\delta_\ell(k)
 = {-}\frac{1}{a_{\ell}} + \frac{r_{\ell}}{2}k^2 + \OO(k^4) \,.
\label{eq:ERE}
\end{equation}
For $S$ waves ($\ell=0$), the parameters $a_0$ and $r_0$ in Eq.~\eqref{eq:ERE}
are called, respectively, the \emph{scattering length} and the \emph{effective
range}.

Effective field theories (EFTs)~\citet{Hammer:2019poc} for systems with
short-range interactions are constructed to map the ERE in the two-particle
sector to a sequence of (regularized) zero-range interactions, relating the ERE
parameters to the \textit{a priori} unknown coupling  strengths of these
interactions, called ``low-energy constants (LECs).''
Of particular interest is the situation where the scattering length $a_0$ is
large (meaning much larger than the typical length scale associated with the
system), as in this phenomenon is linked to the occurence of shallow (meaning
``low energy'') poles in the $T$ matrix.
For example, if $a_0$ is large and positive, one can neglect the quadratic term
in Eq.~\eqref{eq:ERE} and find that $T_0(k)$ will have a pole at a purely
imaginary momentum $k=\ii\kappa$ with $\kappa = 1/a$.
In nuclear physics, the deuteron bound state is a famous example of such a
low-energy pole.
If $a_0$ is large but \emph{negative}, there is an analogous pole on the
negative imaginary axis.
This is the virtual-state phenomenon that was mentioned previously
(Sec.~\ref{sec:ResonanceTypes}), and it occurs in nuclear physics for example in
the singlet $S$-wave channel of nucleon-nucleon scattering.

If one keeps the quadratic term in Eq.~\eqref{eq:ERE}, the denominator in
Eq.~\eqref{eq:T-K} can support a richer set of zeros:
\begin{equation}
 {-}\frac{1}{a_{\ell}} + \frac{r_{\ell}}{2}k^2 - \ii k^{2\ell+1} = 0 \,.
 \label{eq:T-K}
\end{equation}
In particular, for appropriate combinations of $a_{\ell}$ and $r_{\ell}$, it
becomes possible to find complex solutions that describe physical resonances
(i.e., occurring a complex-conjugate pairs with one solution in the fourth
quadrant of the complex $k$ plane).
For example, for $\ell=0$ one can easily solve the quadratic equation and find
the pair of zeros
\begin{equation}
 k = \pm\sqrt{\frac{2r_0}{a_0} - 1} + \frac{\ii}{r_0} \,.
\end{equation}
For these to be in the lower-half $k$ plane, one needs $r_0<0$, and then, for
the root to be real, it is required that $2r_0/a_0 > 1$, which can be achieved
if the scattering length $a_0$ is negative and sufficiently large.

In the EFT context, this scenario was first investigated by
\citet{Bedaque:2003wa}, with particular focus on the implications for the EFT
\emph{power counting}, that is, the assignment of interactions terms to orders
in a systematic expansion.
Important applications of the general approach are primarily in the EFT
description of so-called \emph{halo nuclei}~\citep{Bertulani:2002sz}, and a
comprehensive overview of the pole structure in this context can be found in the
review of \citet{Hammer:2017tjm}, covering different partial waves.
Further details, and in particular the important extension to systems of charged
particles, can be found in Refs.~\citep{Higa:2008dn} and~\citep{Gelman:2009be},
while \citet{Habashi:2020qgw} carefully constructed the EFT for narrow $S$-wave
resonance, including subleading orders in strict perturbation theory.

\subsection{Stabilization methods}

``Stabilization Methods'' is a term that refers to a collection of similar
methods that can be used to infer resonance properties from a Hermitian setup,
using only real energy eigenvalues.
The basic idea, first introduced by \citet{Hazi:1970aa}, is to study the energy
spectrum as a function of a parameter as this parameter is varied.
For a fixed Hamiltonian $H$ that describes the physical system of interest,
assumed here not to have any inherent parametric dependence, such a dependence
can be introduced by means of representing it in a truncated Hilbert space.

The perhaps cleanest way to do so is by considering the system in a box, with
the edge length $L$ imposing an infrared cutoff on the states that can be
represented.
As $L$ is varied, resonance states become manifest as avoided crossing between
states with identical quantum numbers, and from the energies where these avoided
level crossings (ALCs) occur one can infer the physical (infinite-volume) energy
of the resonance.
At least for simple systems, one can also extract information about the
resonance width by relating it to the ``sharpness'' of the avoided crossing.

\begin{figure}[tbp]
 \centering
 \includegraphics[width=0.75\textwidth]{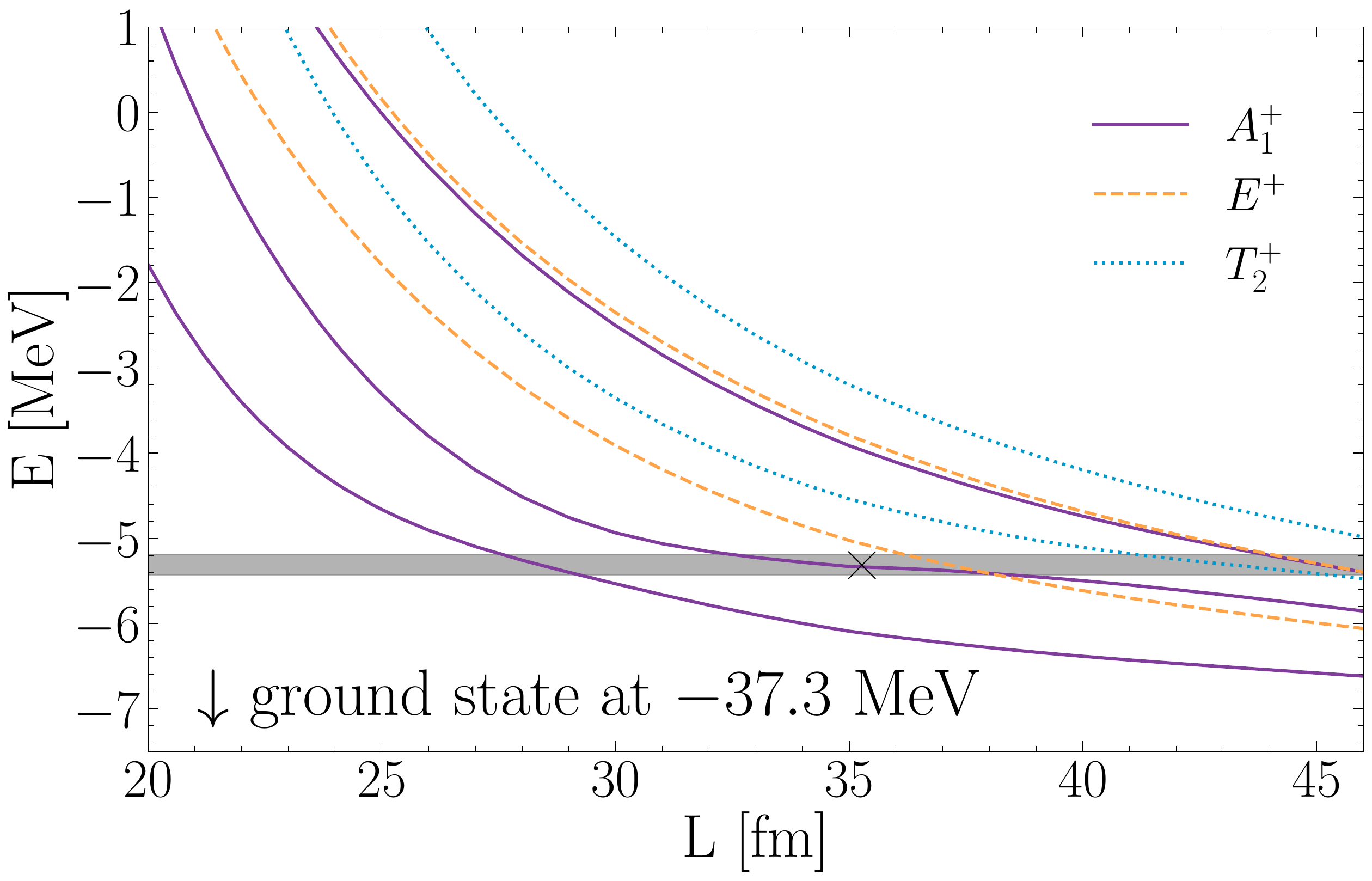}
 \caption{%
Energy spectrum of three bosons in finite volume for different box sizes $L$
interacting via the potential modeled as a sum of an attractive and a (shifted)
repulsive Gaussian term that together generate a shape resonance, following
\citet{Blandon:2007aa}.
As this calculation was performed in a cubic box, spherical symmetry is broken
and energy levels are labeled by irreducible representations of the cubic group.
States corresponding to the $A_1$ cubic representation, roughly corresponding to
$S$-wave states in infinite volume, are shown as solid lines, whereas $E^+$ and
$T_2^+$ states are indicated as dashed and dotted lines, respectively.
The shaded area indicates the resonance position as calculated in the reference
paper~\citep{Blandon:2007aa}, which is nicely matched by the avoided crossings
in the energy spectrum.
The cross marks the resonance energy extracted via the inflection-point
prescription of \citet{Klos:2018sen} (this figure has been adapted from that
work).
}
\label{fig:En-3b-Blandon-Pp}
\end{figure}

For the case of a periodic boundary condition, \citet{Wiese:1988qy} provides a
nice illustration, based on the finite-volume quantization condition (also known
as a ``Lüscher formula''~\citep{Luscher:1985dn,Luscher:1986pf,Luscher:1990ux}),
for how the ALC is related to the behavior of the scattering phase shift near
the resonance energy.
While this explanation in terms of the phase shift is clear for two-body
resonances~\citep{Rummukainen:1995vs}, it has also been demonstrated that
genuine few-body resonances can be identified via ALCs~\citep{Klos:2018sen}.
An example of such a calculation, for the specific example of a three-boson
system that supports a resonance in addition to a bound ground state, is shown
in Fig.~\ref{fig:En-3b-Blandon-Pp}.
While in this case the avoided crossings are relatively subtle, the example
illustrates how the method works not only for two-body resonances, but also in
the few-body sector, with further cases discussed by \citet{Klos:2018sen}.

These finite-volume calculations with periodic boundary conditions are of course
only one particular example of a stabilization calculation.
\citet{Moiseyev:2011} discusses the method generally, using a box with hard-wall
boundary condition as specific example, and also relates the ALC phenomenon to
the density of states in the continuum spectrum, with a localized increase
resulting from a resonance as a ``bound state embedded within the continuum.''

\section{Concluding Remarks}

At the outset of this chapter, we emphasized the central role of resonances in
nuclear physics, a claim that we hope the reader agrees with upon reaching
this point.
Starting from the classical resonance phenomenon, we reviewed the
quantum-mechanical description of resonances and discussed the various ways in
which they manifest in nuclear experiments, particularly at the interface
between nuclear structure and reaction dynamics.
Constructing a unified theoretical framework that consistently encompasses these
aspects remains one of the major challenges in nuclear theory, and progress
toward this objective critically depends on treating resonance states on the
same conceptual and computational footing as bound states and scattering
observables.
This consideration applies broadly, and it is especially important in the
vicinity of the nuclear drip lines,  where the nature of nuclei as open quantum
systems becomes prominently exposed.

Resonances are a fascinating topic because they pose challenges along multiple
directions: experimentally, conceptually, and computationally.
This means that many open questions remain to be explored, and in closing this
exposition, it is our hope that the reader feels inspired to contribute to this
in their own research.

\begin{ack}{}{}
We thank Nuwan Yapa for giving us permission to adapt his original work for
creating figures~\ref{fig:s_matrix-k} and~\ref{fig:ResonanceWF}.
This work was supported in part by the U.S.\ National Science Foundation under
Grants No.~PHY--2044632 and PHY--2238752.
This material is based upon work supported by the U.S.\ Department of Energy,
Office of Science, Office of Nuclear Physics, under the FRIB Theory Alliance,
Award No.~DE-SC0013617.
\end{ack}

\begin{ack}[Declaration of Generative AI and AI-assisted technologies in the
 writing process]{}
 Gemini (Google) has been used to review parts of the manuscript during the
 writing process and to assist with literature research in some cases.
 All text has been written by the authors, with only limited use of AI -- via
 Overleaf and Claude (Anthropic) -- to streamline the wording in some sections.
\end{ack}

\bibliographystyle{Harvard}
\begin{thebibliography*}{112}
\providecommand{\bibtype}[1]{}
\providecommand{\natexlab}[1]{#1}
{\catcode`\|=0\catcode`\#=12\catcode`\@=11\catcode`\\=12
|immediate|write|@auxout{\expandafter\ifx\csname
  natexlab\endcsname\relax\gdef\natexlab#1{#1}\fi}}
\renewcommand{\url}[1]{{\tt #1}}
\providecommand{\urlprefix}{URL }
\expandafter\ifx\csname urlstyle\endcsname\relax
  \providecommand{\doi}[1]{doi:\discretionary{}{}{}#1}\else
  \providecommand{\doi}{doi:\discretionary{}{}{}\begingroup
  \urlstyle{rm}\Url}\fi
\providecommand{\bibinfo}[2]{#2}
\providecommand{\eprint}[2][]{\url{#2}}

\bibtype{Article}%
\bibitem[Aguilar and Combes(1971)]{Aguilar:1971ve}
\bibinfo{author}{Aguilar J} and  \bibinfo{author}{Combes JM}
  (\bibinfo{year}{1971}).
\bibinfo{title}{A class of analytic perturbations for one-body
  {{Schr{\"o}dinger Hamiltonians}}}.
\bibinfo{journal}{{\em Comm. Math. Phys.}} \bibinfo{volume}{22}
  (\bibinfo{number}{4}): \bibinfo{pages}{269--279}.
ISSN \bibinfo{issn}{1432-0916}. \bibinfo{doi}{\doi{10.1007/BF01877510}}.

\bibtype{Article}%
\bibitem[{Al Kalanee} et al.(2013)]{kalanee13_1909}
\bibinfo{author}{{Al Kalanee} T}, \bibinfo{author}{Gibelin J},
  \bibinfo{author}{{Roussel-Chomaz} P}, \bibinfo{author}{Keeley N},
  \bibinfo{author}{Beaumel D}, \bibinfo{author}{Blumenfeld Y},
  \bibinfo{author}{{Fern\'andez-Dom\'inguez} B}, \bibinfo{author}{Force C},
  \bibinfo{author}{Gaudefroy L}, \bibinfo{author}{Gillibert A},
  \bibinfo{author}{Guillot J}, \bibinfo{author}{Iwasaki H},
  \bibinfo{author}{Krupko S}, \bibinfo{author}{Lapoux V},
  \bibinfo{author}{Mittig W}, \bibinfo{author}{Mougeot X},
  \bibinfo{author}{Nalpas L}, \bibinfo{author}{Pollacco E},
  \bibinfo{author}{Rusek K}, \bibinfo{author}{Roger T},
  \bibinfo{author}{Savajols H}, \bibinfo{author}{{de S\'er\'eville} N},
  \bibinfo{author}{Sidorchuk S}, \bibinfo{author}{Suzuki D},
  \bibinfo{author}{Strojek I} and  \bibinfo{author}{Orr NA}
  (\bibinfo{year}{2013}).
\bibinfo{title}{Structure of unbound neutron-rich ${ {}^{9}\text{He} }$ studied
  using single-neutron transfer}.
\bibinfo{journal}{{\em Phys. Rev. C}} \bibinfo{volume}{88}:
  \bibinfo{pages}{034301}.
\bibinfo{url}{\url{http://doi.org/10.1103/PhysRevC.88.034301}}.

\bibtype{Article}%
\bibitem[Auerbach and Zelevinsky(2011)]{auerbach11_879}
\bibinfo{author}{Auerbach N} and  \bibinfo{author}{Zelevinsky V}
  (\bibinfo{year}{2011}).
\bibinfo{title}{Super-radiant dynamics, doorways and resonances in nuclei and
  other open mesoscopic systems}.
\bibinfo{journal}{{\em Rep. Prog. Phys.}} \bibinfo{volume}{74}:
  \bibinfo{pages}{106301}.
\bibinfo{url}{\url{https://doi.org/10.1088/0034-4885/74/10/106301}}.

\bibtype{Article}%
\bibitem[Baldwin and Klaiber(1947)]{baldwin1947photo}
\bibinfo{author}{Baldwin GC} and  \bibinfo{author}{Klaiber GS}
  (\bibinfo{year}{1947}).
\bibinfo{title}{Photo-fission in heavy elements}.
\bibinfo{journal}{{\em Physical Review}} \bibinfo{volume}{71}
  (\bibinfo{number}{1}): \bibinfo{pages}{3}.

\bibtype{Article}%
\bibitem[Balslev and Combes(1971)]{Balslev:1971vb}
\bibinfo{author}{Balslev E} and  \bibinfo{author}{Combes JM}
  (\bibinfo{year}{1971}).
\bibinfo{title}{Spectral properties of many-body {{Schr{\"o}dinger}} operators
  with dilatation-analytic interactions}.
\bibinfo{journal}{{\em Commun.Math. Phys.}} \bibinfo{volume}{22}
  (\bibinfo{number}{4}): \bibinfo{pages}{280--294}.
ISSN \bibinfo{issn}{1432-0916}. \bibinfo{doi}{\doi{10.1007/BF01877511}}.

\bibtype{Article}%
\bibitem[Baz'(1967)]{baz1967}
\bibinfo{author}{Baz' AI} (\bibinfo{year}{1967}).
\bibinfo{title}{A quantum mechanical calculation of the collision time}.
\bibinfo{journal}{{\em Sov. J. Nucl. Phys.}}  (\bibinfo{number}{5}):
  \bibinfo{pages}{161}.

\bibtype{Book}%
\bibitem[Baz' et al.(1969)]{baz69_b3}
\bibinfo{author}{Baz' AI}, \bibinfo{author}{Zel'dovich YB} and
  \bibinfo{author}{Perelomov AM} (\bibinfo{year}{1969}).
\bibinfo{title}{Scattering, {R}eactions and {D}ecay in {N}onrelativistic
  {Q}uantum {M}echanics}, \bibinfo{edition}{first} ed.,
  \bibinfo{publisher}{Israel Program for Scientific Translations, Jerusalem}.

\bibtype{Article}%
\bibitem[Bedaque et al.(2003)]{Bedaque:2003wa}
\bibinfo{author}{Bedaque PF}, \bibinfo{author}{Hammer HW} and
  \bibinfo{author}{{van Kolck} U} (\bibinfo{year}{2003}).
\bibinfo{title}{Narrow resonances in effective field theory}.
\bibinfo{journal}{{\em Phys. Lett. B}} \bibinfo{volume}{569}
  (\bibinfo{number}{3}): \bibinfo{pages}{159--167}.
ISSN \bibinfo{issn}{0370-2693}.
  \bibinfo{doi}{\doi{10.1016/j.physletb.2003.07.049}}.

\bibtype{Article}%
\bibitem[Bennaceur et al.(1999)]{bennaceur99_46}
\bibinfo{author}{Bennaceur K}, \bibinfo{author}{Nowacki F},
  \bibinfo{author}{Oko{\l}owicz J} and  \bibinfo{author}{P{\l}oszajczak M}
  (\bibinfo{year}{1999}).
\bibinfo{title}{Study of the ${ {}^{7}\text{Be} ( p , \gamma ) {}^{8}\text{B}
  }$ and ${ {}^{7}\text{Li} ( n , \gamma ) {}^{8}\text{Li} }$ capture reactions
  using the shell model embedded in the continuum}.
\bibinfo{journal}{{\em Nucl. Phys. A}} \bibinfo{volume}{651}:
  \bibinfo{pages}{289}.
\bibinfo{url}{\url{https://doi.org/10.1016/S0375-9474(99)00133-5}}.

\bibtype{Article}%
\bibitem[Bennaceur et al.(2000)]{bennaceur00_44}
\bibinfo{author}{Bennaceur K}, \bibinfo{author}{Nowacki F},
  \bibinfo{author}{Oko{\l}owicz J} and  \bibinfo{author}{P{\l}oszajczak M}
  (\bibinfo{year}{2000}).
\bibinfo{title}{Analysis of the ${ {}^{16}\text{O} ( p , \gamma )
  {}^{17}\text{F} }$ capture reaction using the shell model embedded in the
  continuum}.
\bibinfo{journal}{{\em Nucl. Phys. A}} \bibinfo{volume}{671}:
  \bibinfo{pages}{203}.
\bibinfo{url}{\url{https://doi.org/10.1016/S0375-9474(99)00851-9}}.

\bibtype{Article}%
\bibitem[Berggren(1968)]{berggren68_32}
\bibinfo{author}{Berggren T} (\bibinfo{year}{1968}).
\bibinfo{title}{On the use of resonant states in eigenfunction expansions of
  scattering and reaction amplitudes}.
\bibinfo{journal}{{\em Nucl. Phys. A}} \bibinfo{volume}{109}:
  \bibinfo{pages}{265}.
\bibinfo{url}{\url{https://doi.org/10.1016/0375-9474(68)90593-9}}.

\bibtype{Article}%
\bibitem[Berggren and Lind(1993)]{berggren93_481}
\bibinfo{author}{Berggren T} and  \bibinfo{author}{Lind P}
  (\bibinfo{year}{1993}).
\bibinfo{title}{Resonant state expansion of the resolvent}.
\bibinfo{journal}{{\em Phys. Rev. C}} \bibinfo{volume}{47}:
  \bibinfo{pages}{768}.
\bibinfo{url}{\url{https://doi.org/10.1103/PhysRevC.47.768}}.

\bibtype{Article}%
\bibitem[Bertulani et al.(2002)]{Bertulani:2002sz}
\bibinfo{author}{Bertulani CA}, \bibinfo{author}{Hammer HW} and
  \bibinfo{author}{{van Kolck} U} (\bibinfo{year}{2002}).
\bibinfo{title}{Effective field theory for halo nuclei: Shallow p-wave states}.
\bibinfo{journal}{{\em Nucl. Phys. A}} \bibinfo{volume}{712}
  (\bibinfo{number}{1}): \bibinfo{pages}{37--58}.
ISSN \bibinfo{issn}{0375-9474}.
  \bibinfo{doi}{\doi{10.1016/S0375-9474(02)01270-8}}.

\bibtype{Article}%
\bibitem[Blandon et al.(2007)]{Blandon:2007aa}
\bibinfo{author}{Blandon J}, \bibinfo{author}{Kokoouline V} and
  \bibinfo{author}{Masnou-Seeuws F} (\bibinfo{year}{2007}).
\bibinfo{title}{{Calculation of three-body resonances using slow-variable
  discretization coupled with a complex absorbing potential}}.
\bibinfo{journal}{{\em Phys. Rev. A}} \bibinfo{volume}{75}:
  \bibinfo{pages}{042508}. \bibinfo{doi}{\doi{10.1103/PhysRevA.75.042508}}.

\bibtype{Article}%
\bibitem[Bohr(1936)]{bohr1936neutroneneinfang}
\bibinfo{author}{Bohr N} (\bibinfo{year}{1936}).
\bibinfo{title}{Neutroneneinfang und bau der atomkerne}.
\bibinfo{journal}{{\em Naturwissenschaften}} \bibinfo{volume}{24}
  (\bibinfo{number}{16}): \bibinfo{pages}{241--245}.

\bibtype{Article}%
\bibitem[Brillouin(1933)]{brillouin33}
\bibinfo{author}{Brillouin L} (\bibinfo{year}{1933}).
\bibinfo{journal}{{\em Acta. Sci. Ind.}} \bibinfo{volume}{71}:
  \bibinfo{pages}{159}.

\bibtype{Article}%
\bibitem[Catara et al.(1984)]{catara84_1880}
\bibinfo{author}{Catara F}, \bibinfo{author}{Insolia A},
  \bibinfo{author}{Maglione E} and  \bibinfo{author}{Vitturi A}
  (\bibinfo{year}{1984}).
\bibinfo{title}{Relation between pairing correlations and two-particle space
  correlations}.
\bibinfo{journal}{{\em Phys. Rev. C}} \bibinfo{volume}{29}:
  \bibinfo{pages}{1091}.
\bibinfo{url}{\url{https://doi.org/10.1103/PhysRevC.29.1091}}.

\bibtype{Article}%
\bibitem[Chan and Sharma(2011)]{chan11_3194}
\bibinfo{author}{Chan GKL} and  \bibinfo{author}{Sharma S}
  (\bibinfo{year}{2011}).
\bibinfo{title}{The density matrix renormalization group in quantum chemistry}.
\bibinfo{journal}{{\em Annu. Rev. Phys. Chem.}} \bibinfo{volume}{62}:
  \bibinfo{pages}{465}.
\bibinfo{url}{\url{https://doi.org/10.1146/annurev-physchem-032210-103338}}.

\bibtype{Article}%
\bibitem[Charity et al.(2023)]{charity23_2960}
\bibinfo{author}{Charity RJ}, \bibinfo{author}{Wylie J}, \bibinfo{author}{Wang
  SM}, \bibinfo{author}{Webb TB}, \bibinfo{author}{Brown KW},
  \bibinfo{author}{Cerizza G}, \bibinfo{author}{Chajecki Z},
  \bibinfo{author}{Elson JM}, \bibinfo{author}{Estee J}, \bibinfo{author}{Hoff
  DEM}, \bibinfo{author}{Kuvin SA}, \bibinfo{author}{Lynch WG},
  \bibinfo{author}{Manfredi J}, \bibinfo{author}{Michel N},
  \bibinfo{author}{{McNeel} DG}, \bibinfo{author}{Morfouace P},
  \bibinfo{author}{Nazarewicz W}, \bibinfo{author}{Pruit CD},
  \bibinfo{author}{Santamaria C}, \bibinfo{author}{Sweany S},
  \bibinfo{author}{Smith J}, \bibinfo{author}{Sobotka LG},
  \bibinfo{author}{Tsang MB} and  \bibinfo{author}{Wuosmaa AH}
  (\bibinfo{year}{2023}).
\bibinfo{title}{Strong evidence for ${ {}^{9}\text{N} }$ and the limits of
  existence of atomic nuclei}.
\bibinfo{journal}{{\em Phys. Rev. Lett.}} \bibinfo{volume}{131}:
  \bibinfo{pages}{172501}.
\bibinfo{url}{\url{https://doi.org/10.1103/PhysRevLett.131.172501}}.

\bibtype{Article}%
\bibitem[Chartier et al.(2001)]{chartier01_2576}
\bibinfo{author}{Chartier M}, \bibinfo{author}{Beene JR},
  \bibinfo{author}{Blank B}, \bibinfo{author}{Chen L},
  \bibinfo{author}{Galonsky A}, \bibinfo{author}{Gan N},
  \bibinfo{author}{Govaert K}, \bibinfo{author}{Hansen PG},
  \bibinfo{author}{Kruse J}, \bibinfo{author}{Maddalena V},
  \bibinfo{author}{Thoennessen M} and  \bibinfo{author}{Varner RL}
  (\bibinfo{year}{2001}).
\bibinfo{title}{Identification of the ${ {}^{10}\text{Li} }$ ground state}.
\bibinfo{journal}{{\em Phys. Lett. B}} \bibinfo{volume}{510}:
  \bibinfo{pages}{24}.
\bibinfo{url}{\url{https://doi.org/10.1016/S0370-2693(01)00602-5}}.

\bibtype{Article}%
\bibitem[Chin et al.(2010)]{Chin:2010crf}
\bibinfo{author}{Chin C}, \bibinfo{author}{Grimm R}, \bibinfo{author}{Julienne
  P} and  \bibinfo{author}{Tiesinga E} (\bibinfo{year}{2010}).
\bibinfo{title}{Feshbach resonances in ultracold gases}.
\bibinfo{journal}{{\em Reviews of Modern Physics}} \bibinfo{volume}{82}
  (\bibinfo{number}{2}): \bibinfo{pages}{1225--1286}.
  \bibinfo{doi}{\doi{10.1103/RevModPhys.82.1225}}.

\bibtype{Article}%
\bibitem[Deltuva(2018)]{deltuva18_2079}
\bibinfo{author}{Deltuva A} (\bibinfo{year}{2018}).
\bibinfo{title}{Tetraneutron: {R}igorous continuum calculation}.
\bibinfo{journal}{{\em Phys. Lett. B}} \bibinfo{volume}{782}:
  \bibinfo{pages}{238}.
\bibinfo{url}{\url{https://doi.org/10.1016/j.physletb.2018.05.041}}.

\bibtype{Article}%
\bibitem[Dobaczewski et al.(2007)]{dobaczewski07_17}
\bibinfo{author}{Dobaczewski J}, \bibinfo{author}{Michel N},
  \bibinfo{author}{Nazarewicz W}, \bibinfo{author}{P{\l}oszajczak M} and
  \bibinfo{author}{Rotureau J} (\bibinfo{year}{2007}).
\bibinfo{title}{Shell structure of exotic nuclei}.
\bibinfo{journal}{{\em Prog. Part. Nucl. Phys.}} \bibinfo{volume}{59}:
  \bibinfo{pages}{432}.
\bibinfo{url}{\url{https://doi.org/10.1016/j.ppnp.2007.01.022}}.

\bibtype{Article}%
\bibitem[Duer et al.(2022)]{duer22_2494}
\bibinfo{author}{Duer M}, \bibinfo{author}{Aumann T},
  \bibinfo{author}{Gernh\"auser R}, \bibinfo{author}{Panin V},
  \bibinfo{author}{Paschalis S}, \bibinfo{author}{Rossi DM},
  \bibinfo{author}{Achouri NL}, \bibinfo{author}{Ahn D}, \bibinfo{author}{Baba
  H}, \bibinfo{author}{Bertulani CA}, \bibinfo{author}{B\"ohmer M},
  \bibinfo{author}{Boretzky K}, \bibinfo{author}{Caesar C},
  \bibinfo{author}{Chiga N}, \bibinfo{author}{Corsi A},
  \bibinfo{author}{{Cortina-Gil} D}, \bibinfo{author}{Douma CA},
  \bibinfo{author}{Dufter F}, \bibinfo{author}{Elekes Z}, \bibinfo{author}{Feng
  J}, \bibinfo{author}{{Fern\'andez-Dom\'inguez} B}, \bibinfo{author}{Forsberg
  U}, \bibinfo{author}{Fukuda N}, \bibinfo{author}{Gasparic I},
  \bibinfo{author}{Ge Z}, \bibinfo{author}{Gheller JM},
  \bibinfo{author}{Gibelin J}, \bibinfo{author}{Gillibert A},
  \bibinfo{author}{Hahn KI}, \bibinfo{author}{Hal\'asz Z},
  \bibinfo{author}{Harakeh MN}, \bibinfo{author}{Hirayama A},
  \bibinfo{author}{Holl M}, \bibinfo{author}{Inabe N}, \bibinfo{author}{Isobe
  T}, \bibinfo{author}{Kahlbow J}, \bibinfo{author}{{Kalantar-Nayestanaki} N},
  \bibinfo{author}{Kim D}, \bibinfo{author}{Kim S}, \bibinfo{author}{Kobayashi
  T}, \bibinfo{author}{Kondo Y}, \bibinfo{author}{K\"orper D},
  \bibinfo{author}{Koseoglou P}, \bibinfo{author}{Kubota Y},
  \bibinfo{author}{Kuti I}, \bibinfo{author}{Li PJ}, \bibinfo{author}{Lehr C},
  \bibinfo{author}{Lindberg S}, \bibinfo{author}{Liu Y},
  \bibinfo{author}{Marqu\'es FM}, \bibinfo{author}{Masuoka S},
  \bibinfo{author}{Matsumoto M}, \bibinfo{author}{Mayer J},
  \bibinfo{author}{Miki K}, \bibinfo{author}{Monteagudo B},
  \bibinfo{author}{Nakamura T}, \bibinfo{author}{Nilsson T},
  \bibinfo{author}{Obertelli A}, \bibinfo{author}{Orr NA},
  \bibinfo{author}{Otsu H}, \bibinfo{author}{Park SY}, \bibinfo{author}{Parlog
  M}, \bibinfo{author}{Potlog PM}, \bibinfo{author}{Reichert S},
  \bibinfo{author}{Revel A}, \bibinfo{author}{Saito AT},
  \bibinfo{author}{Sasano M}, \bibinfo{author}{Scheit H},
  \bibinfo{author}{Schindler F}, \bibinfo{author}{Shimoura S},
  \bibinfo{author}{Simon H}, \bibinfo{author}{Stuhl L}, \bibinfo{author}{Suzuki
  H}, \bibinfo{author}{Symochko D}, \bibinfo{author}{Takeda H},
  \bibinfo{author}{Tanaka J}, \bibinfo{author}{Togano Y},
  \bibinfo{author}{Tomai T}, \bibinfo{author}{T\"ornqvist HT},
  \bibinfo{author}{Tscheuschner J}, \bibinfo{author}{Uesaka T},
  \bibinfo{author}{Wagner V}, \bibinfo{author}{Yamada H}, \bibinfo{author}{Yang
  B}, \bibinfo{author}{Yang L}, \bibinfo{author}{Yang ZH},
  \bibinfo{author}{Yasuda M}, \bibinfo{author}{Yoneda K},
  \bibinfo{author}{Zanetti L}, \bibinfo{author}{Zenihiro J} and
  \bibinfo{author}{Zhukov MV} (\bibinfo{year}{2022}).
\bibinfo{title}{Observation of a correlated free four-neutron system}.
\bibinfo{journal}{{\em Nature}} \bibinfo{volume}{606}: \bibinfo{pages}{678}.
\bibinfo{url}{\url{https://doi.org/10.1038/s41586-022-04827-6}}.

\bibtype{Article}%
\bibitem[Duguet et al.(2024)]{Duguet:2023wuh}
\bibinfo{author}{Duguet T}, \bibinfo{author}{Ekstr{\"o}m A},
  \bibinfo{author}{Furnstahl RJ}, \bibinfo{author}{K{\"o}nig S} and
  \bibinfo{author}{Lee D} (\bibinfo{year}{2024}).
\bibinfo{title}{{Colloquium: Eigenvector continuation and projection-based
  emulators}}.
\bibinfo{journal}{{\em Rev. Mod. Phys.}} \bibinfo{volume}{96}
  (\bibinfo{number}{3}): \bibinfo{pages}{031002}.
  \bibinfo{doi}{\doi{10.1103/RevModPhys.96.031002}}.
\eprint{2310.19419}.

\bibtype{Article}%
\bibitem[Dukelsky and Dussel(1999)]{dukelsky99_2004}
\bibinfo{author}{Dukelsky J} and  \bibinfo{author}{Dussel GG}
  (\bibinfo{year}{1999}).
\bibinfo{title}{Application of the density matrix renormalization group to the
  two level pairing model}.
\bibinfo{journal}{{\em Phys. Rev. C}} \bibinfo{volume}{59}:
  \bibinfo{pages}{R3005(R)}.
\bibinfo{url}{\url{https://doi.org/10.1103/PhysRevC.59.R3005}}.

\bibtype{Article}%
\bibitem[Dukelsky and Pittel(2001)]{dukelsky01_2003}
\bibinfo{author}{Dukelsky J} and  \bibinfo{author}{Pittel S}
  (\bibinfo{year}{2001}).
\bibinfo{title}{New approach to large-scale nuclear structure calculations}.
\bibinfo{journal}{{\em Phys. Rev. C}} \bibinfo{volume}{63}:
  \bibinfo{pages}{061303(R)}.
\bibinfo{url}{\url{https://doi.org/10.1103/PhysRevC.63.061303}}.

\bibtype{Article}%
\bibitem[Dukelsky et al.(2002)]{dukelsky02_1568}
\bibinfo{author}{Dukelsky J}, \bibinfo{author}{Pittel S},
  \bibinfo{author}{Dimitrova SS} and  \bibinfo{author}{Stoitsov MV}
  (\bibinfo{year}{2002}).
\bibinfo{title}{Density matrix renormalization group method and large-scale
  nuclear shell-model calculations}.
\bibinfo{journal}{{\em Phys. Rev. C}} \bibinfo{volume}{65}:
  \bibinfo{pages}{054319}.
\bibinfo{url}{\url{http://doi.org/10.1103/PhysRevC.65.054319}}.

\bibtype{Article}%
\bibitem[Efimov(1970)]{efimov1970energy}
\bibinfo{author}{Efimov V} (\bibinfo{year}{1970}).
\bibinfo{title}{Energy levels arising from resonant two-body forces in a
  three-body system}.
\bibinfo{journal}{{\em Physics Letters B}} \bibinfo{volume}{33}
  (\bibinfo{number}{8}): \bibinfo{pages}{563--564}.

\bibtype{Article}%
\bibitem[Epelbaum et al.(2009)]{epelbaum09_866}
\bibinfo{author}{Epelbaum E}, \bibinfo{author}{Hammer HW} and
  \bibinfo{author}{Mei{\ss}ner U} (\bibinfo{year}{2009}).
\bibinfo{title}{Modern theory of nuclear forces}.
\bibinfo{journal}{{\em Rev. Mod. Phys.}} \bibinfo{volume}{81}:
  \bibinfo{pages}{1773}.
\bibinfo{url}{\url{https://doi.org/10.1103/RevModPhys.81.1773}}.

\bibtype{Article}%
\bibitem[Feshbach(1962)]{feshbach1962unified}
\bibinfo{author}{Feshbach H} (\bibinfo{year}{1962}).
\bibinfo{title}{A unified theory of nuclear reactions. ii}.
\bibinfo{journal}{{\em Annals of Physics}} \bibinfo{volume}{19}
  (\bibinfo{number}{2}): \bibinfo{pages}{287--313}.

\bibtype{Book}%
\bibitem[Feshbach(1992)]{Feshbach:1992}
\bibinfo{author}{Feshbach H} (\bibinfo{year}{1992}).
\bibinfo{title}{Theoretical Nuclear Physics: Nuclear Reactions},
  \bibinfo{publisher}{{Wiley}}, \bibinfo{address}{{New York}}.
\bibinfo{comment}{ISBN} \bibinfo{isbn}{978-0-471-05750-5}.

\bibtype{Article}%
\bibitem[Fliessbach and Walliser(1982)]{fliessbach82_638}
\bibinfo{author}{Fliessbach T} and  \bibinfo{author}{Walliser H}
  (\bibinfo{year}{1982}).
\bibinfo{title}{The structure of the resonating group equation}.
\bibinfo{journal}{{\em Nucl. Phys. A}} \bibinfo{volume}{377}:
  \bibinfo{pages}{84}.
\bibinfo{url}{\url{https://doi.org/10.1016/0375-9474(82)90322-0}}.

\bibtype{Article}%
\bibitem[Fossez and Rotureau(2022)]{fossez22_2540}
\bibinfo{author}{Fossez K} and  \bibinfo{author}{Rotureau J}
  (\bibinfo{year}{2022}).
\bibinfo{title}{Density matrix renormalization group description of the island
  of inversion isotopes ${ {}^{28-33}\text{F} }$}.
\bibinfo{journal}{{\em Phys. Rev. C}} \bibinfo{volume}{106}:
  \bibinfo{pages}{034312}.
\bibinfo{url}{\url{https://doi.org/10.1103/PhysRevC.106.034312}}.

\bibtype{Article}%
\bibitem[Fossez et al.(2017)]{fossez17_1916}
\bibinfo{author}{Fossez K}, \bibinfo{author}{Rotureau J},
  \bibinfo{author}{Michel N} and  \bibinfo{author}{P{\l}oszajczak M}
  (\bibinfo{year}{2017}).
\bibinfo{title}{Can tetraneutron be a narrow resonance?}
\bibinfo{journal}{{\em Phys. Rev. Lett.}} \bibinfo{volume}{119}:
  \bibinfo{pages}{032501}.
\bibinfo{url}{\url{https://doi.org/10.1103/PhysRevLett.119.032501}}.

\bibtype{Article}%
\bibitem[Gamow(1928)]{Gamow:1928zz}
\bibinfo{author}{Gamow G} (\bibinfo{year}{1928}).
\bibinfo{title}{Zur {Quantentheorie} des {Atomkernes}}.
\bibinfo{journal}{{\em Zeitschrift für Physik}} \bibinfo{volume}{51}
  (\bibinfo{number}{3}): \bibinfo{pages}{204--212}.
ISSN \bibinfo{issn}{0044-3328}. \bibinfo{doi}{\doi{10.1007/BF01343196}}.

\bibtype{Article}%
\bibitem[Gelman(2009)]{Gelman:2009be}
\bibinfo{author}{Gelman BA} (\bibinfo{year}{2009}).
\bibinfo{title}{Narrow resonances and short-range interactions}.
\bibinfo{journal}{{\em Phys. Rev. C}} \bibinfo{volume}{80}
  (\bibinfo{number}{3}): \bibinfo{pages}{034005}.
  \bibinfo{doi}{\doi{10.1103/PhysRevC.80.034005}}.

\bibtype{Article}%
\bibitem[Glockle(1978)]{Glockle:1978zz}
\bibinfo{author}{Glockle W} (\bibinfo{year}{1978}).
\bibinfo{title}{{S-matrix pole trajectory in a three-neutron model}}.
\bibinfo{journal}{{\em Phys. Rev. C}} \bibinfo{volume}{18}:
  \bibinfo{pages}{564--572}. \bibinfo{doi}{\doi{10.1103/PhysRevC.18.564}}.

\bibtype{Book}%
\bibitem[Gl{\"o}ckle(1983)]{Gloeckle:1983}
\bibinfo{author}{Gl{\"o}ckle W} (\bibinfo{year}{1983}).
\bibinfo{title}{The Quantum Mechanical Few-Body Problem},
  \bibinfo{publisher}{{Springer}}, \bibinfo{address}{{Berlin; New York}}.

\bibtype{Article}%
\bibitem[Gu et al.(2023)]{gu23_3201}
\bibinfo{author}{Gu G}, \bibinfo{author}{Sun ZH}, \bibinfo{author}{Hagen G} and
   \bibinfo{author}{Papenbrock T} (\bibinfo{year}{2023}).
\bibinfo{title}{Entanglement entropy of nuclear systems}.
\bibinfo{journal}{{\em Phys. Rev. C}} \bibinfo{volume}{108}:
  \bibinfo{pages}{054309}.
\bibinfo{url}{\url{https://doi.org/10.1103/PhysRevC.108.054309}}.

\bibtype{Article}%
\bibitem[Habashi et al.(2020)]{Habashi:2020qgw}
\bibinfo{author}{Habashi JB}, \bibinfo{author}{Sen S}, \bibinfo{author}{Fleming
  S} and  \bibinfo{author}{van Kolck U} (\bibinfo{year}{2020}).
\bibinfo{title}{{Effective Field Theory for Two-Body Systems with Shallow
  S-Wave Resonances}}.
\bibinfo{journal}{{\em Annals Phys.}} \bibinfo{volume}{422}:
  \bibinfo{pages}{168283}. \bibinfo{doi}{\doi{10.1016/j.aop.2020.168283}}.
\eprint{2007.07360}.

\bibtype{Article}%
\bibitem[Hammer et al.(2017)]{Hammer:2017tjm}
\bibinfo{author}{Hammer HW}, \bibinfo{author}{Ji C} and
  \bibinfo{author}{Phillips DR} (\bibinfo{year}{2017}).
\bibinfo{title}{Effective field theory description of halo nuclei}.
\bibinfo{journal}{{\em J. Phys. G}} \bibinfo{volume}{44}
  (\bibinfo{number}{10}): \bibinfo{pages}{103002}.
ISSN \bibinfo{issn}{0954-3899}. \bibinfo{doi}{\doi{10.1088/1361-6471/aa83db}}.

\bibtype{Article}%
\bibitem[Hammer et al.(2020)]{Hammer:2019poc}
\bibinfo{author}{Hammer HW}, \bibinfo{author}{K{\"o}nig S} and
  \bibinfo{author}{van Kolck U} (\bibinfo{year}{2020}).
\bibinfo{title}{{Nuclear effective field theory: status and perspectives}}.
\bibinfo{journal}{{\em Rev. Mod. Phys.}} \bibinfo{volume}{92}
  (\bibinfo{number}{2}): \bibinfo{pages}{025004}.
  \bibinfo{doi}{\doi{10.1103/RevModPhys.92.025004}}.
\eprint{1906.12122}.

\bibtype{Article}%
\bibitem[Hazi and Taylor(1970)]{Hazi:1970aa}
\bibinfo{author}{Hazi AU} and  \bibinfo{author}{Taylor HS}
  (\bibinfo{year}{1970}).
\bibinfo{title}{Stabilization {{Method}} of {{Calculating Resonance Energies}}:
  {{Model Problem}}}.
\bibinfo{journal}{{\em Phys. Rev. A}} \bibinfo{volume}{1}
  (\bibinfo{number}{4}): \bibinfo{pages}{1109--1120}.
  \bibinfo{doi}{\doi{10.1103/PhysRevA.1.1109}}.

\bibtype{Article}%
\bibitem[Hebborn et al.(2023)]{Hebborn:2022vzm}
\bibinfo{author}{Hebborn C} and  et al. (\bibinfo{year}{2023}).
\bibinfo{title}{{Optical potentials for the rare-isotope beam era}}.
\bibinfo{journal}{{\em J. Phys. G}} \bibinfo{volume}{50} (\bibinfo{number}{6}):
  \bibinfo{pages}{060501}. \bibinfo{doi}{\doi{10.1088/1361-6471/acc348}}.
\eprint{2210.07293}.

\bibtype{Article}%
\bibitem[Higa et al.(2008)]{Higa:2008dn}
\bibinfo{author}{Higa R}, \bibinfo{author}{Hammer HW} and  \bibinfo{author}{van
  Kolck U} (\bibinfo{year}{2008}).
\bibinfo{title}{{alpha alpha Scattering in Halo Effective Field Theory}}.
\bibinfo{journal}{{\em Nucl. Phys. A}} \bibinfo{volume}{809}:
  \bibinfo{pages}{171--188}.
  \bibinfo{doi}{\doi{10.1016/j.nuclphysa.2008.06.003}}.
\eprint{0802.3426}.

\bibtype{Article}%
\bibitem[Higgins et al.(2020)]{higgins20_2495}
\bibinfo{author}{Higgins MD}, \bibinfo{author}{Greene CH},
  \bibinfo{author}{Kievsky A} and  \bibinfo{author}{Viviani M}
  (\bibinfo{year}{2020}).
\bibinfo{title}{Nonresonant density of states enhancement at low energies for
  three or four neutrons}.
\bibinfo{journal}{{\em Phys. Rev. Lett.}} \bibinfo{volume}{125}:
  \bibinfo{pages}{052501}.
\bibinfo{url}{\url{https://doi.org/10.1103/PhysRevLett.125.052501}}.

\bibtype{Article}%
\bibitem[Hiyama et al.(2016)]{hiyama16_1624}
\bibinfo{author}{Hiyama E}, \bibinfo{author}{Lazauskas R},
  \bibinfo{author}{Carbonell J} and  \bibinfo{author}{Kamimura M}
  (\bibinfo{year}{2016}).
\bibinfo{title}{Possibility of generating a 4-neutron resonance with a ${ T =
  3/2 }$ isospin 3-neutron force}.
\bibinfo{journal}{{\em Phys. Rev. C}} \bibinfo{volume}{93}:
  \bibinfo{pages}{044004}.
\bibinfo{url}{\url{http://doi.org/10.1103/PhysRevC.93.044004}}.

\bibtype{Article}%
\bibitem[Kisamori et al.(2016)]{kisamori16_1463}
\bibinfo{author}{Kisamori K}, \bibinfo{author}{Shimoura S},
  \bibinfo{author}{Miya H}, \bibinfo{author}{Michimasa S}, \bibinfo{author}{Ota
  S}, \bibinfo{author}{Assie M}, \bibinfo{author}{Baba H},
  \bibinfo{author}{Baba T}, \bibinfo{author}{Beaumel D},
  \bibinfo{author}{Donozo M}, \bibinfo{author}{Fujii T},
  \bibinfo{author}{Fukuda N}, \bibinfo{author}{Go S}, \bibinfo{author}{Hammache
  F}, \bibinfo{author}{Ideguchi E}, \bibinfo{author}{Inabe N},
  \bibinfo{author}{Itoh M}, \bibinfo{author}{Kameda D}, \bibinfo{author}{Kawase
  S}, \bibinfo{author}{Kawabata T}, \bibinfo{author}{Kobayashi M},
  \bibinfo{author}{Kondo Y}, \bibinfo{author}{Kubo T}, \bibinfo{author}{Kubota
  Y}, \bibinfo{author}{{Kurata-Nishimura} M}, \bibinfo{author}{Lee CS},
  \bibinfo{author}{Maeda Y}, \bibinfo{author}{Matsubara H},
  \bibinfo{author}{Miki K}, \bibinfo{author}{Nishi T}, \bibinfo{author}{Noji
  S}, \bibinfo{author}{Sakaguchi S}, \bibinfo{author}{Sakai H},
  \bibinfo{author}{Sasamoto Y}, \bibinfo{author}{Sasano M},
  \bibinfo{author}{Sato H}, \bibinfo{author}{Shimizu Y}, \bibinfo{author}{Stolz
  A}, \bibinfo{author}{Suzuki H}, \bibinfo{author}{Takaki M},
  \bibinfo{author}{Takeda H}, \bibinfo{author}{Takeuchi S},
  \bibinfo{author}{Tamii A}, \bibinfo{author}{Tang L}, \bibinfo{author}{Tokieda
  H}, \bibinfo{author}{Tsumura M}, \bibinfo{author}{Uesaka T},
  \bibinfo{author}{Yako K}, \bibinfo{author}{Yanagisawa Y},
  \bibinfo{author}{Yokoyama R} and  \bibinfo{author}{Yoshida K}
  (\bibinfo{year}{2016}).
\bibinfo{title}{Candidate resonant tetraneutron state populated by the ${
  {}^{4}\text{He} ({}^{8}\text{He} , {}^{8}\text{Be}) }$ reaction}.
\bibinfo{journal}{{\em Phys. Rev. Lett.}} \bibinfo{volume}{116}:
  \bibinfo{pages}{052501}.
\bibinfo{url}{\url{http://doi.org/10.1103/PhysRevLett.116.052501}}.

\bibtype{Article}%
\bibitem[Klos et al.(2018)]{Klos:2018sen}
\bibinfo{author}{Klos P}, \bibinfo{author}{K{\"o}nig S},
  \bibinfo{author}{Hammer HW}, \bibinfo{author}{Lynn JE} and
  \bibinfo{author}{Schwenk A} (\bibinfo{year}{2018}).
\bibinfo{title}{Signatures of few-body resonances in finite volume}.
\bibinfo{journal}{{\em Phys. Rev. C}} \bibinfo{volume}{98}
  (\bibinfo{number}{3}): \bibinfo{pages}{034004}.
  \bibinfo{doi}{\doi{10.1103/PhysRevC.98.034004}}.

\bibtype{Article}%
\bibitem[Kok(1980)]{kok80_1994}
\bibinfo{author}{Kok LP} (\bibinfo{year}{1980}).
\bibinfo{title}{Accurate determination of the ground-state level of the ${
  {}^{2}\text{He} }$ nucleus}.
\bibinfo{journal}{{\em Phys. Rev. Lett.}} \bibinfo{volume}{45}:
  \bibinfo{pages}{427}.
\bibinfo{url}{\url{https://doi.org/10.1103/PhysRevLett.45.427}}.

\bibtype{Article}%
\bibitem[Kondo et al.(2023)]{kondo23_2923}
\bibinfo{author}{Kondo Y}, \bibinfo{author}{Achouri NL}, \bibinfo{author}{{Al
  Falou} H}, \bibinfo{author}{Atar L}, \bibinfo{author}{Aumann T},
  \bibinfo{author}{Baba H}, \bibinfo{author}{Boretzky K},
  \bibinfo{author}{Caesar C}, \bibinfo{author}{Calvet D}, \bibinfo{author}{Chae
  H}, \bibinfo{author}{Chiga N}, \bibinfo{author}{Corsi A},
  \bibinfo{author}{Delaunay F}, \bibinfo{author}{Delbart A},
  \bibinfo{author}{Deshayes Q}, \bibinfo{author}{Dombr\'adi Z},
  \bibinfo{author}{Douma CA}, \bibinfo{author}{Ekstr\"om A},
  \bibinfo{author}{Elekes Z}, \bibinfo{author}{Forss\'en C},
  \bibinfo{author}{Ga\v{s}pari\'c I}, \bibinfo{author}{Gheller JM},
  \bibinfo{author}{Gibelin J}, \bibinfo{author}{Gillibert A},
  \bibinfo{author}{Hagen G}, \bibinfo{author}{Harakeh MN},
  \bibinfo{author}{Hirayama A}, \bibinfo{author}{Hoffman CR},
  \bibinfo{author}{Holl M}, \bibinfo{author}{Horvat A},
  \bibinfo{author}{Horv\'ath A}, \bibinfo{author}{Hwang JW},
  \bibinfo{author}{Isobe T}, \bibinfo{author}{Jiang WG},
  \bibinfo{author}{Kahlbow J}, \bibinfo{author}{{Kalantar-Nayestanaki } N},
  \bibinfo{author}{Kawase S}, \bibinfo{author}{Kim S},
  \bibinfo{author}{Kisamori K}, \bibinfo{author}{Kobayashi T},
  \bibinfo{author}{K\"orper D}, \bibinfo{author}{Koyama S},
  \bibinfo{author}{Kuti I}, \bibinfo{author}{Lapoux V},
  \bibinfo{author}{Lindberg S}, \bibinfo{author}{Marqu\'es FM},
  \bibinfo{author}{Masuoka S}, \bibinfo{author}{Mayer J}, \bibinfo{author}{Miki
  K}, \bibinfo{author}{Murakami T}, \bibinfo{author}{Najafi M},
  \bibinfo{author}{Nakamura T}, \bibinfo{author}{Nakano K},
  \bibinfo{author}{Nakatsuka N}, \bibinfo{author}{Nilsson T},
  \bibinfo{author}{Obertelli A}, \bibinfo{author}{Ogata K},
  \bibinfo{author}{Orr FdNA}, \bibinfo{author}{Otsu H}, \bibinfo{author}{Otsuka
  T}, \bibinfo{author}{Ozaki T}, \bibinfo{author}{Panin V},
  \bibinfo{author}{Papenbrock T}, \bibinfo{author}{Paschalis S},
  \bibinfo{author}{Revel A}, \bibinfo{author}{Rossi D}, \bibinfo{author}{Saito
  AT}, \bibinfo{author}{Saito TY}, \bibinfo{author}{Sasano M},
  \bibinfo{author}{Sato H}, \bibinfo{author}{Satou Y}, \bibinfo{author}{Scheit
  H}, \bibinfo{author}{Schindler F}, \bibinfo{author}{Schrock P},
  \bibinfo{author}{Shikata M}, \bibinfo{author}{Shimizu N},
  \bibinfo{author}{Shimizu Y}, \bibinfo{author}{Simon H},
  \bibinfo{author}{Sohler D}, \bibinfo{author}{Sorlin O},
  \bibinfo{author}{Stuhl L}, \bibinfo{author}{Sun ZH},
  \bibinfo{author}{Takeuchi S}, \bibinfo{author}{Tanaka M},
  \bibinfo{author}{Thoennessen M}, \bibinfo{author}{T\"ornqvist H},
  \bibinfo{author}{Togano Y}, \bibinfo{author}{Tomai T},
  \bibinfo{author}{Tscheuschner J}, \bibinfo{author}{Tsubota J},
  \bibinfo{author}{Tsunoda N}, \bibinfo{author}{Uesaka T},
  \bibinfo{author}{Utsuno Y}, \bibinfo{author}{Vernon I}, \bibinfo{author}{Wang
  H}, \bibinfo{author}{Yang Z}, \bibinfo{author}{Yasuda M},
  \bibinfo{author}{Yoneda K} and  \bibinfo{author}{Yoshida S}
  (\bibinfo{year}{2023}).
\bibinfo{title}{First observation of ${ {}^{28}\text{O} }$}.
\bibinfo{journal}{{\em Nature}} \bibinfo{volume}{620}: \bibinfo{pages}{965}.
\bibinfo{url}{\url{https://doi.org/10.1038/s41586-023-06352-6}}.

\bibtype{Article}%
\bibitem[Kukulin and Krasnopol'sky(1977)]{kukulin1977description}
\bibinfo{author}{Kukulin V} and  \bibinfo{author}{Krasnopol'sky V}
  (\bibinfo{year}{1977}).
\bibinfo{title}{Description of few-body systems via analytical continuation in
  coupling constant}.
\bibinfo{journal}{{\em Journal of physics A: Mathematical and general}}
  \bibinfo{volume}{10} (\bibinfo{number}{2}): \bibinfo{pages}{L33--L37}.

\bibtype{Book}%
\bibitem[Kukulin et al.(1989)]{Kukulin:1989}
\bibinfo{author}{Kukulin VI}, \bibinfo{author}{Krasnopol'sky VM} and
  \bibinfo{author}{Hor{\'a}{\v c}ek J} (\bibinfo{year}{1989}).
\bibinfo{title}{Theory of Resonances: Principles and Applications},
  \bibinfo{edition}{springer reprint} ed., \bibinfo{series}{Reidel {{Texts}} in
  the {{Mathematical Sciences}}}, \bibinfo{publisher}{{Kluwer}}.
\bibinfo{comment}{ISBN} \bibinfo{isbn}{978-90-481-8432-3}.

\bibtype{Article}%
\bibitem[Lazauskas(2018)]{lazauskas18_2032}
\bibinfo{author}{Lazauskas R} (\bibinfo{year}{2018}).
\bibinfo{title}{Description of five-nucleon systems using
  {F}addeev-{Y}akubovsky equations}.
\bibinfo{journal}{{\em Few-Body Syst.}} \bibinfo{volume}{59}:
  \bibinfo{pages}{13}.
\bibinfo{url}{\url{https://doi.org/10.1007/s00601-018-1333-7}}.

\bibtype{Article}%
\bibitem[Lazauskas et al.(2019)]{lazauskas19_2363}
\bibinfo{author}{Lazauskas R}, \bibinfo{author}{Hiyama E} and
  \bibinfo{author}{Carbonell J} (\bibinfo{year}{2019}).
\bibinfo{title}{\textit{Ab initio} calculations of ${ {}^{5}\text{H} }$
  resonant states}.
\bibinfo{journal}{{\em Phys. Lett. B}} \bibinfo{volume}{791}:
  \bibinfo{pages}{335}.
\bibinfo{url}{\url{https://doi.org/10.1016/j.physletb.2019.02.047}}.

\bibtype{Article}%
\bibitem[Lazauskas et al.(2023)]{lazauskas23_2744}
\bibinfo{author}{Lazauskas R}, \bibinfo{author}{Hiyama E} and
  \bibinfo{author}{Carbonell J} (\bibinfo{year}{2023}).
\bibinfo{title}{Low energy structures in nuclear reactions with ${ 4n }$ in the
  final state}.
\bibinfo{journal}{{\em Phys. Rev. Lett.}} \bibinfo{volume}{130}:
  \bibinfo{pages}{102501}.
\bibinfo{url}{\url{https://doi.org/10.1103/PhysRevLett.130.102501}}.

\bibtype{Article}%
\bibitem[Li et al.(2019)]{li19_2634}
\bibinfo{author}{Li JG}, \bibinfo{author}{Michel N}, \bibinfo{author}{Hu BS},
  \bibinfo{author}{Zuo W} and  \bibinfo{author}{Xu FR} (\bibinfo{year}{2019}).
\bibinfo{title}{\textit{Ab initio} no-core {G}amow shell-model calculations of
  multineutron systems}.
\bibinfo{journal}{{\em Phys. Rev. C}} \bibinfo{volume}{100}:
  \bibinfo{pages}{054313}.
\bibinfo{url}{\url{https://doi.org/10.1103/PhysRevC.100.054313}}.

\bibtype{Article}%
\bibitem[L\"owdin(1955)]{lowdin55_2499}
\bibinfo{author}{L\"owdin PO} (\bibinfo{year}{1955}).
\bibinfo{title}{Quantum theory of many-particle systems. {I}. {P}hysical
  interpretations by means of density matrices, natural spin-orbitals, and
  convergence problems in the method of configuration interaction}.
\bibinfo{journal}{{\em Phys. Rev.}} \bibinfo{volume}{97}:
  \bibinfo{pages}{1474}.
\bibinfo{url}{\url{https://doi.org/10.1103/PhysRev.97.1474}}.

\bibtype{Article}%
\bibitem[L\"owdin and Shull(1956)]{lowdin56_2498}
\bibinfo{author}{L\"owdin PO} and  \bibinfo{author}{Shull H}
  (\bibinfo{year}{1956}).
\bibinfo{title}{Natural orbitals in the quantum theory of two-electron
  systems}.
\bibinfo{journal}{{\em Phys. Rev.}} \bibinfo{volume}{101}:
  \bibinfo{pages}{1730}.
\bibinfo{url}{\url{https://doi.org/10.1103/PhysRev.101.1730}}.

\bibtype{Book}%
\bibitem[L{\"u}ders and Pohl(2017)]{Pohl:2017}
\bibinfo{author}{L{\"u}ders K} and  \bibinfo{author}{Pohl RO}
  (\bibinfo{year}{2017}).
\bibinfo{title}{Pohl's {{Introduction}} to {{Physics}} : {{Volume}} 1:
  {{Mechanics}}, {{Acoustics}} and {{Thermodynamics}}},
  \bibinfo{publisher}{Springer International Publishing}.
\bibinfo{comment}{ISBN} \bibinfo{isbn}{978-3-319-40046-4}.

\bibtype{Article}%
\bibitem[Lüscher(1986{\natexlab{a}})]{Luscher:1985dn}
\bibinfo{author}{Lüscher M} (\bibinfo{year}{1986}{\natexlab{a}}).
\bibinfo{title}{Volume dependence of the energy spectrum in massive quantum
  field theories - {I}. {Stable} particle states}.
\bibinfo{journal}{{\em Comm. Math. Phys.}} \bibinfo{volume}{104}
  (\bibinfo{number}{2}): \bibinfo{pages}{177--206}.
ISSN \bibinfo{issn}{1432-0916}. \bibinfo{doi}{\doi{10.1007/BF01211589}}.

\bibtype{Article}%
\bibitem[Lüscher(1986{\natexlab{b}})]{Luscher:1986pf}
\bibinfo{author}{Lüscher M} (\bibinfo{year}{1986}{\natexlab{b}}).
\bibinfo{title}{Volume dependence of the energy spectrum in massive quantum
  field theories - {II}. {Scattering} states}.
\bibinfo{journal}{{\em Comm. Math. Phys.}} \bibinfo{volume}{105}
  (\bibinfo{number}{2}): \bibinfo{pages}{153--188}.
ISSN \bibinfo{issn}{1432-0916}. \bibinfo{doi}{\doi{10.1007/BF01211097}}.

\bibtype{Article}%
\bibitem[Lüscher(1991)]{Luscher:1990ux}
\bibinfo{author}{Lüscher M} (\bibinfo{year}{1991}).
\bibinfo{title}{Two-particle states on a torus and their relation to the
  scattering matrix}.
\bibinfo{journal}{{\em Nucl. Phys. B}} \bibinfo{volume}{354}
  (\bibinfo{number}{2-3}): \bibinfo{pages}{531--578}.
ISSN \bibinfo{issn}{0550-3213}.
  \bibinfo{doi}{\doi{10.1016/0550-3213(91)90366-6}}.

\bibtype{Article}%
\bibitem[Machleidt and Entem(2011)]{machleidt11_414}
\bibinfo{author}{Machleidt R} and  \bibinfo{author}{Entem DR}
  (\bibinfo{year}{2011}).
\bibinfo{title}{Chiral effective field theory and nuclear forces}.
\bibinfo{journal}{{\em Phys. Rep.}} \bibinfo{volume}{503}: \bibinfo{pages}{1}.
\bibinfo{url}{\url{https://doi.org/10.1016/j.physrep.2011.02.001}}.

\bibtype{Article}%
\bibitem[Mai et al.(2023)]{Mai:2022eur}
\bibinfo{author}{Mai M}, \bibinfo{author}{Mei{\ss}ner UG} and
  \bibinfo{author}{Urbach C} (\bibinfo{year}{2023}).
\bibinfo{title}{{Towards a theory of hadron resonances}}.
\bibinfo{journal}{{\em Phys. Rept.}} \bibinfo{volume}{1001}:
  \bibinfo{pages}{1--66}. \bibinfo{doi}{\doi{10.1016/j.physrep.2022.11.005}}.
\eprint{2206.01477}.

\bibtype{Article}%
\bibitem[Manolopoulos(2002)]{Manolopoulos_ABP}
\bibinfo{author}{Manolopoulos DE} (\bibinfo{year}{2002}), \bibinfo{month}{12}.
\bibinfo{title}{Derivation and reflection properties of a transmission-free
  absorbing potential}.
\bibinfo{journal}{{\em The Journal of Chemical Physics}} \bibinfo{volume}{117}
  (\bibinfo{number}{21}): \bibinfo{pages}{9552--9559}.
ISSN \bibinfo{issn}{0021-9606}. \bibinfo{doi}{\doi{10.1063/1.1517042}}.
\bibinfo{url}{\url{https://doi.org/10.1063/1.1517042}}.

\bibtype{Article}%
\bibitem[Marqu\'es et al.(2002)]{marques02_1460}
\bibinfo{author}{Marqu\'es FM}, \bibinfo{author}{Labiche M},
  \bibinfo{author}{Orr NA}, \bibinfo{author}{Ang\'elique JC},
  \bibinfo{author}{Axelsson L}, \bibinfo{author}{Benoit B},
  \bibinfo{author}{Bergmann UC}, \bibinfo{author}{Borge MJG},
  \bibinfo{author}{Catford WN}, \bibinfo{author}{Chappell SPG},
  \bibinfo{author}{Clarke NM}, \bibinfo{author}{Costa G},
  \bibinfo{author}{Curtis N}, \bibinfo{author}{{D'Arrigo} A},
  \bibinfo{author}{Brennand Ed}, \bibinfo{author}{{de Oliveira Santos} F},
  \bibinfo{author}{Dorvaux O}, \bibinfo{author}{Fazio G},
  \bibinfo{author}{Freer M}, \bibinfo{author}{Fulton BR},
  \bibinfo{author}{Giardina G}, \bibinfo{author}{Gr\'evy S},
  \bibinfo{author}{Guillemaud-Mueller D}, \bibinfo{author}{Hanappe F},
  \bibinfo{author}{Heusch B}, \bibinfo{author}{Jonson B}, \bibinfo{author}{{Le
  Brun} C}, \bibinfo{author}{Leenhardt S}, \bibinfo{author}{Lewitowicz M},
  \bibinfo{author}{L\'opez MJ}, \bibinfo{author}{Markenroth K},
  \bibinfo{author}{Mueller AC}, \bibinfo{author}{Nilsson T},
  \bibinfo{author}{Ninane A}, \bibinfo{author}{Nyman G},
  \bibinfo{author}{Piqueras I}, \bibinfo{author}{Riisager K},
  \bibinfo{author}{{{Saint Laurent}} MG}, \bibinfo{author}{Sarazin F},
  \bibinfo{author}{Singer SM}, \bibinfo{author}{Sorlin O} and
  \bibinfo{author}{Stuttg\'e L} (\bibinfo{year}{2002}).
\bibinfo{title}{Detection of neutron clusters}.
\bibinfo{journal}{{\em Phys. Rev. C}} \bibinfo{volume}{65}:
  \bibinfo{pages}{044006}.
\bibinfo{url}{\url{10.1103/PhysRevC.65.044006}}.

\bibtype{Article}%
\bibitem[Martin et al.(2016)]{martin16_2292}
\bibinfo{author}{Martin D}, \bibinfo{author}{Arcones A},
  \bibinfo{author}{Nazarewicz W} and  \bibinfo{author}{Olsen E}
  (\bibinfo{year}{2016}).
\bibinfo{title}{Impact of nuclear mass uncertainties on the ${ r }$ process}.
\bibinfo{journal}{{\em Phys. Rev. Lett.}} \bibinfo{volume}{116}:
  \bibinfo{pages}{121101}.
\bibinfo{url}{\url{https://doi.org/10.1103/PhysRevLett.116.121101}}.

\bibtype{Article}%
\bibitem[Matsuo et al.(2005)]{matsuo05_3139}
\bibinfo{author}{Matsuo M}, \bibinfo{author}{Mizuyama K} and
  \bibinfo{author}{Serizawa Y} (\bibinfo{year}{2005}).
\bibinfo{title}{Di-neutron correlation and soft dipole excitation in medium
  mass neutron-rich nuclei near drip line}.
\bibinfo{journal}{{\em Phys. Rev. C}} \bibinfo{volume}{71}:
  \bibinfo{pages}{064326}.
\bibinfo{url}{\url{https://doi.org/10.1103/PhysRevC.71.064326}}.

\bibtype{Book}%
\bibitem[Michel and P{\l}oszajczak(2021)]{michel21_b260}
\bibinfo{author}{Michel N} and  \bibinfo{author}{P{\l}oszajczak M}
  (\bibinfo{year}{2021}).
\bibinfo{title}{Gamow {S}hell {M}odel}, \bibinfo{edition}{first} ed.,
  \bibinfo{publisher}{Springer}.
\bibinfo{url}{\url{https://doi.org/10.1007/978-3-030-69356-5}}.

\bibtype{Article}%
\bibitem[Michel et al.(2002)]{michel02_8}
\bibinfo{author}{Michel N}, \bibinfo{author}{Nazarewicz W},
  \bibinfo{author}{P{\l}oszajczak M} and  \bibinfo{author}{Bennaceur K}
  (\bibinfo{year}{2002}).
\bibinfo{title}{{G}amow shell model description of neutron-rich nuclei}.
\bibinfo{journal}{{\em Phys. Rev. Lett.}} \bibinfo{volume}{89}:
  \bibinfo{pages}{042502}.
\bibinfo{url}{\url{https://doi.org/10.1103/PhysRevLett.89.042502}}.

\bibtype{Article}%
\bibitem[Michel et al.(2003)]{michel03_10}
\bibinfo{author}{Michel N}, \bibinfo{author}{Nazarewicz W},
  \bibinfo{author}{P{\l}oszajczak M} and  \bibinfo{author}{Oko{\l}owicz J}
  (\bibinfo{year}{2003}).
\bibinfo{title}{Gamow shell-model description of weakly bound nuclei and
  unbound nuclear states}.
\bibinfo{journal}{{\em Phys. Rev. C}} \bibinfo{volume}{67}:
  \bibinfo{pages}{054311}.
\bibinfo{url}{\url{https://doi.org/10.1103/PhysRevC.67.054311}}.

\bibtype{Article}%
\bibitem[Michel et al.(2009)]{michel09_2}
\bibinfo{author}{Michel N}, \bibinfo{author}{Nazarewicz W},
  \bibinfo{author}{P{\l}oszajczak M} and  \bibinfo{author}{Vertse T}
  (\bibinfo{year}{2009}).
\bibinfo{title}{Shell model in the complex energy plane}.
\bibinfo{journal}{{\em J. Phys. G: Nucl. Part. Phys.}} \bibinfo{volume}{36}:
  \bibinfo{pages}{013101}.
\bibinfo{url}{\url{https://doi.org/10.1088/0954-3899/36/1/013101}}.

\bibtype{Article}%
\bibitem[Moiseyev(1998)]{moiseyev98_92}
\bibinfo{author}{Moiseyev N} (\bibinfo{year}{1998}).
\bibinfo{title}{Quantum theory of resonances: calculating energies, widths and
  cross-sections by complex scaling}.
\bibinfo{journal}{{\em Phys. Rep.}} \bibinfo{volume}{302}:
  \bibinfo{pages}{212}.
\bibinfo{url}{\url{https://doi.org/10.1016/S0370-1573(98)00002-7}}.

\bibtype{Book}%
\bibitem[Moiseyev(2011)]{Moiseyev:2011}
\bibinfo{author}{Moiseyev N} (\bibinfo{year}{2011}).
\bibinfo{title}{Non-{{Hermitian Quantum Mechanics}}},
  \bibinfo{publisher}{{Cambridge University Press}}.
\bibinfo{comment}{ISBN} \bibinfo{isbn}{978-1-139-49699-5}.

\bibtype{Article}%
\bibitem[Mumpower et al.(2016)]{mumpower16_2386}
\bibinfo{author}{Mumpower MR}, \bibinfo{author}{Surman R},
  \bibinfo{author}{{McLaughlin} GC} and  \bibinfo{author}{Aprahamian A}
  (\bibinfo{year}{2016}).
\bibinfo{title}{The impact of individual nuclear properties on ${ r }$-process
  nucleosynthesis}.
\bibinfo{journal}{{\em Prog. Part. Nucl. Phys.}} \bibinfo{volume}{86}:
  \bibinfo{pages}{86}.
\bibinfo{url}{\url{http://doi.org/10.1016/j.ppnp.2015.09.001}}.

\bibtype{Article}%
\bibitem[Myo and Kat{\=o}(2020)]{Myo:2020rni}
\bibinfo{author}{Myo T} and  \bibinfo{author}{Kat{\=o} K}
  (\bibinfo{year}{2020}).
\bibinfo{title}{Complex scaling: {{Physics}} of unbound light nuclei and
  perspective}.
\bibinfo{journal}{{\em Prog. Th. Exp. Phys.}} \bibinfo{volume}{2020}
  (\bibinfo{number}{12A101}).
ISSN \bibinfo{issn}{2050-3911}. \bibinfo{doi}{\doi{10.1093/ptep/ptaa101}}.

\bibtype{Article}%
\bibitem[Navr\'atil et al.(2016)]{navratil16_1956}
\bibinfo{author}{Navr\'atil P}, \bibinfo{author}{Quaglioni S},
  \bibinfo{author}{Hupin G}, \bibinfo{author}{{Romero-Redondo} C} and
  \bibinfo{author}{Calci A} (\bibinfo{year}{2016}).
\bibinfo{title}{Unified \textit{ab initio} approaches to nuclear structure and
  reactions}.
\bibinfo{journal}{{\em Phys. Scr.}} \bibinfo{volume}{91}:
  \bibinfo{pages}{053002}.
\bibinfo{url}{\url{http://doi.org/10.1088/0031-8949/91/5/053002}}.

\bibtype{Article}%
\bibitem[Neuhauser and Baer(1989)]{Neuhauser_ABP}
\bibinfo{author}{Neuhauser D} and  \bibinfo{author}{Baer M}
  (\bibinfo{year}{1989}), \bibinfo{month}{10}.
\bibinfo{title}{The application of wave packets to reactive atom–diatom
  systems: A new approach}.
\bibinfo{journal}{{\em The Journal of Chemical Physics}} \bibinfo{volume}{91}
  (\bibinfo{number}{8}): \bibinfo{pages}{4651--4657}.
ISSN \bibinfo{issn}{0021-9606}. \bibinfo{doi}{\doi{10.1063/1.456755}}.
\bibinfo{url}{\url{https://doi.org/10.1063/1.456755}}.

\bibtype{Book}%
\bibitem[Newton(1982)]{newton82_b6}
\bibinfo{author}{Newton RG} (\bibinfo{year}{1982}).
\bibinfo{title}{Scattering {T}heory of {W}aves and {P}articles},
  \bibinfo{edition}{second} ed., \bibinfo{publisher}{Springer-Verlag, New
  York}.

\bibtype{Book}%
\bibitem[Nussenzveig(1972)]{Nussenzveig:1972}
\bibinfo{author}{Nussenzveig HM} (\bibinfo{year}{1972}).
\bibinfo{title}{{Causality and dispersion relations}}, \bibinfo{volume}{95},
  \bibinfo{publisher}{Academic Press}, \bibinfo{address}{New York, London}.
\bibinfo{comment}{ISBN} \bibinfo{isbn}{978-0-12-523050-6, 978-0-08-095604-6}.

\bibtype{Article}%
\bibitem[Offermann and Gl{\"o}ckle(1979)]{Offermann:1979wbx}
\bibinfo{author}{Offermann R} and  \bibinfo{author}{Gl{\"o}ckle W}
  (\bibinfo{year}{1979}).
\bibinfo{title}{{Is there a three-neutron resonance?}}
\bibinfo{journal}{{\em Nucl. Phys. A}} \bibinfo{volume}{318}:
  \bibinfo{pages}{138--144}. \bibinfo{doi}{\doi{10.1016/0375-9474(79)90475-5}}.

\bibtype{Article}%
\bibitem[Oko{\l}owicz et al.(2003)]{okolowicz03_21}
\bibinfo{author}{Oko{\l}owicz J}, \bibinfo{author}{P{\l}oszajczak M} and
  \bibinfo{author}{Rotter I} (\bibinfo{year}{2003}).
\bibinfo{title}{Dynamics of quantum systems embedded in a continuum}.
\bibinfo{journal}{{\em Phys. Rep.}} \bibinfo{volume}{374}:
  \bibinfo{pages}{271}.
\bibinfo{url}{\url{https://doi.org/10.1016/S0370-1573(02)00366-6}}.

\bibtype{Article}%
\bibitem[Pazy(2023)]{pazy23_3230}
\bibinfo{author}{Pazy E} (\bibinfo{year}{2023}).
\bibinfo{title}{Entanglement entropy between short range correlations and the
  {F}ermi sea in nuclear structure}.
\bibinfo{journal}{{\em Phys. Rev. C}} \bibinfo{volume}{107}:
  \bibinfo{pages}{054308}.
\bibinfo{url}{\url{https://doi.org/10.1103/PhysRevC.107.054308}}.

\bibtype{Book}%
\bibitem[Peschel et al.(1999)]{peschel99_b245}
\bibinfo{author}{Peschel I}, \bibinfo{author}{Wang X}, \bibinfo{author}{Kaulke
  M} and  \bibinfo{author}{Hallberg K} (\bibinfo{year}{1999}).
\bibinfo{title}{Density-{M}atrix {R}enormalization - {A} {N}ew {N}umerical
  {M}ethod in {P}hysics}, \bibinfo{edition}{first} ed.,
  \bibinfo{publisher}{Springer-Verlag Berlin Heidelberg}.
\bibinfo{url}{\url{https://doi.org/10.1007/BFb0106062}}.

\bibtype{Article}%
\bibitem[Pf\"utzner et al.(2023)]{pfutzner23_2859}
\bibinfo{author}{Pf\"utzner M}, \bibinfo{author}{Mukha I} and
  \bibinfo{author}{Wang SW} (\bibinfo{year}{2023}).
\bibinfo{title}{Two-proton emission and related phenomena}.
\bibinfo{journal}{{\em Prog. Part. Nucl. Phys.}} \bibinfo{volume}{132}.
\bibinfo{url}{\url{https://doi.org/10.1016/j.ppnp.2023.104050}}.

\bibtype{Article}%
\bibitem[Pieper(2003)]{pieper03_1461}
\bibinfo{author}{Pieper SC} (\bibinfo{year}{2003}).
\bibinfo{title}{Can modern nuclear {H}amiltonians tolerate a bound
  tetraneutron?}
\bibinfo{journal}{{\em Phys. Rev. Lett.}} \bibinfo{volume}{90}:
  \bibinfo{pages}{252501}.
\bibinfo{url}{\url{https://doi.org/10.1103/PhysRevLett.90.252501}}.

\bibtype{Article}%
\bibitem[Pillet et al.(2007)]{pillet07_1879}
\bibinfo{author}{Pillet N}, \bibinfo{author}{Sandulescu N} and
  \bibinfo{author}{Schuck P} (\bibinfo{year}{2007}).
\bibinfo{title}{Generic strong coupling behavior of {C}ooper pairs on the
  surface of superfluid nuclei}.
\bibinfo{journal}{{\em Phys. Rev. C}} \bibinfo{volume}{76}:
  \bibinfo{pages}{024310}.
\bibinfo{url}{\url{http://doi.org/10.1103/PhysRevC.76.024310}}.

\bibtype{Article}%
\bibitem[Pittel and Dukelsky(2001)]{pittel01_2008}
\bibinfo{author}{Pittel S} and  \bibinfo{author}{Dukelsky J}
  (\bibinfo{year}{2001}).
\bibinfo{title}{The density matrix renormalization group: a new approach to
  large-scale nuclear structure calculations}.
\bibinfo{journal}{{\em Rev. Mex. Fis.}} \bibinfo{volume}{47 Suppl. 2}:
  \bibinfo{pages}{42}.

\bibtype{Article}%
\bibitem[Rommer and \"Ostlund(1997)]{rommer97_3246}
\bibinfo{author}{Rommer S} and  \bibinfo{author}{\"Ostlund S}
  (\bibinfo{year}{1997}).
\bibinfo{title}{Class of ansatz wave functions for one-dimensional spin systems
  and their relation to the density matrix renormalization group}.
\bibinfo{journal}{{\em Phys. Rev. B}} \bibinfo{volume}{55}:
  \bibinfo{pages}{2164}.
\bibinfo{url}{\url{https://doi.org/10.1103/PhysRevB.55.2164}}.

\bibtype{Article}%
\bibitem[Rotter(1991)]{rotter91_448}
\bibinfo{author}{Rotter I} (\bibinfo{year}{1991}).
\bibinfo{title}{A continuum shell model for the open quantum mechanical nuclear
  system}.
\bibinfo{journal}{{\em Rep. Prog. Phys.}} \bibinfo{volume}{54}:
  \bibinfo{pages}{635}.
\bibinfo{url}{\url{https://doi.org/10.1088/0034-4885/54/4/003}}.

\bibtype{Article}%
\bibitem[Rotter and Bird(2015)]{rotter15_2464}
\bibinfo{author}{Rotter I} and  \bibinfo{author}{Bird JP}
  (\bibinfo{year}{2015}).
\bibinfo{title}{A review of progress in the physics of open quantum systems:
  {T}heory and experiment}.
\bibinfo{journal}{{\em Rep. Prog. Phys.}} \bibinfo{volume}{78}:
  \bibinfo{pages}{114001}.
\bibinfo{url}{\url{http://doi.org/10.1088/0034-4885/78/11/114001}}.

\bibtype{Article}%
\bibitem[Rotureau et al.(2006)]{rotureau06_15}
\bibinfo{author}{Rotureau J}, \bibinfo{author}{Michel N},
  \bibinfo{author}{Nazarewicz W}, \bibinfo{author}{P{\l}oszajczak M} and
  \bibinfo{author}{Dukelsky J} (\bibinfo{year}{2006}).
\bibinfo{title}{Density matrix renormalisation group approach for many-body
  open quantum systems}.
\bibinfo{journal}{{\em Phys. Rev. Lett.}} \bibinfo{volume}{97}:
  \bibinfo{pages}{110603}.
\bibinfo{url}{\url{https://doi.org/10.1103/PhysRevLett.97.110603}}.

\bibtype{Article}%
\bibitem[Rotureau et al.(2009)]{rotureau09_140}
\bibinfo{author}{Rotureau J}, \bibinfo{author}{Michel N},
  \bibinfo{author}{Nazarewicz W}, \bibinfo{author}{P{\l}oszajczak M} and
  \bibinfo{author}{Dukelsky J} (\bibinfo{year}{2009}).
\bibinfo{title}{Density matrix renormalization group approach to two-fluid open
  many-fermion systems}.
\bibinfo{journal}{{\em Phys. Rev. C}} \bibinfo{volume}{79}:
  \bibinfo{pages}{014304}.
\bibinfo{url}{\url{https://doi.org/10.1103/PhysRevC.79.014304}}.

\bibtype{Article}%
\bibitem[Rummukainen and Gottlieb(1995)]{Rummukainen:1995vs}
\bibinfo{author}{Rummukainen K} and  \bibinfo{author}{Gottlieb S}
  (\bibinfo{year}{1995}).
\bibinfo{title}{Resonance scattering phase shifts on a non-rest-frame lattice}.
\bibinfo{journal}{{\em Nucl. Phys. B}} \bibinfo{volume}{450}
  (\bibinfo{number}{1}): \bibinfo{pages}{397--436}.
ISSN \bibinfo{issn}{0550-3213}.
  \bibinfo{doi}{\doi{10.1016/0550-3213(95)00313-H}}.

\bibtype{Book}%
\bibitem[Sakurai(1994)]{Sakurai:1994}
\bibinfo{author}{Sakurai JJ} (\bibinfo{year}{1994}).
\bibinfo{title}{Modern {Quantum} {Mechanics}},
  \bibinfo{publisher}{Addison-Wesley}, \bibinfo{address}{Reading, Mass.}
\bibinfo{comment}{ISBN} \bibinfo{isbn}{978-0-201-53929-5}.
\bibinfo{note}{OCLC: 28065703}.

\bibtype{Article}%
\bibitem[Schollw\"ock(2005)]{schollwock05_479}
\bibinfo{author}{Schollw\"ock U} (\bibinfo{year}{2005}).
\bibinfo{title}{The density-matrix renormalization group}.
\bibinfo{journal}{{\em Rev. Mod. Phys.}} \bibinfo{volume}{77}:
  \bibinfo{pages}{259}.
\bibinfo{url}{\url{https://doi.org/10.1103/RevModPhys.77.259}}.

\bibtype{Article}%
\bibitem[Shirokov et al.(2016)]{shirokov16_1791}
\bibinfo{author}{Shirokov AM}, \bibinfo{author}{Papadimitriou G},
  \bibinfo{author}{Mazur AI}, \bibinfo{author}{Mazur IA}, \bibinfo{author}{Roth
  R} and  \bibinfo{author}{Vary JP} (\bibinfo{year}{2016}).
\bibinfo{title}{Prediction for a four-neutron resonance}.
\bibinfo{journal}{{\em Phys. Rev. Lett.}} \bibinfo{volume}{117}:
  \bibinfo{pages}{182502}.
\bibinfo{url}{\url{https://doi.org/10.1103/PhysRevLett.117.182502}}.

\bibtype{Book}%
\bibitem[Taylor(1972)]{Taylor:1972}
\bibinfo{author}{Taylor JR} (\bibinfo{year}{1972}).
\bibinfo{title}{Scattering {{Theory}}: {{The Quantum Theory}} of
  {{Nonrelativistic Collisions}}}, \bibinfo{publisher}{{Dover}},
  \bibinfo{address}{{Newburyport}}.
\bibinfo{comment}{ISBN} \bibinfo{isbn}{0-471-84900-6}.

\bibtype{Book}%
\bibitem[Taylor(2004)]{Taylor:2004}
\bibinfo{author}{Taylor JR} (\bibinfo{year}{2004}), \bibinfo{month}{Sep.}
\bibinfo{title}{Classical {Mechanics}}, \bibinfo{publisher}{MIT Press},
  \bibinfo{address}{Cambridge, MA, USA}.
\bibinfo{comment}{ISBN} \bibinfo{isbn}{978-1-891389-22-1}.

\bibtype{Article}%
\bibitem[Thoennessen(2004)]{thoennessen04_1165}
\bibinfo{author}{Thoennessen M} (\bibinfo{year}{2004}).
\bibinfo{title}{Reaching the limits of nuclear stability}.
\bibinfo{journal}{{\em Rep. Prog. Phys.}} \bibinfo{volume}{67}:
  \bibinfo{pages}{1187}.
\bibinfo{url}{\url{https://doi.org/10.1088/0034-4885/67/7/R04}}.

\bibtype{Article}%
\bibitem[Volya and Zelevinsky(2006)]{volya06_94}
\bibinfo{author}{Volya A} and  \bibinfo{author}{Zelevinsky V}
  (\bibinfo{year}{2006}).
\bibinfo{title}{Continuum shell model}.
\bibinfo{journal}{{\em Phys. Rev. C}} \bibinfo{volume}{74}:
  \bibinfo{pages}{064314}.
\bibinfo{url}{\url{https://doi.org/10.1103/PhysRevC.74.064314}}.

\bibtype{Article}%
\bibitem[Wang et al.(2026)]{wang26_3366}
\bibinfo{author}{Wang JL}, \bibinfo{author}{Xie MR}, \bibinfo{author}{Li KH},
  \bibinfo{author}{Wang PY}, \bibinfo{author}{Michel N}, \bibinfo{author}{Yuan
  Q} and  \bibinfo{author}{Li JG} (\bibinfo{year}{2026}).
\bibinfo{title}{Gamow shell model predictions for six-proton unbound nucleus ${
  {}^{20}\text{Si} }$}.
\bibinfo{journal}{{\em Phys. Lett. B}} \bibinfo{volume}{872}:
  \bibinfo{pages}{140030}.
\bibinfo{url}{\url{https://doi.org/10.1016/j.physletb.2025.140030}}.

\bibtype{Article}%
\bibitem[Wheeler(1937)]{wheeler37_2315}
\bibinfo{author}{Wheeler JA} (\bibinfo{year}{1937}).
\bibinfo{title}{Molecular viewpoints in nuclear structure}.
\bibinfo{journal}{{\em Phys. Rev.}} \bibinfo{volume}{52}:
  \bibinfo{pages}{1083}.
\bibinfo{url}{\url{https://doi.org/10.1103/PhysRev.52.1083}}.

\bibtype{Article}%
\bibitem[White(1992)]{white92_488}
\bibinfo{author}{White SR} (\bibinfo{year}{1992}).
\bibinfo{title}{Density matrix formulation for quantum renormalization groups}.
\bibinfo{journal}{{\em Phys. Rev. Lett.}} \bibinfo{volume}{69}:
  \bibinfo{pages}{2863}.
\bibinfo{url}{\url{https://doi.org/10.1103/PhysRevLett.69.2863}}.

\bibtype{Article}%
\bibitem[White(1993)]{white93_491}
\bibinfo{author}{White SR} (\bibinfo{year}{1993}).
\bibinfo{title}{Density-matrix algorithms for quantum renormalization groups}.
\bibinfo{journal}{{\em Phys. Rev. B}} \bibinfo{volume}{48}:
  \bibinfo{pages}{10345}.
\bibinfo{url}{\url{https://doi.org/10.1103/PhysRevB.48.10345}}.

\bibtype{Article}%
\bibitem[Wiese(1989)]{Wiese:1988qy}
\bibinfo{author}{Wiese UJ} (\bibinfo{year}{1989}).
\bibinfo{title}{Identification of resonance parameters from the finite volume
  energy spectrum}.
\bibinfo{journal}{{\em Nucl. Phys. B Proc. Suppl.}} \bibinfo{volume}{9}:
  \bibinfo{pages}{609--613}.
ISSN \bibinfo{issn}{0920-5632}.
  \bibinfo{doi}{\doi{10.1016/0920-5632(89)90171-0}}.

\bibtype{Article}%
\bibitem[Yaghi et al.(2025)]{Yaghi:2025ftv}
\bibinfo{author}{Yaghi O}, \bibinfo{author}{Hupin G} and
  \bibinfo{author}{Navr{\'a}til P} (\bibinfo{year}{2025}), \bibinfo{month}{7}.
\bibinfo{title}{{Ab Initio Complex Scaling and Similarity Renormalization Group
  for Continuum Properties of Nuclei}} \eprint{2507.01595}.

\bibtype{Article}%
\bibitem[Yapa et al.(2023)]{Yapa:2023xyf}
\bibinfo{author}{Yapa N}, \bibinfo{author}{Fossez K} and
  \bibinfo{author}{K{\"o}nig S} (\bibinfo{year}{2023}).
\bibinfo{title}{{Eigenvector continuation for emulating and extrapolating
  two-body resonances}}.
\bibinfo{journal}{{\em Phys. Rev. C}} \bibinfo{volume}{107}
  (\bibinfo{number}{6}): \bibinfo{pages}{064316}.
  \bibinfo{doi}{\doi{10.1103/PhysRevC.107.064316}}.
\eprint{2303.06139}.

\bibtype{Article}%
\bibitem[Yu et al.(2024)]{yu24_3045}
\bibinfo{author}{Yu H}, \bibinfo{author}{Yapa N} and  \bibinfo{author}{K\"onig
  S} (\bibinfo{year}{2024}).
\bibinfo{title}{Complex scaling in finite volume}.
\bibinfo{journal}{{\em Phys. Rev. C}} \bibinfo{volume}{109}:
  \bibinfo{pages}{014316}.
\bibinfo{url}{\url{https://doi.org/10.1103/PhysRevC.109.014316}}.

\bibtype{Book}%
\bibitem[Zelevinsky and Volya(2023)]{zelevinsky23_b337}
\bibinfo{author}{Zelevinsky V} and  \bibinfo{author}{Volya A}
  (\bibinfo{year}{2023}).
\bibinfo{title}{Mesoscopic {N}uclear {P}hysics}, \bibinfo{edition}{first} ed.,
  \bibinfo{publisher}{World Scientific}.
\bibinfo{url}{\url{https://doi.org/10.1142/13049}}.

\end{thebibliography*}

\end{document}